\documentclass[graybox]{svmult}

\usepackage{mathptmx}       
\usepackage{helvet}         
\usepackage{courier}        
\usepackage{type1cm}        
\usepackage{makeidx}         
\usepackage{graphicx}        
\usepackage{multicol}        
\usepackage[bottom]{footmisc}
\usepackage{hyperref}        
\usepackage{soul}            
\hypersetup{colorlinks=true,urlcolor=blue}
\usepackage[square,numbers]{natbib}
\makeindex             

\begin{document}

\title*{Coded Mask Instruments for Gamma-Ray Astronomy}
\titlerunning{Coded Mask Instruments for $\gamma$-Ray Astronomy} 
\author{Andrea Goldwurm
and Aleksandra Gros}
\authorrunning{A. Goldwurm and A. Gros} 
\urlstyle{same}
\institute{Andrea Goldwurm (corresponding author)\at 
Universit\'e Paris Cit\'e, CNRS, CEA, Astroparticule et Cosmologie, F-75013 Paris, France and \\
D\'epartement d'Astrophysique /IRFU/DRF, CEA-Saclay, 91191 Gif-sur-Yvette, France, \\
\email{andrea.goldwurm@cea.fr} 
\and Aleksandra Gros \at 
Universit\'e Paris-Saclay, Universit\'e Paris Cit\'e, CEA, CNRS, AIM, 91191 Gif-sur-Yvette, France,
\email{aleksandra.gros@cea.fr}
%
\\
\\
\urlstyle{same}
{In: \it Handbook of X-ray and Gamma-ray Astrophysics}, Springer 2023, Singapore, 
eds. C. Bambi, A. Santangelo, ISBN: 978-981-16-4544-0 \\
\url{https://doi.org/10.1007/978-981-16-4544-0_44-1}
}
%
%
\maketitle
\abstract{
Coded mask instruments have been used in high-energy astronomy for the last forty years now and designs for future hard X-ray/low gamma-ray telescopes are still based on this technique when they need to reach moderate angular resolutions over large field of views, particularly for observations dedicated to the, now flourishing, field of time domain astrophysics. However these systems are somehow unfamiliar to the general astronomers as they actually are two-step imaging devices where the recorded picture is very different from the imaged object and the data processing takes a crucial part in the reconstruction of the sky image. Here we present the concepts of these optical systems applied to high-energy astronomy, the basic reconstruction methods including some useful formulae and the trend of the expected and observed performances as function of the system designs. We review the historical developments and recall the flown space-borne coded mask instruments along with the description of a few relevant examples of major successful implementations and future projects in space astronomy.
}
\textbf{
\\
Keywords}
\\
Coded Masks; Coded Apertures; Imaging Systems; Gamma-Ray Astronomy; Image Decoding; 
Image Processing; Data Analysis.
\\
\\

~\\ 
\\ 
\\
\\
\large{\textbf{Table of Contents}}
\begingroup
\let\clearpage\relax
\\
\normalsize
\renewcommand\contentsname{\small{Coded Mask Instruments for $\gamma$-Ray Astronomy - A. Goldwurm \& A. Gros}}
{  \hypersetup{linkcolor=blue}
  \tableofcontents{} }
\endgroup
\newpage
\normalsize
\hypersetup{linkcolor=red, citecolor=green, urlcolor=blue}

\section{Introduction} \label{sec:Intro}

\textbf{Coded aperture mask imaging systems}, 
in short Coded Mask Instruments hereinafter \textbf{CMI}, are multiplexing optical devices
that allow, through a proper spatial modulation of the incident radiation
and its following recording by a position sensitive detector, 
the simultaneous measurement of flux and position of multiple sources in the field of view.

The basic idea of these systems is to couple a \textbf{position sensitive photon-detector (PSD)}
to a filter or a \textbf{mask} that absorbs part of the incident radiation in such a way 
to operate a spatial modulation of the recorded photon flux that is dependent on the 
angular distribution of the sources in the field of view.
From the recorded image, the input sky image is reconstructed through 
a specific data post-processing
that takes into account the modulation that the mask operates.

CMIs are employed when conventional focusing systems based on lenses or reflectors 
cannot be used. Nowadays their major application is in high-energy astronomy, 
and in particular in the hard X-ray (10--100~keV) and soft gamma-ray (100~keV--10~MeV) domains 
where conventional focusing techniques, commonly employed at energies lower than 15~keV,
are not easily implemented because the radiation wavelengths are comparable or shorter than 
the typical inter-atomic distances. 
At energies higher than a few MeV the mask material becomes more and more transparent, 
the Compton scattering dominant and the geometrical spatial modulation, 
based on the photoelectric absorption process, less efficient. 
Absorption increases again at energies $>$10~MeV thanks to the pair-production effect, 
but there it is more efficient to use the intrinsic directional properties of this 
interaction on the few detected photons rather than the collective effect of 
shadow projection by many rays in order to measure the source direction. 
X/gamma-ray astronomy is also the domain where the high and 
variable background becomes dominant over the source contributions, 
which drastically limits the performance of standard on/off monitoring techniques
and where the simultaneous measurement of source and background is crucial 
even for the simple source detection.

Coded masks were conceived in the 1970s-1980s and employed successfully in the past 40-30 
years in high-energy astronomy, on balloon-borne instruments first 
and then onboard space missions like Spacelab~2, GRANAT, and BeppoSAX.
They have been chosen as imaging systems for experiments on a number of major missions
presently in operation, the European INTEGRAL, the American Swift
and the Indian ASTROSAT, and for some future projects like the Chinese-French 
SVOM mission.
Today the NuSTAR and the Hitomi missions have successfully pushed up to 80 keV 
the technique of grazing incidence X-ray mirrors \cite{Har13} \cite{Tak14}.
However the limited field of view (few arcmin) achieved by these telescopes and 
the variability of the sky at these energies make the coded mask systems still the best 
options to search for bright transient or variable events in wide field of views.

Coded aperture systems have been employed also in medicine and in monitoring nuclear 
plants and implementations in nuclear security programs are also envisaged \cite{Cie16}\cite{Acc01}. 
Even if basic concepts are still valid for these systems, certain conditions, specific 
to gamma-ray astronomy, can be relaxed (e.g., source at infinity, high level and 
variable background, etc.),
and therefore designs and data analysis for CMI for terrestrial studies can take 
avery different form. In particular for close sources (the so-called near-field condition), 
the system can actually provide three-dimensional imaging because of the
intrinsic enlarging effect of shadow projection as the source distance decreases. This interesting
property is not applicable in astronomy and we will not discuss it here.

In spite of the large literature on the topic, few comprehensive reviews were dedicated 
to these systems; the most complete is certainly the one by Caroli et al. 1987 \cite{Car87}, 
which however was compiled before the extensive use of CMI in actual missions. 
In this paper we review the basic concepts, the general characteristics 
and specific terminology
(troughout the paper key terms are written in \textbf{bold} when first defined)
of coded mask imaging for gamma-ray astronomy (\S~\ref{sec:Conc}), with a historical presentation 
of the studies dedicated to the search of the optimum mask patterns and best system designs.
We present (\S~\ref{sec:Anal}) in a simple way the standard techniques of the image reconstruction 
based on cross-correlation of the detector image with a function derived from the mask pattern,
providing the explicit formulae for this analysis and for the associated statistical errors, 
and the further processing of the images which usually involves iterative noise cleaning.
We will discuss the performance of the systems (\S~\ref{sec:Perf}), 
in particular the sensitivity and localization accuracy, under some reasonable assumptions on the background, 
and their relation with the instrument design.

We cannot be exhaustive in all topics and references of this vast subject.
Clearly the analysis of coded mask system data relies, as for any other telescope, 
on a careful calibration of the instrument, the understanding of systematic effects
of the detector and the measurement and proper modeling of the background. 
For these aspects these telescopes are not different from any other one 
and we will not treat these topics, 
apart from the specific question of non-uniform background shape over the PSD,
because they are specific to detectors, satellites, space operations 
and environment of the individual missions. 
Also we will not discuss detailed characteristics of the PSDs
and we will neglect description of one-dimensional (1-d) aperture designs 
and systems that couple spatial and time modulation (like rotational collimators),
as we are mainly interested in the overall 2-d coded aperture imaging system.

We include a section (\S~\ref{sec:Inst}) on the application of CMI in high--energy astronomy with 
a presentation of the historical developments from rocket-borne to space-borne projects
and mentioning all the experiments that were successfully flown up to today 
on space missions.
Finally we dedicate specific sub-sections (\S~\ref{subsec:Sigm}-\ref{subsec:Ecla}) 
to three gamma-ray CMI telescopes: 
SIGMA, that flew on the GRANAT space mission in the 1990s,
IBIS currently operating on the INTEGRAL gamma-ray observatory 
and ECLAIRs, planned for launch in the next years on board the SVOM mission. 
These experiments are used to illustrate the different concepts and issues presented
and to show some of the most remarkable "imaging" results obtained with CMI,
in high-energy astronomy in the past 30 years.

\begin{figure}[h]
	\centering
 \includegraphics[width=0.45\textwidth]{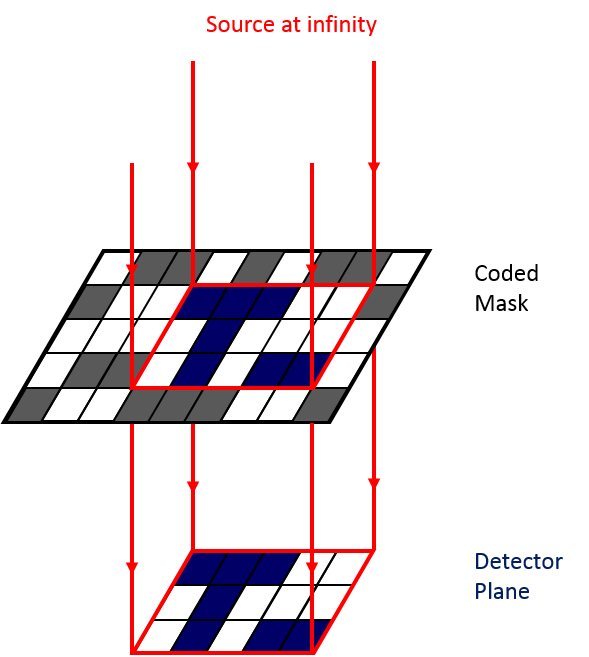}
 ~~~~~~~~~
 \includegraphics[width=0.34\textwidth]{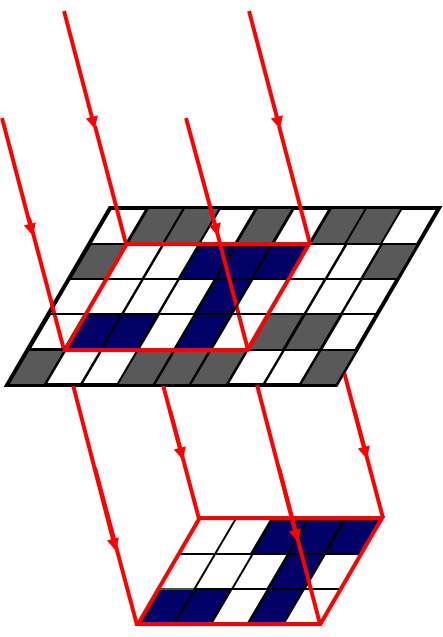}
	\caption{Coded aperture principle. Two sources at infinity project 
	a different pattern of the mask on the detector plane (shadow-gram). 
	For a cyclic system, here a mask with a URA basic pattern of 5$\times$3 
	replicated to 9$\times$5, 
	it is the same pattern but shifted according to the source position.
	}
	\label{fig:cm-prin}
\end{figure}
\begin{figure}[h]
	\centering
 \includegraphics[width=1.0\textwidth]{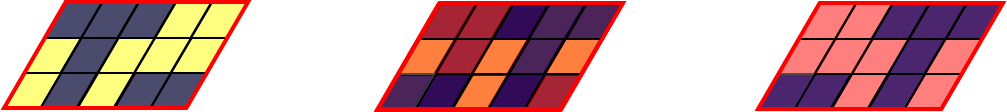}
	\caption{Coded aperture principle. The resulting images recorded by 
	the position sensitive detector for the configuration 
	of Fig.~\ref{fig:cm-prin} 
	for the two sources separately (left and right),
	and combined (center).
    }
	\label{fig:cm-prin-b}
\end{figure}

\section{Basics Principles of Coded Mask Imaging}  \label{sec:Conc}

\subsection{Definitions and Main Properties}  \label{subsec:Defi}

In coded aperture telescopes the source radiation is spatially modulated by a \textbf{mask}, 
a screen of \textbf{opaque and transparent elements}, 
usually of the same shape and size,
ordered in some specific way (\textbf{mask pattern}), 
before being recorded by a position sensitive detector.
For each source, 
the detector will simultaneously measure its flux together with background flux
in the detector area corresponding to the projected transparent mask elements, 
and background flux alone in detector area corresponding 
to the projected opaque elements (Fig.~\ref{fig:cm-prin}).
From the recorded \textbf{detector image} (Fig.~\ref{fig:cm-prin-b}), 
which includes the shadows of parts of the mask (\textbf{shadow-grams}) 
projected by the sources in the field of view onto the detector plane 
and using the mask pattern itself 
an image of the sky can, under certain conditions, be reconstructed.

Mask patterns must be designed to allow each source in the field of view 
to cast a unique shadow-gram on the detector, in order to avoid ambiguities 
in the reconstruction of the sky image. In fact each source shadow-gram shall
be as different as possible from those of the other sources.
The simplest aperture that fulfills this condition is of course 
the one that has only one hole, the well-known \textbf{pinhole camera}. 
The response to a point source of this system is a peak
of the projected size of the hole and null values elsewhere.
The overall resulting image on the detector is a blurred and inverted image of the object.
However the sensitive area and angular resolution, 
for given mask-detector distance, are inversely proportional to each other: effective area 
can only be increased by increasing the hole size, which worsens the angular resolution
(increases the blurring).

A practical alternative is to design a mask with several small transparent elements 
of the same size (Fig.~\ref{fig:cm-prin}), a \textbf{multi-hole camera}.
The resolution still depends on the dimension of one individual hole 
but the sensitive area can be increased by increasing the number of transparent elements.
In this case however the disposition of the holes is important since when more than one 
hole is used, ambiguity can rise regarding which part of the sky is contributing to 
the recorded image. 
For example with a regular chess board pattern mask, different sources would 
project identical shadows and disentangling their contributions would be impossible.
Mask patterns that have good imaging properties exist (\S~\ref{subsec:Patt}).
With the use of a properly designed multiple-hole mask system an image of the sky 
can be recovered from the recorded shadow-gram through a convenient computation.
In general the sky image reconstruction, or \textbf{image deconvolution} as it is often called, 
is based on a correlation procedure between the recorded detector image
and a decoding array derived from the mask pattern.
Such correlation will provide a maximum value for the decoding array "position" 
corresponding to the source position, where the match between the source shadow-gram 
and the mask pattern is optimum, and generally lower values elsewhere. 
Note that, unlike focusing telescopes,
individual recorded events are not uniquely positioned in the sky:
each event is either background or coming from any of the sky areas 
which project an open mask element at the event detector position.
The sky areas compatible with a single recorded event will draw a mask pattern in the sky. 
It is rather the mask shadow, collectively projected by many source rays, 
that can be "positioned" in the sky.

Assuming a perfect detector (infinite spatial resolution) and a perfect mask 
(infinitely thin, totally opaque closed elements, totally transparent open elements), 
the angular resolution of such a system is then defined by the angle subtended
by one hole at the detector. The sensitive area depends instead on
the  total surface of transparent mask elements viewed by the detector. 
So, reducing hole size or increasing mask to detector distance 
while increasing accordingly the number of holes 
improves the angular resolution without loss of sensitivity.
Increasing the aperture area will increase the effective surface but, since
the estimation of the background is also a crucial element, this does not mean 
that the best sensitivity would increase monotonically with the increase of 
the \textbf{mask open fraction} (the ratio between transparent mask area 
and total mask area, also sometimes designed as \textbf{aperture or transparent fraction}).
In the gamma-ray domain where the count rate is dominated by the background,
the optimum aperture is actually one-half.
In the X-ray domain instead, the optimum value rather depends on the
expected sky to be imaged even if in general, because of the Cosmic X-ray Background (CXB)
which dominates at low energies, the optimal aperture is somewhat less than 0.5.

\begin{figure}[h]
	\centering
    \includegraphics[width=0.50 \textwidth]{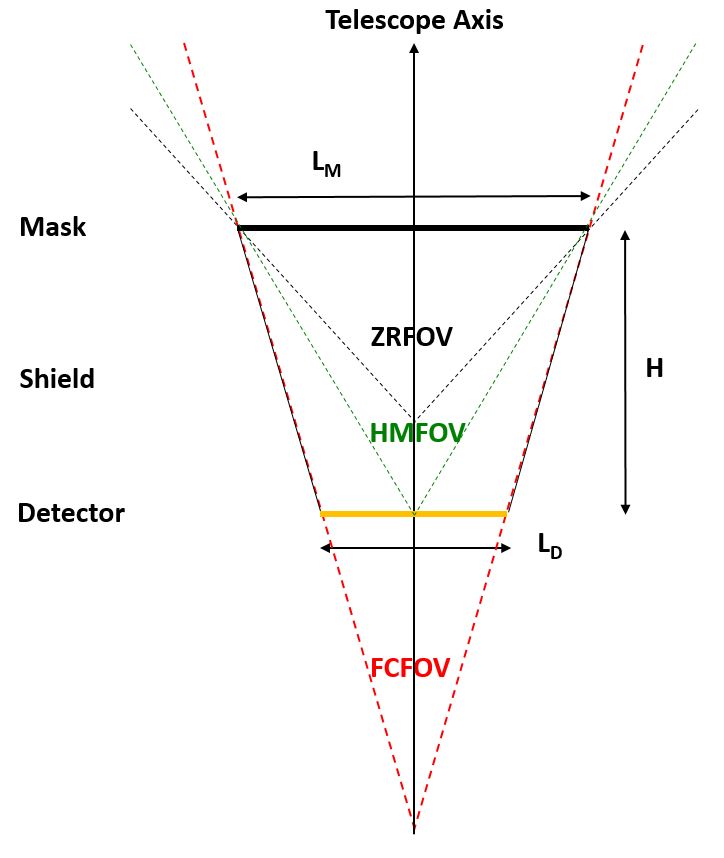}
    \includegraphics[width=0.49\textwidth]{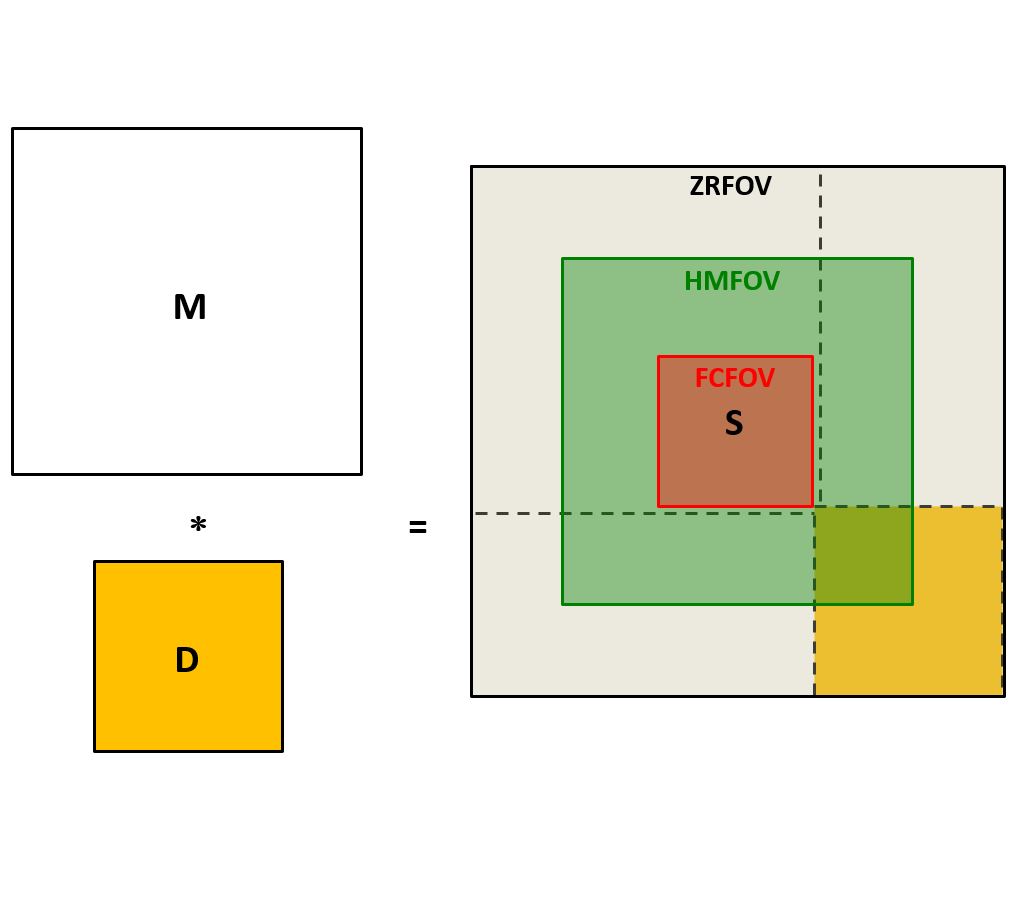}
	\caption{
	Left: A coded aperture telescope geometry with a mask larger 
	than the detector and a shield connecting them. 
	The field of views around the telescope axis are shown: 
	the FCFOV (red), the Half Modulation EXFOV (green)
	and the Zero Response EXFOV (black).
	Right: Relation, for a square CMI, between the array sizes of mask ($M$), detector ($D$), 
    and sky ($S$), with indication, in the sky, of the FOVs (same color code than left panel).
	}
	\label{fig:cm-geo}
\end{figure}

The field of view (FOV) of the instrument is defined as the set of sky directions 
from which the source radiation is modulated by the mask and its angular dimension
is determined by the mask and the detector sizes and by their respective distance,
in the absence of collimators.
Since only the modulated radiation can be reconstructed, 
in order to optimize the sensitive area of the detector and 
have a large FOV, masks larger than the detector plane are usually employed, 
even if equal dimensions (for the so-called \textbf{simple} or \textbf{box type} CMI systems) 
have also been used.
The FOV is thus divided in two parts: the \textbf{fully coded (FC) FOV} for which all 
source radiation directed toward the detector plane 
is modulated by the mask and the \textbf{partially coded (PC) FOV} 
for which only a fraction of it is modulated by the mask (Fig.~\ref{fig:cm-geo}).
The non-modulated source radiation, even if detected, 
cannot be distinguished from the (non-modulated) background.
In order to reduce its statistical noise and background radiation, 
collimators on the PSD or
an absorbing tube connecting the mask and detector are used.

The typical CMI geometry and its FOVs are shown in Fig.~\ref{fig:cm-geo}.
If holes are uniformly distributed over the mask, the sensitivity 
is approximately constant in the FCFOV and decreases in the PCFOV
linearly because the mask modulation decreases to zero.
The total FOV (FC+PC) is often called \textbf{extended (EX) FOV} and
can be characterized by the level of modulation of the PC.
Figure~\ref{fig:cm-geo} right shows the relative sizes of the 
(ZR)EXFOV, detector and mask. For simple systems the FCFOV is limited 
to the on-axis direction and all the EXFOV is PCFOV.

Table~\ref{tab:cm-prop} reports the approximate imaging characteristics 
provided by a coded aperture system (as illustrated in Fig.~\ref{fig:cm-geo})
as functions of its geometrical parameters along one direction.
Values of the IBIS/ISGRI system (\S~\ref{subsec:Ibis}) for which the design 
parameters are given in the notes are reported as example.
The EXFOV dimensions are given for half modulation level and for zero response.
Both angular resolution (\S~\ref{subsec:Anre}) and localization power (\S~\ref{subsec:Psla}), 
which are at the first
order linked to the angle subtended by the mask element size to the detector
and the detector pixel (or spatial resolution) to the mask, 
depend actually also on the reconstruction method and even on the distribution 
of holes in the mask pattern as described below.

\begin{table}
\caption{Expected Imaging Properties of a Coded Aperture System}
\begin{center}
\begin{tabular}{lll}
\hline
\textbf{Quantity}~~~~~~~~~~~~~~~~~~~~~~~~~~~~~~~~~~~~~~~~~~~~~~~~~~~~~~~~~~~~~~

&  \textbf{Angular value} 
~~~~~~~~~~~~~~~~~~~~~~~~
&  \textbf{IBIS/ISGRI}
\\
\hline
FCFOV (100~$\%$ sensitivity)   &  $2\cdot \arctan{(L_M - L_D) \over 2\cdot H}$~~~~~~ 
& 8.2°
\\
EXFOV (50~$\%$ sensitivity)   &  $2\cdot \arctan{L_M \over 2\cdot H}$         
& 18.9°
\\
EXFOV (0~$\%$ sensitivity)    &  $2\cdot \arctan{(L_M + L_D) \over 2\cdot H}$     
& 29.2°
\\
Angular resolution on-axis (FWHM)     &  $\approx$ $\arctan{\sqrt{m^2 + d^2} \over H}$       &  13$'$ 
\\
Localization error radius on-axis (90\% c.l.)     &   $\approx \arctan {
\left({
{1.52 \over SNR}
{d \over H} \sqrt{ {m \over d} - {1 \over 3}} 
}\right)}
$   
&   22$''$ at~SNR=30
\\
\hline
\end{tabular}
\label{tab:cm-prop}
\end{center}
Notes : $L_M$ mask linear size, $L_D$ detector linear size, H detector-mask distance, 
$m$ mask element linear size ($m>d$), $d$ detector pixel linear size for pixelated detector,
$d = 2 \sqrt{3 }\sigma _D$ where $\sigma_D$ is the linear detector resolution (in $\sigma$)
for continuous detector. SNR here is the "imaging signal~to~noise~ratio" $SNR_I$
for known source position (\S~\ref{subsec:Sens}). IBIS/ISGRI approximate parameters: 
$L_M$~=~1064~mm, $L_D$~=~600~mm, $H$~=~3200~mm, $m$~=~11.2~mm, $d$~=~4.6~mm.
\end{table}
\subsection{Coding and Decoding: The Case of Optimum Systems}  \label{subsec:CoDe}
To analyze the properties of coded mask systems we first simplify the 
treatment by considering an \textbf{optimum coded mask system}
which provides after the image reconstruction a shift invariant and side-lobes-free 
spatial response to a point source, the so called \textbf{System Point Spread 
Function (SPSF)}, in the FCFOV (e.g., \citep{FC78}).

We assume a fully absorbing infinitely thin mask, 
a perfectly defined infinitely thin PSD with infinite spatial resolution 
and perfect detection efficiency.
The object, the sky image, described by the term $S$ 
is viewed by the imaging system composed by a mask described by the function $M$ and 
a detector that provides an image $D$.
$S$, $M$ and $D$ are then continuous real functions of two real variables.
$M$ assumes values of 1 in correspondence to transparent elements and 0 for opaque elements, 
the detector array $D$ is given by the correlation\footnote{The two dimension integral correlation 
between two functions, say A and B, is indicated by the symbol $\star$ and is given by
$A \star B = C(s,t) = \int{ \bar{A}(x,y)\cdot B(x+s,y+t) dx dy}$ where $\bar{A}$ is the complex conjugate of the function $A$.}
of the sky image $S$ with $M$ plus an un-modulated background array term $B$
$$D = S \star M + B$$

If $M$ admits a so-called \textbf{correlation inverse} function, say $G$, 
such that $M \star G = \delta$-function, 
then we can reconstruct the sky by performing 
$$S' = D \star G = S \star M \star G + B \star G = 
S \star \delta + B \star G = S + B \star G$$
and $S'$ differs from $S$ only by the $B \star G$ term.

\begin{figure}[h]
	\centering
 \includegraphics[width=0.26\textwidth]{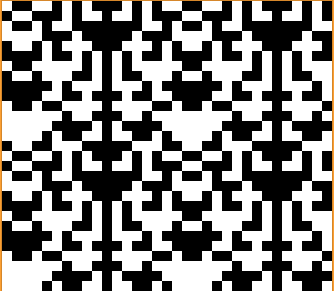}
 ~
  \includegraphics[width=0.26\textwidth]{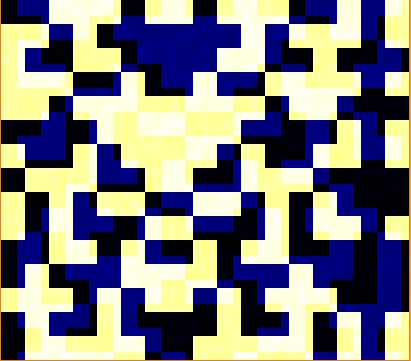}
 ~
  \includegraphics[width=0.44\textwidth]{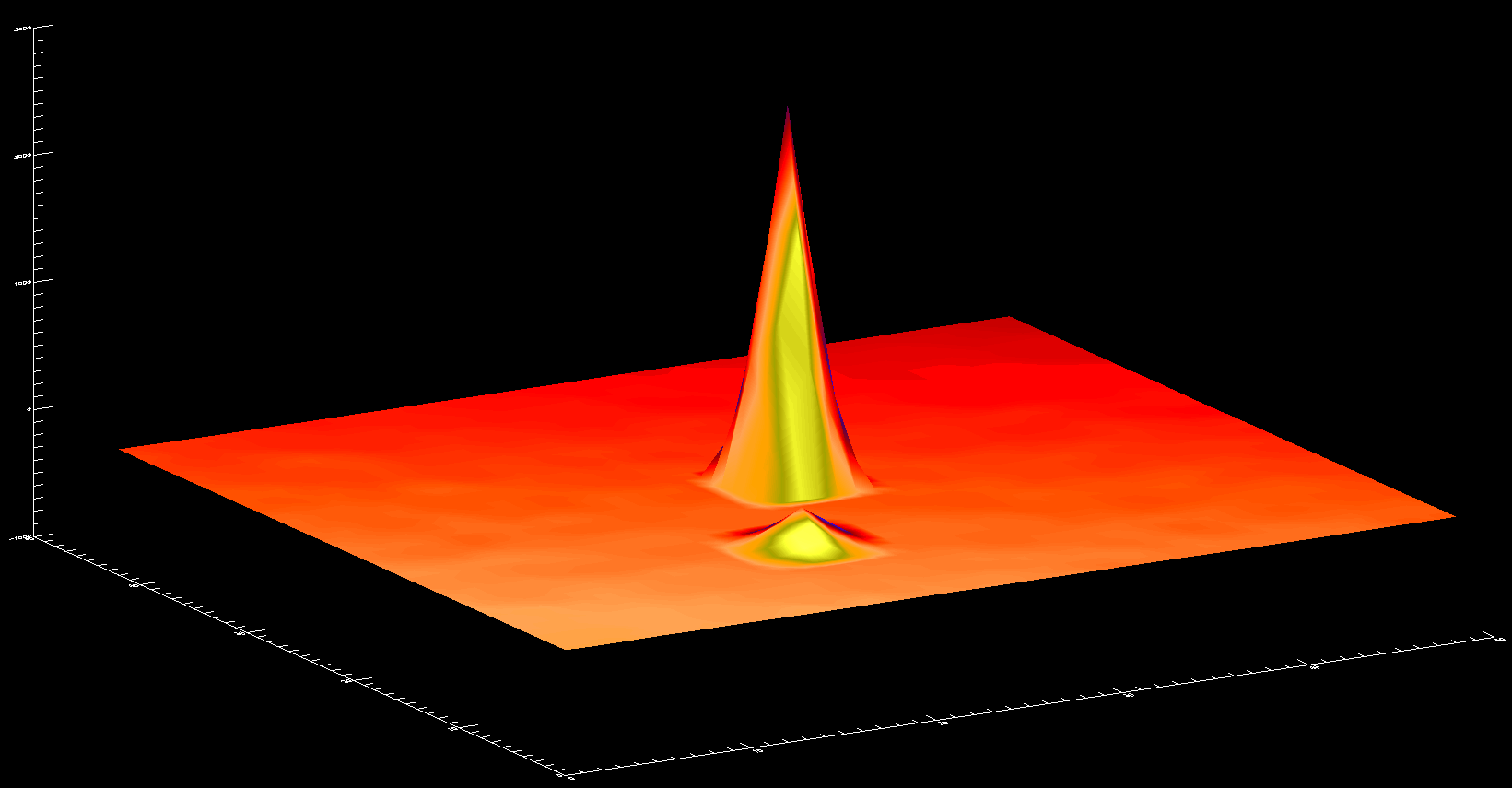}
	\caption{Mask, detector, FC sky of an optimum CMI. 
	Left: A $33\times29$ replicated mask 
	of basic pattern $17\times15$ (Hadamard from an m-sequence CDS 
	with $N=255$ and $m=8$ disposed along the diagonal, see \S~\ref{subsec:Patt}).
	Center: Simulated detector image of two sources in the FCFOV
	and low background for a CMI with the mask on the left 
	and a detector of same size of its basic pattern,
	with $3\times3$ pixels per mask element. 
	Right: The relative decoded SNR sky image in the FCFOV. 
	}
	\label{fig:cm-concept}
\end{figure}

In certain cases, when, for example, the mask $M$ is derived from a cyclic
replication of an optimum basic pattern, 
if the background is flat then the term $B \star G$ is also constant and can be removed.
Since even for very high detector resolution the information must be digitally 
treated in the form of images with a finite number of image pixels, the problem 
must be generally considered in its discrete form.
We can formulate the same process in digital form by substituting the continuous functions 
with discrete arrays and considering discrete operators instead of integral operators.
$S$, $M$, $G$, $D$, and $B$ terms will therefore be finite real 2-d arrays,
and the delta function the delta symbol of Kronecker. 
The discrete correlation is obtained by finite summations and 
the reconstructed sky $S'$ by
$$ S'_{i,j} = \sum^{ }_{kl}{D_{k,l} G_{i+k,j+l}} $$
with $i,j$ indices that run over the sky image pixels and $k,l$ over the detector pixels.
Mask patterns that admit a correlation inverse array exist (\S~\ref{subsec:Patt}), 
and can be used to design the so-called optimum systems.  
For instance, for masks $M$ that have an auto-correlation given by a delta function,
the decoding array constructed posing $G = 2 \cdot  M - 1$ 
(i.e., $G = +1$ for $M = 1$ and $G=-1$ for $M=0$) is then a correlation inverse.
To have such a side-lobe-free response in an optimum system, a source must 
however be able to cast on the detector a whole basic pattern.
To make use of all the detector area and to allow more than one source
to be fully coded, the mask basic pattern is normally taken to be
the same size and shape as the detector and the total mask made by a cyclic repetition 
of the basic pattern 
(in general up to a maximum of 2 times minus 1 in each dimension to avoid ambiguities)
(Fig.~\ref{fig:cm-concept}).
For such optimum systems, a FCFOV source will always project a 
cyclically shifted version of the basic pattern, and correlating 
the detector image with the $G$ decoding array within the limits of the FCFOV
will provide a flat side-lobe peak with position-invariant shape at the 
source position (Fig.~\ref{fig:cm-concept} right).

A source outside the FCFOV but within the PCFOV will instead cast an 
incomplete pattern, and its contribution cannot be a priori automatically subtracted 
by the correlation with the decoding array at other positions than its own, 
and it will produce secondary lobes over all the reconstructed EXFOV including the FCFOV.
Following a standard nomenclature of CMI, we refer to these secondary lobes 
as \textbf{coding noise}.

\begin{figure}[h]
	\centering
  \includegraphics[width=1\textwidth]{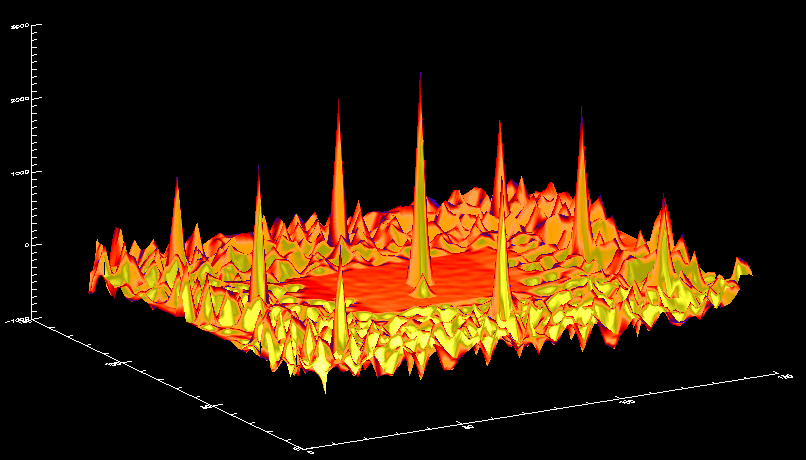}
	\caption{Reconstructed SNR sky image in the full EXFOV for
	the CMI and simulation of Fig.~\ref{fig:cm-concept} left/center.
	The central part corresponds to the FCFOV (same as Fig.~\ref{fig:cm-concept} right) 
	and shows optimal properties: the shift invariant peaks and flat side-lobes
	of the two simulated FC sources.
	In the PCFOV coding noise and the strong eight ghosts of the main source 
	are clearly visible.
	}
	\label{fig:cm-concept-b}
\end{figure}

On the other hand the modulated radiation from PC sources can be
reconstructed by extending with a proper normalization the correlation 
procedure to the PCFOV.
The reconstructed sky in the total field (EXFOV)
is therefore composed by the central region (FCFOV) 
of constant sensitivity and optimum image properties, 
i.e., a position-invariant and flat side-lobes SPSF, 
surrounded by the PCFOV of decreasing sensitivity and non-perfect SPSF 
(Fig.~\ref{fig:cm-concept-b}).
In the PCFOV, the SPSF includes coding noise,
the sensitivity decreases, and the relative variance increases toward the 
edge of the field.
Also, even FCFOV sources will produce coding noise in the PCFOV, while 
sources outside the EXFOV are not modulated by the mask and 
simply contribute to the background level on the detector.
When a complete mask is made of a cyclic repetition of a basic pattern, then each 
source in the FOV will produce eight large secondary lobes 
(in rectangular geometry) at the positions which are symmetrical 
with respect to the real source position at distances given by the basic pattern: 
these spurious peaks of large coding noise are usually called \textbf{ghosts} 
or \textbf{artifacts} (Fig.~\ref{fig:cm-concept-b}). 

These optimum masks also minimize the statistical errors associated
to the reconstructed peaks and make it uniform along the FCFOV.
Since $G$ is two-valued and made of +1s or --1s the variance associated to the reconstructed image
in the FCFOV is given by $V = G^2 \star D = \Sigma D $, the variance associated to 
each reconstructed sky image pixel is constant in the FCFOV 
and equal to the total counts recorded by the detector.
This implies that the source {\bf signal to noise ratio (SNR)} is simply
$$
SNR = {C_S \over { \sqrt{C_S + C_B} } } = 
{Reconstructed~Source~Counts \over \sqrt{~Total~Detected~Counts}}
$$
where $C_S$ and $C_B$ are source and background recorded counts.
The deconvolution is then equivalent to summing up counts from all the detector 
open elements (source and background counts) and subtracting counts from the closed 
ones for that source (background counts only).


\subsection{Historical Developments and Mask Patterns} \label{subsec:Patt}

Following the first idea to modulate incident radiation using 
Fresnel plates, formulated by Mertz and Young \cite{MY61}, 
the concept of a pinhole camera with multiple holes for high energy astronomy
(the {\it multiple-pinhole camera})
was proposed by Ables \cite{Abl68} and  Dicke \cite{Dic68} at the end of the 1960s.
In these designs multiple holes of the same dimensions are distributed randomly 
on the absorbing plate and in spite of the inherent production of side-lobes in the SPSF,
the increase in the aperture fraction compared to the single pinhole design 
highly improves the sensitivity of the system, at the same time maintaining 
the angular resolution.

Toward the end of the 1970s it was realized that special mask patterns 
could provide optimum imaging properties for coded aperture systems, 
and then a large number of the early works focused on
the search for these optimal or nearly optimal aperture patterns.
Most of these are built using binary sets called 
{\bf cyclic different sets (CDS)} \cite{Bau71} which have the remarkable property 
that their cyclic {\bf auto-correlation function (ACF)} is two-valued and approximates 
a delta function modulo a constant term.
Certain of these sets can be disposed (following certain prescriptions) to form 2-d arrays,
the so-called {\bf basic patterns},
which also have the property of having 2-d cyclic auto-correlations which are bi-dimensional 
delta functions, thus allowing design of coded aperture systems where a correlation inverse 
is directly derived from the mask pattern.
Thus by disposing, in a rectangular geometry, 2$\times$2 such 2-d basic pattern side by side 
(actually less than 2 times in each direction in order to avoid full repetition 
of the pattern and then ambiguity in reconstruction)
at a certain distance from a detector plane of the same dimension as the basic pattern
one obtains an optimum system with maximum FCFOV, free of peak repetitions and coding noise.
Care must be taken on how the mosaic of the basic pattern is done in order for
a source to project on the detector a shifted, but complete, version of the basic pattern.

\subsubsection{Patterns Based on Cyclic Different Sets}
A cyclic different set $D(N,k,{\lambda})$ is a collection of $k$ distinct integer residues 
$d{_1}, ... , d{_k}$ modulo $N$, 
for which the congruence $d{_i} - d{_j} = q~mod(N)$ has exactly $\lambda$ distinct solution 
pairs ($d{_i}$,$d{_j}$) in $D$ for every residue $q \ne 0~mod(N)$. 
If such a different set $D$ exists, then $\lambda = k(k-1)/(N-1)$.
This mathematical definition simply means that for these sets, a cyclic 
(over the dimension $N$ of the larger set to which they belong) displacement vector 
between any two elements of the set occurs exactly a constant number of times, 
given by the parameter $\lambda$, which is called the "redundancy" of the set.
For this reason binary arrays based on CDS are also called uniformly redundant arrays (URA).
From this property immediately follows that a 1-d binary sequence $M$ of dimension $N$ 
built from a CDS $D(N,k,{\lambda})$ by the following prescription
$$ m_i = \cases{ 1 ~~\rm{~~if~} i \in D \cr
0 ~~\rm{~~if~} i \notin D \cr}
~~~~~~~~\rm{has~an~ACF}~~~~
a_{i} = \sum_j{m_{j}m_{j+i}}= \cases{ k ~~\rm{~~for~} i = 0~mod~N \cr
\lambda ~~\rm{~~for~} i \ne 0~mod~N \cr}
$$
that is a $\delta$ function.
The parameter $k - \lambda$ is also an important characteristic of the set since 
it determines the difference between the peak and the plateau of the ACF.
The higher this value the better is the signal to noise response to a point source of the 
derived imaging system.

Several types of CDS exist and early studies on the subject were focused to find as many 
such sequences as possible and establish the way to build them.
A class that was already well known from the coding theory was the 
Non Redundant Arrays (NRA), which are in fact CDS with redundancy = 1. 
These have however densities of elements very small ($<$ 10$\%$) and therefore 
provide apertures far from the ideal 30$\%$ to 50$\%$ open fraction needed by X/gamma-ray astronomy.  

The most interesting CDS for 2-d astronomical imaging are those 
called m-sequences or also pseudo-noise (PN), pseudo-random, shift-register or Hadamard-Singer sets. 
These are part of the more general class of Hadamard sets, which are CDS with
$N=4t-1$ and $t$ integer ($k=2t-1$, ${\lambda}=t-1$) and are related to the 
rows of Hadamard matrices 
(which are matrices with mutually orthogonal rows).
The m-sequence sets that exist for $N=2^{m}-1$ with $m$ integer $>$~1
are particularly interesting because
they have nearly 50$\%$ open fraction
and when $m$ is even they can be factorized in $p~\times~q$ arrays
with $p=2^{m/2}+1$ and $q=2^{m/2}-1$ in order to form rectangular 
(quasi-square) arrays.

The first to propose to use different sets for building 2-d imaging optimum systems for X-ray 
astronomy were independently Gunson and Polichronopulos \cite{GP76} and Miyamoto \cite{Miy77} in 1976. 
They both identified the m-sequences as the original set to use for the design, 
but with different mapping in 2-d arrays to obtain the basic pattern.
The second author actually started from the Hadamard arrays that were studied 
in particular for spectroscopy by Harwit, Sloane, and collaborators \cite{HS79}.
The proposed pattern is equivalent to the one 
obtained by filling up the array with the PN sequence row by row.
The first authors instead proposed to build basic patterns from m-sequences
filling the array along extended diagonals 
(this further requires that $p$ and $q$ are mutually primes).
In the two cases, the mosaic of cyclic repetition of the basic pattern
must be performed in a different way in order to preserve the $\delta$-function ACF property. 
For the diagonal prescription, the basic patterns can just set adjacent to each other;
for the row by row construction, those on the side must be shifted vertically by one row 
(see \cite{Car87} for the details). 
We call these masks \textbf{Hadamard} masks to distinguish them from the URA described below, 
even if both can be considered URA.
A more complete discussion of the way PN-sequences are used for an imaging coded mask instrument 
of the type proposed by \cite{GP76}, including the way of filling the 2-d array by extended diagonal,
was provided soon after by Proctor et al. \cite{Pro79} who also discussed the 
implementation in the SL1501 experiment \cite{Pro78} (\S~\ref{subsec:Rock}). 
Examples of Hadamard masks are shown in Figs.~\ref{fig:cm-concept} left, \ref{fig:Mask-pat} left.

A particular subset of Hadamard CDS, the {\it twin prime} CDS, are those for which $p$ and $q$ 
are primes and differ by 2 ($q=p-2$). These sets can be directly mapped in $p~\times~q$ arrays using the 
prescription proposed by Fenimore and Cannon \cite{FC78} in 1977. 
In a series of other seminal papers these authors and collaborators improved the description 
of coded aperture imaging using these URA arrays and discussed their performances 
and a number of other associated topics \cite{Fen78}\cite{Fen80}\cite{FC81}\cite{FW81}.
These \textbf{URA} masks, as we will call them following \cite{FC78}, are generated 
from quadratic residue sequences of order $p$ and $q$ ($p=q+2$) according 
to the following prescription:
$$ M_{ij} = \cases{ 0 ~~\rm{if} ~i=0 \cr
                    1 ~~\rm{if} ~j=0, ~i\ne0    \cr
                    1 ~~\rm{if} ~C_i^p \cdot C_j^q = 1 ~~~~\rm{where}
                 ~ C_i^p = \cases{ +1 ~~\rm{~if~ \exists ~k \in \textbf{Z},~1 \leq k < p, ~i=k^2~mod~p} \cr
                 -1 ~\rm{~~otherwise}  \cr} \cr
                    0 ~~\rm{~~otherwise}  \cr}
$$
Other 2-d rectangular arrays presenting delta function ACF were identified as Perfect Binary Arrays (PBA). 
Again they are a generalization in 2-d of CDS, include the URAs, and are based on different set group theory \cite{KS94}.

Early designs of CMI assumed rectangular geometry but in 1985 mask patterns for hexagonal geometry
were proposed by Finger and Prince \cite{FP85}. 
These are based on Skew-Hadamard sequences 
(Hadamard sequences with order $N$ prime and constructed from quadratic residues) 
that, for dimensions $N=12t+7$ where $t$ is an integer, can be mapped onto hexagonal lattices, 
with axes at 60$^\circ$ from each other, to form hexagonal URA, the \textbf{HURA}. 
In addition to be optimum arrays (they have a $\delta$-function ACF), they are also anti-symmetric with respect 
to their central element (complete inversion of the pattern) under 60$^\circ$ rotation. 
This property allows one to use them to subtract a non-uniform background, if a rotation of the mask of 60$^\circ$ can be implemented, 
and even to smear out the ghosts created by a replicated pattern if a continuous rotation can be performed.  
The hexagonal geometry is also particularly adapted to circular detectors. 
The complications induced by moving elements in satellites have limited 
the use of such mask/anti-mask concept based on mask rotation with respect to the detector plane. 
A rotating HURA mask (Fig.~\ref{fig:Mask-pat} center) was successfully implemented 
in the GRIP balloon borne experiment \cite{Alt85} 
and operated during a few flights allowing for an efficient removal 
of the background non-uniformity (\S~\ref{subsec:Rock}). 
A fixed non-replicated HURA of 127 elements has been implemented for the SPI/INTEGRAL instrument (\S~\ref{subsec:Sate}).

\subsubsection{Other Optimum Patterns}
The limited number of dimensions for which CDS exist coupled to the additional 
limitation that $N$ must be factorized in two integers for a rectangular geometry 
or comply with more stringent criteria for the hexagonal one
implies that a small number of sequences can actually be used for optimum masks.
This led several authors to look for other optimum patterns, and 
several new designs were proposed in the 1980s and 1990s, 
even if somehow related to PN sequences.
Even though for these patterns the ACF is not exactly a delta function, 
it is close enough that a simple modification of the decoding arrays 
from the simple mask patterns allows recovery of a shift invariant and side-lobe-free SPSF.
For these masks therefore an inverse correlation array exists
and an optimum imaging system can be designed.

\begin{figure}[h]
	\centering
 \includegraphics[width=0.3\textwidth]{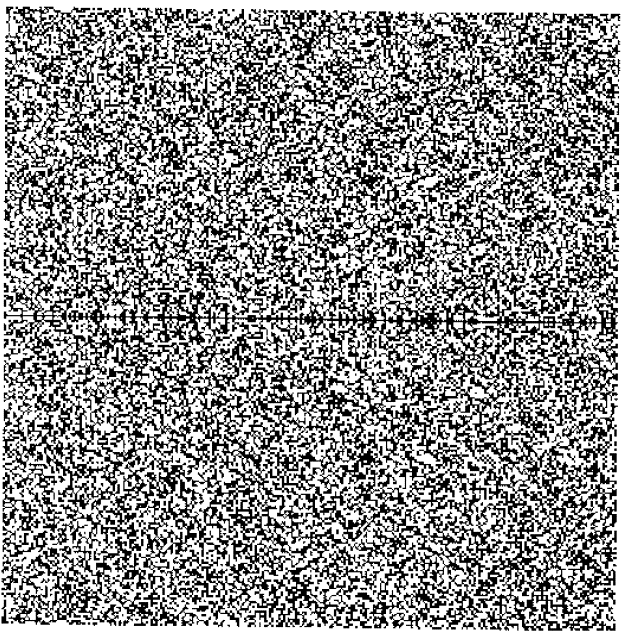}
 ~~~~~
 \includegraphics[width=0.3\textwidth]{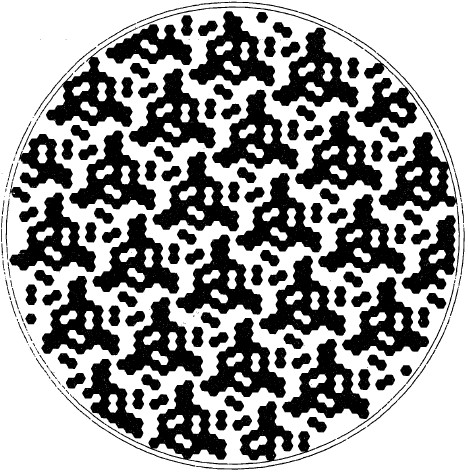}
 ~~~~~
 \includegraphics[width=0.3\textwidth]{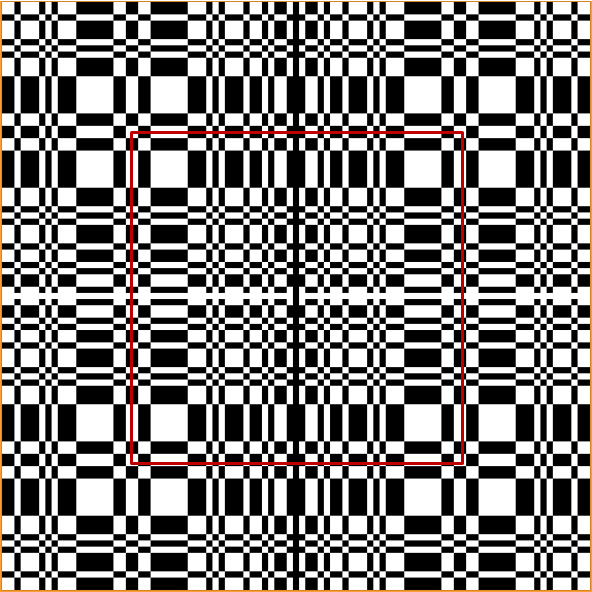}
	\caption{
	Optimum coded aperture patterns, note the symmetry of URAs 
	with respect to the "more random" Hadamard ones (also Fig.~\ref{fig:cm-concept} left).
	Left: The non-replicated 255$\times$257 Hadamard pattern 
	of the COMIS/TTM mask (from \cite{Int92}), 
	from an m-sequence CDS ordered along the extended diagonal. 
	Center: Replicated HURA mask of 127 basic pattern used 
	in the GRIP experiment (from \cite{Alt85}, © NASA).
	Right: The IBIS 95$\times$95 mask, replication of the MURA 
	53$\times$53 basic pattern (central red square) (see \S~\ref{subsec:Ibis}).
	}
	\label{fig:Mask-pat}
\end{figure}

The most used of them was certainly the {\bf Modified URA} or {\bf MURA} (Fig.~\ref{fig:Mask-pat} right) 
of Gottesman and Fenimore \cite{GF89}.
Square MURAs exist for all prime number linear dimensions and this increases 
by about a factor 3 the number of rectangular optimal arrays with respect to the URA and Hadamard sets.
They are basically built like URA on quadratic residues but for the 
first element (or central element for a 2-d pattern) which is defined as not part of the set. 
The MURAs also have symmetric properties with respect to the central element which permits a MURA 
using the complement of the pattern (but keeping the central element to 0 value). 
The correlation inverse is built like in URAs ($+1$ for mask open 
elements and $-1$ for opaque ones) apart from the central element, and
its replications, if any, which are set to $+1$ even if the element is opaque.
With this simple change from the mask pattern, the derived decoding 
array $G$ is a correlation inverse and the system is optimum.

Other optimum rectangular designs for which a correlation inverse can be defined were obtained 
from the product of 1-d PN sequences, the Pseudo Noise Product (PNP), or 1-d MURAs
(MP and MM products patterns). 

\subsubsection{Real Systems and Random Patterns}

More recent studies of mask patterns have focused on more practical issues 
such as how to have opaque elements all connected between them by at least one side
in order to build robust self-supporting masks 
able to resist, without (absorbing) support structures, 
to the vibration levels of rocket launches.
As explained above, even for an optimum mask pattern, any source in the PCFOV 
will produce coding noise and spurious peaks also in the FCFOV.
In order to obtain a pure optimum system one has then to implement a collimator 
which reduces the response to 0 at the edge of the FCFOV.
This solution was proposed by \cite{GP76} who also suggested to include the collimator 
directly into the mask rather than in the detector plane.
However the total loss of the PCFOV (even if affected by noise) 
and the loss of efficiency also for FC sources not perfectly on-axis 
are too big a price to pay to obtain a clean system and led to the abandonment 
of the collimator solution in favor of a shield between the mask edges and detector borders 
in order to reduce background and out of FOV source contributions.

In addition the geometry of optimum systems cannot be, in practice, perfectly realized.
Effects like dead area or noisy pixels of the detector plane, 
missing data from telemetry errors, 
not perfect alignment, tilt or rotation of the mask with respect to the
detector, absorption and scattering effects of supporting structures of the mask 
or of the detector plane and several other systematic effects directly increase the
coding noise and ghosts and degrade the imaging quality of the system.

Since the imperfect design of real instrument generally breaks down the
optimum imaging properties of the cyclic optimum mask patterns, 
today these patterns are not anymore considered essential 
for a performing coded mask system and there is a clear revival of random patterns.
Indeed for the typical scientific requirements of CMI 
(detection/localization of sources in large FOV)
one prefers to have some low level of coding noise spread over a large FOV 
rather than few large ghosts produced by the needed cyclic repetition of the optimum patterns
giving strong ambiguities in source location.
This is why the most recent instruments were designed using random or quasi-random patterns.
The drawback is that, for practical reasons, like the need to have solid self-supporting masks, 
pure random distributions are also difficult to implement and then for these "quasi-random masks"
the inherent coding noise becomes less diffuse and more structured.
The issue then becomes how to optimize the choice of these quasi-random patterns
in order to get best performance in terms of coding noise, 
SPSF, sensitivity and localization.

\section{Image Reconstruction and Analysis} \label{sec:Anal}

\subsection{Reconstruction Methods} \label{subsec:Meth}

A coded mask telescope is a two-step imaging system where a specific processing 
of the recorded data is needed in order to reconstruct the sky image 
over the field of view of the instrument.

The reconstruction is usually based on a correlation procedure,
however, in principle, other methods can be envisaged.
Indeed from the simple formulae that describe the image formation in a CMI (\S~\ref{subsec:CoDe})
and which give the relations between the input sky $S$, the mask $M$ and the detector $D$, 
it follows that $S$ can be derived by the simple inversion technique, by means of the 
Fourier transform (FT) of $M$ and $D$, with 
$S' = IFT(FT(D)/FT(M)) = S + IFT(FT(B)/FT(M))$,
where IFT stands for the inverse FT.
However this direct inverse method usually produces a large amplification of the noise in 
the reconstructed image, since the FT of $M$ always contains very small or even null terms, 
and the operation on the background component, which is always present, diverges 
and leads to very large terms.

A way to overcome this problem is to apply a Wiener filter as a reconstruction method
\citep{Sim80}\citep{Willi84}
in order to reduce the frequencies where the noise is dominant over the signal
when performing the inverse deconvolution.
It consists in convolving the recorded image $D$ with a filter $WF$ whose FT is 
$$FT(WF) = FT(M)/[(FT(M))^2 + (FT(SNR))^{-1}]$$
The filter showed to be efficient to recover the input sky image especially when 
a non-optimal system is employed,  
but it requires an estimate of the spectral density of the signal to noise 
ratio ($SNR$) which is not in principle known a priori. 
A simple application using a constant $SNR$ value with spatial frequency was used and compared 
well to correlation and also to Bayesian methods. 

Indeed Bayesian methods have also been specifically applied to CMI in particular in the form
of an iterative Maximum Entropy Method (MEM) algorithm \citep{Sim80}\citep{Willi84}.
The results with MEM are not very different from those obtained by the correlation techniques.
The heavy implementation of MEM compared to the latter ones, 
and the problems linked to how to establish the criteria for stopping the iterative procedure 
to avoid over-fitting the data
have made these techniques less popular than correlation coupled to iterative cleaning.

Most of these data processes are heavy and time-consuming, especially when images are large, 
and the issue of computation time is relevant in CMI analysis,
in particular when iterative algorithms need to compute several times the sky image or a model, 
like in MEM.
Some studies in the past have concentrated on fast algorithms for the deconvolution. 
Systems based on pseudo-noise mask patterns and Hadamard arrays could exploit the 
Fast Hadamard Transform (FHT) which reduces the convolution processing from an order of
$N^2$ to one proportional to $N~logN$ \citep{FW81}. 
Another method exploits the URA/MURA symmetry 
(large part of these arrays are given by the multiplication of the first line with the first column)
in order to reduce significantly the number of operations \citep{Roq87}\citep{Gol95}.
However today, at least for astronomical applications, the use of
the highly optimized routines of 2-d discrete fast Fourier transform (FFT)
available in most software packages for any kind of array order, 
is usually sufficient for the required implementations based on correlation. 
The search for fast algorithms or for specific patterns that allow fast decoding is 
therefore, these days, somehow less crucial.

Recently deep learning methods mainly based on
convolutional neural networks were proposed to improve the performance of 
image reconstruction from data of CMI in condition of near-field observations.
The tests performed for these specific conditions of terrestrial applications,
with their additional complexity of the source distance-dependent image magnification,
show that these novel techniques provide enhanced results compared to the simple correlation 
analysis \citep{ZhaR19}. Further developments in this direction can be expected in the near future.

\subsection{Deconvolution by Correlation in the Extended FOV} \label{subsec:Exde}

The cross-correlation deconvolution described in \S~\ref{subsec:CoDe}
for the FCFOV 
can be applied to the PCFOV, by extending the correlation of the decoding array $G$
with the detector array $D$ in a non-cyclic form to the whole field (EXFOV)
\citep{Gol95}\citep{Gol03}. 
To perform this a FOV-size $G$ array is derived from the mask array $M$ following a prescription 
that we describe below, and by padding the array with 0 elements outside $M$ 
in order to complete the matrix for the correlation. 

Since only the detector section modulated by the PC source is 
used to reconstruct the signal, the statistical error at the
source position and also the significance of the ghost peaks, if any, are minimized.
To ensure a flat image in the absence of sources, detector pixels
which for a given sky position correspond to mask opaque elements
must be balanced, before subtraction, with a proper ratio of the number of transparent 
to opaque elements for that reconstructed sky pixel.
This normalization factor is stored in a FOV-size array, called here $Bal$, 
and its use in decoding is equivalent to the so-called balanced
deconvolution for the FCFOV \citep{FC78}.

In order to correctly account for detector pixel contributions or even attitude drifts or 
other effects, a weighting array $W$ of the size of the detector array 
and with values comprised between 0 and 1 is defined 
and multiplied with the array $D$ before correlation \citep{Gol95}. 
It is used to neglect the detector areas which are not relevant (e.g., for bad, 
noisy, or dead area pixels) by setting the corresponding entries to 0. 
If one is interested in studying weak sources when a bright one is also present in FOV, 
$W$ may be used to suppress the bright source contamination by setting to 0 the $W$ entries 
corresponding to detector pixels illuminated by the bright source above some given fraction. 
The array $W$ is also used to give different weights to parts of the detector, for example 
when pixels have different efficiencies, e.g., due to different dead times or energy thresholds.
The balance array $Bal$ is built using $W$ to properly normalize the balance considering 
the weights given to the detector pixels.
Obviously when $W$ contains some zero values, it means that there is not 
complete uniform coding of the basic pattern, and this will break the perfect character 
of an optimum system, introducing coding noise. 
In case a small fraction of pixels is concerned the effect will be however small.

In order to insure the best imaging sensitivity, $G$ is built from the mask $M$ by:
$$ G =  \frac{1}{a} \cdot M - \textbf{1}$$
where the factor $a$ gives the aperture of the mask.
For $a$ = 0.5 (like in URAs) $G=2\cdot M-\textbf{1}$ and assumes
values +1 or --1 as in the standard prescriptions \citep{Fen78}.

Defining the two arrays $G^+$ and $G^-$ such that  
$$ G^+ = \cases{ G \rm{~~for~} G \ge 0 \cr
0 ~~ \rm{elsewhere} \cr}
~~~~~~G^- = \cases{ G \rm{~~for~} G \le 0  \cr
0 ~~\rm{elsewhere} \cr}
$$
where of course $G = G^+ + G^-$,
we obtain the reconstructed sky count image from
\begin{equation}
    S = {G^+ \star (D \cdot W) - Bal\cdot (G^- \star (D\cdot W)) \over A }
    \label{equ:anal-sky}
\end{equation}
where dot operator or division applied to matrices indicates here element-by-element matrix multiplication or division.
The balance array used to account for the different open to closed mask element ratios 
is given by
$$Bal = {G^+ \star W \over G^- \star W }$$
and ensures a flat image with 0 mean in absence of sources. 
The normalization array
$$A = (G^+ \cdot M) \star W - Bal \cdot ((G^- \cdot M) \star W ) $$
allows a correct source flux reconstruction which takes into account the partial modulation.
With this normalization the sky reconstruction gives at the source peak the 
mean recorded source counts within one totally illuminated detector pixel.
Note that source flux shall not be computed by integrating the signal around the peak, 
as this is a correlation image.
An additional correction for off-axis effects 
(including, e.g., variations of material transparency etc.) may have 
to be included, once the reconstruction, including ghost cleaning, has been carried out. 

The normalized variance, which is approximately constant in the FCFOV for optimum or pure random
masks, and whose relative value increases outside the FCFOV going towards the edges, 
is computed accordingly\footnote{Here $X^{2} = X \cdot X$}:
\begin{equation}
V = {(G^{+})^{2} \star (D \cdot W^2) + Bal^2 \cdot ( (G^{-})^{2} \star (D \cdot W^2) ) \over A^2 }
   \label{equ:anal-var}
\end{equation}
since the cross-terms $G^+ \cdot G^-$ vanish. 

Here it is assumed that the variance in the detector image is just given by the 
detector image itself (assumption of Poisson noise and not processing of the image); 
however if it is not the case the $D$ array in this last expression shall be substituted 
by the estimated detector image variance.
The signal to noise image is given by the ratio 
${S \over \sqrt{V}}$ 
and is used to search for significant excesses.
The deconvolution procedure can be explicitly expressed by discrete summations 
over sky and detector indices of the type given in \S~\ref{subsec:CoDe} for $S_{ij}$ \citep{Gol03}.

Different normalizations may be applied in the reconstruction \citep{SkiPon94}; 
for example one can normalize in order to have in the sky image the total number of counts 
in the input detector image.
However the basic properties of the reconstructed sky image do not change. 
In particular with the presence of a detector background there are more unknowns than 
measurements and therefore reconstructed sky pixels are correlated.
It is possible to show \citep{SkiPon94} that, at least for optimum masks the level of correlation 
is of the order of $1/N$ (where $N$ is again the number of elements in the basic pattern).
Clearly if binning is introduced then the level of correlation increases, depending
on the reconstruction algorithm employed as discussed below.

All the previous calculations can be performed in an efficient and fast way using the 
discrete fast Fourier transform algorithm because all operations involved are either 
element-by-element products or summations or array correlations for which 
we can use the correlation theorem\footnote{For which
$A \star B = IFT(\overline {FT(A)} \cdot FT(B))$
where FT is the Fourier transform, IFT the inverse Fourier transform
and the bar indicates complex conjugate.}.
\begin{figure}[h]
	\centering
 \includegraphics[width=1\textwidth]{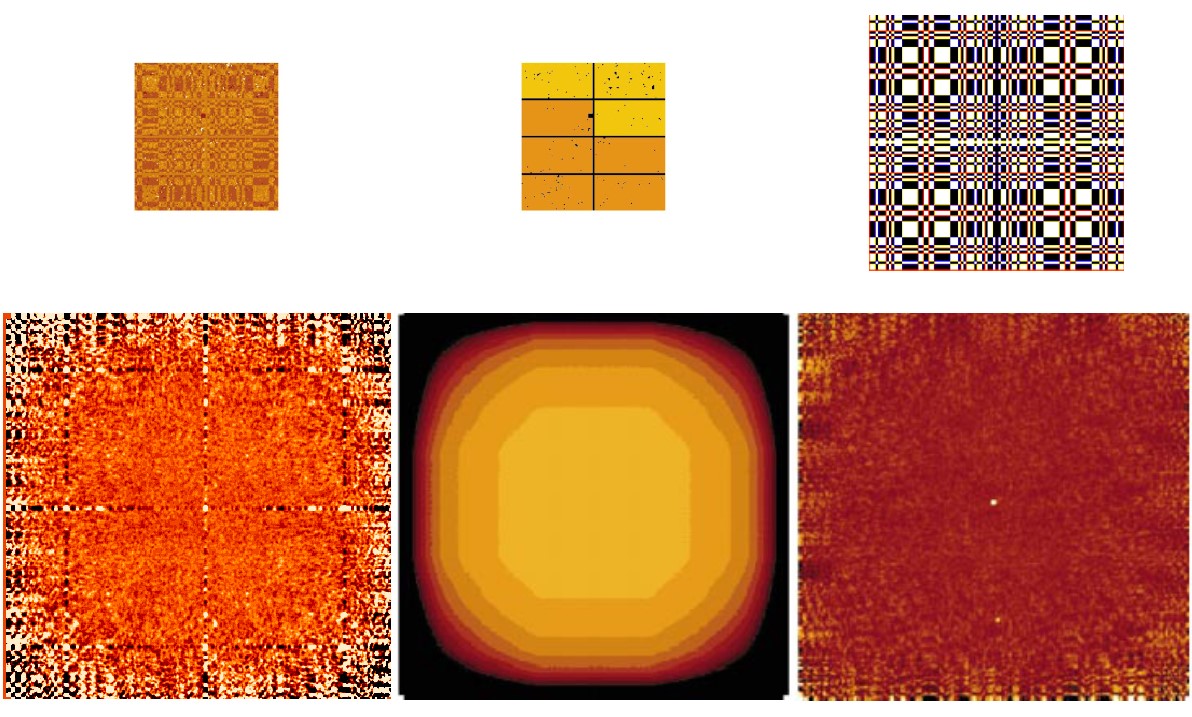}
	\caption{IBIS images showing the sky reconstruction process from data of the Cygnus region.
	From left/top to right/bottom: binned corrected detector image (130$\times$134 pixels) including dead zones ($D^C$), 
	associated efficiency image ($W$), rebinned MURA mask on a detector pixel grid (233$\times$233 pixels) ($M^R$), 
	decoded intensity sky image ($S$) (358$\times$362 pixels), associated variance ($V$) and final SNR sky image after 
	cleaning of coding noise of the two detected sources (Cyg~X-1 and Cyg~X-3).
	}
	\label{fig:decon}
\end{figure}

\subsection{Detector Binning and Resolution: Fine, Delta and Weighted Decoding} \label{subsec:Binn}

We have until now implicitly assumed to have a detector of infinite spatial resolution 
and data digitization for which images are recorded in detector elements (pixels) 
with the same shape and pitch as the mask elements 
and that sources are located in the center of a sky pixel, allowing for 
perfect detector recording of the projected mask shadow.
These approximations are of course not verified in a real system, 
which implies a degradation of the imaging performance.
Recorded photons are either collected in discrete detector elements (for pixelated detectors) 
or recorded by a continuous detector (like an Anger camera) subject to a localization error 
described by the detector point spread function (PSF), and where the measured positions are 
digitally recorded in discrete steps (pixels). In both cases we will have detector pixels
with a finite {\bf detector spatial resolution} characterized by the detector pixel size $d$ 
or the $\sigma_D$ of a Gaussian describing the detector PSF and the digitization.

Pixels may have sizes and pitches different from those of the mask elements,
but for a good recording of the mask shadow, resolution and digital pixels must be equal or smaller 
than the mask element size, otherwise the shadow boundary is poorly measured and there is a
large loss of sensitivity and in source localization.
One can define the {\bf resolution parameter $r$} as the ratio $r=m/d$ in each direction 
of the linear sizes of the mask element and the detector pixel 
(where pixel size means pixel pitch, since the physical pixel may be smaller with some 
dead area around it). 

Fenimore and Cannon \citep{FC81} considered the case of $r$ integer in both directions 
and showed that the same procedure of cross-correlation reconstruction 
can be carried out just by binning the array $M$ with the same pixel grid as the detector,
which will give the rebinned mask $M^R$, 
by assigning to all its pixels corresponding to one given mask element the value of that element,
defining $G$ accordingly ($G=2\cdot M^R-1$, for a=0.5) and then by carrying out the correlation over all pixels.
So, for example, for $r\times r$ detector pixels (square geometry) per mask element, 
each element of the mask is divided in $r\times r$ mask pixels.
To each of them, one assigns the value of the element 
and then carries out the G-definition and correlations accordingly. 
This is the {\bf fine cross-correlation deconvolution}. 

Another way, when building the decoding array $G$, 
is to assign the value of the mask element to one pixel 
from the $r \times r$ ones that bin this mask element
while the others are set to the aperture $a$. 
For a URA ($a$=0.5), the $G$ array has +1 or --1 for one pixel per mask element
and the others are set to 0 (and do not intervene in the correlation). 
This is the so-called {\bf delta-decoding} \citep{Fen80}\citep{FC81}.
This implies that the reconstructed adjacent $r\times r$ sky pixels are built using different 
pixels of the detector and therefore they are statistically independent. 
Of course a delta-decoding reconstruction can be transformed in fine decoded image 
by convolving the delta-decoded image with a $ r \times r$ box-function of 1s.
The delta-decoding also allows one to use the FHT in the case of detector
binning finer than the mask element
(if $M$ is a Hadamard array, the rebinned array $M^R$ build for the fine decoding is not) \citep{FW81}. 
As discussed above, FHT is not relevant anymore as the FFT can do the job, 
but the relative independence of delta-decoded sky image pixel over sizes of the SPSF peak 
was found useful in order to apply standard methods of chi-square fitting directly on
the reconstructed sky images including parameter uncertainty estimations \citep{Gol95}.

When there is a non-integer number of pixels per mask element, which is a typical, and sometimes
desirable\footnote{The exact integer ratio, when pixels are surrounded by dead zones, leads to 
incompressible localization uncertainty, given by the angle subtended by the dead area, 
for source positions for which mask element borders are projected within the dead zones.},
condition of pixelated detectors,
then the mask is rebinned by projecting the $M$ array on a regular grid with same
pixel pitch as in the detector and by assigning to the mask pixels the fraction of open 
element area projected in the pixel.
The same decoding array definition and correlation operation given above 
(Eqs.~\ref{equ:anal-sky}--\ref{equ:anal-var} and all associated definitions) are then 
applied using the {\bf rebinned mask array $M^R$} at the place of $M$.
$M^R$ can take (for non-integer $r$) fractional values 
between 0 and 1, and the decoding $G$ array also can have different
fractional values accordingly.
Weighing the inverse correlation using a filtered mask
describing the not-integer binning or the finite detector resolution 
optimizes the SNR of point sources \citep{Coo84}\citep{Gol95}\citep{Bou01} and 
is usually implemented ({\bf weighted decoding}) even if this implies a further 
smearing of the source peak. 
Fig.~\ref{fig:decon} shows some of the image arrays involved in the weighted sky
reconstruction process described above and applied to IBIS data (\S~\ref{subsec:Ibis}) 
of a Cygnus region observation \citep{Gol03}.

\subsection{Image Analysis} \label{subsec:Iana}

Following the prescriptions given above, one obtains a reconstructed sky
in the EXFOV of the instrument, composed of an intensity and a variance image. 
They are “correlation” images; each sky image pixel value is built by a linear operation 
on all, or part of, the detector pixels.
Sky image pixels are therefore highly correlated in particular within 
an area of one mask element.
Statistical properties of these images are different from standard astronomical images
and their analysis, including fine derivation of source parameters, 
error estimation, the various steps to reduce systematic noise from 
background or source coding noise and final combination of cleaned 
images in large mosaics must take into account their characteristics.

\subsubsection{Significance of Detection} \label{subsubsec:Sign}

\begin{figure}[h]
	\centering
 \includegraphics[width=0.50\textwidth]{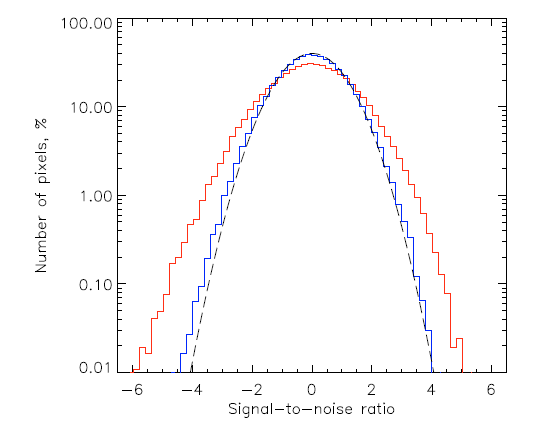}
 \includegraphics[width=0.49\textwidth]{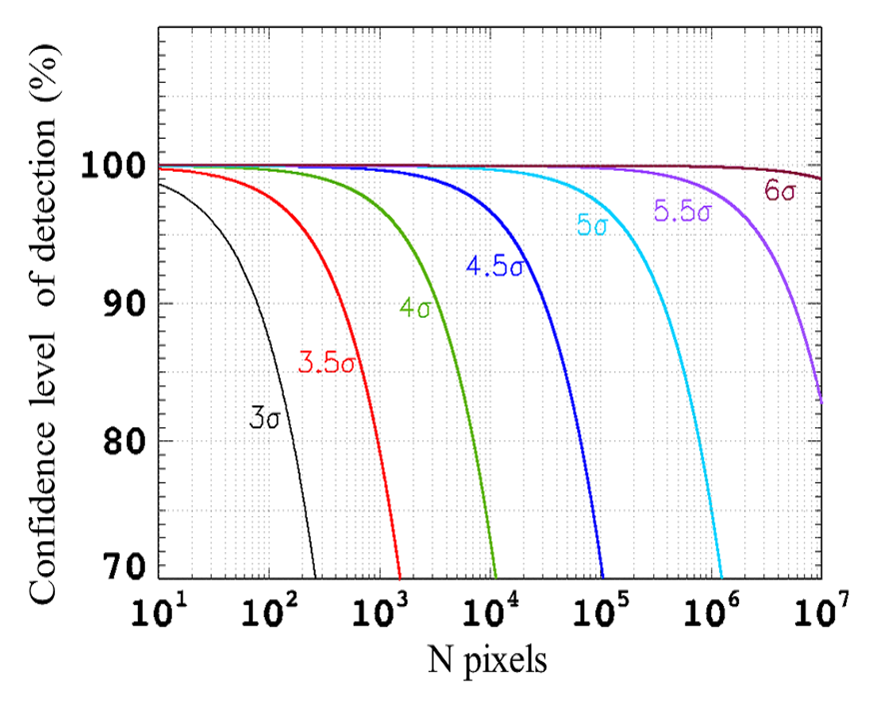}
	\caption{
	Left: SNR distribution of a CMI (IBIS/ISGRI) reconstructed sky image without sources before (red) 
	and after (blue) correction of some systematic effects, compared to the expected normal distribution 
	(dashed line) (reproduced with permission from \citep{Kri10}, © ESO).
	Right: Variation of the confidence level of detection (in \%),
	for no a priori knowledge of source position, as function of number of sky pixels 
	in a CMI sky image for different SNR levels of detection.
	}
	\label{fig:cmi-SNR}
\end{figure}

The reconstructed and normalized sky image shall be searched for 
significant peaks by looking for excesses over the average value, that should be, 
by construction (and neglecting the effect of non uniform background) close to zero.
This is done by searching for relevant peaks in the SNR image.
In the absence of systematic effects the distribution of this SNR image shall follow 
the standard normal distribution. 
Deviations from such distribution indicates residual systematic effects or 
presence of sources and their ghosts (Fig.~\ref{fig:cmi-SNR} left).

Excesses in signal to noise larger than a certain threshold are considered as sources.
However the concept of significance level in such a decoded image 
where each sky pixel is built by correlating all, or part of, the pixels of the 
detector image needs to be carefully considered.
If we are interested to know if one or few sources at given specified positions
are detected, then we can use the standard rule of the 3 sigma excess
that will give a 99.7$\%$ probability that the detected excess at that precise position
under test is not a background fluctuation.
If, instead, we search all over the whole image for a significant excess, then 
the confidence level must take into account that we perform a large number of 
trials (in fact different linear combinations of nearly the same data set) 
to search for such excess.

Assuming standard normal distribution for the noise fluctuations,
the probability that an excess (in $\sigma$) larger than $\alpha$ is produced by noise is 
$$P(\alpha) = {1 \over \sqrt{2 \pi}} \int _{\alpha}^{+\infty} {e^{-x^2 \over {2}} dx} = {1 \over 2} erfc({\alpha \over \sqrt{2}})$$
The confidence level of a detection (not a noise fluctuation) is then $1 - P(\alpha) $
in a single trial. 
Assuming that we have $N$ independent measurements then the confidence level for such excess 
to be a source will be reduced to 
$$[1 - P(\alpha)]^{N} \sim 1 - N P(\alpha)   ~~~~~~for ~NP(\alpha) \ll 1 $$
For a given confidence level and $N$, the value of $\alpha$ is found from this relation.
Curves of $\alpha$ as a function of $N$ can be calculated 
(see Fig.~\ref{fig:cmi-SNR} right and \citep{Car87})
and it is found that to have a confidence level of 99$\%$ for number of pixels N=$10^4-10^5$, 
the excess must be in the range 4.5-6.0.
For coded mask however it is difficult to evaluate $N$, since it does not simply correspond 
to the number of pixels in the sky reconstructed image, unless this refers to the FCFOV
of an optimum system with one detector pixel per mask element.
The reason is that in general sky pixels are not fully independent 
and are highly correlated over areas of the size of the typical SPSF.
The best way to evaluate the threshold is therefore through simulations.
A value of 5.5-6.5 $\sigma$ is typically assumed for a secure (may be conservative) source 
detection threshold in reconstructed images of 200-300 pixel linear size.
%
%
\subsubsection{System Point Spread Function} \label{subsubsec:Spsf}

\begin{figure}[h]
	\centering
 \includegraphics[width=0.40\textwidth]{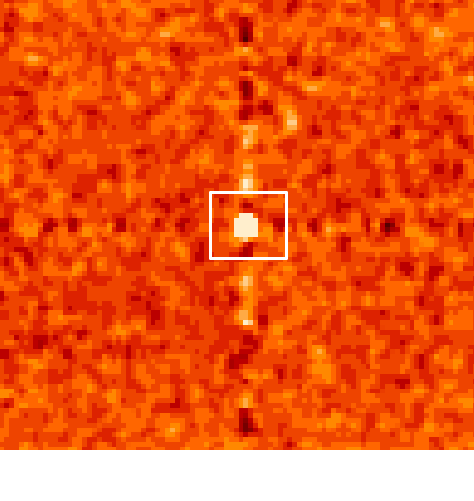}
 ~~~~~
 \includegraphics[width=0.48\textwidth]{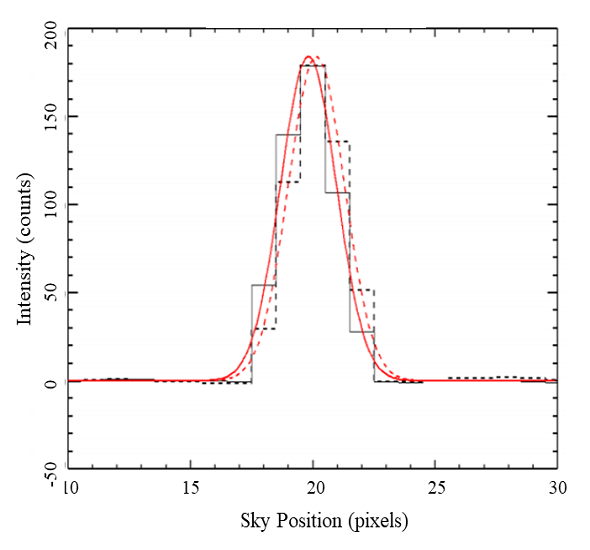}
	\caption{IBIS/ISGRI SPSF from data of a point source observation. 
	Left: Source peak in the center of the FCFOV of a decoded sky image. 
	Color code highlights the, rather low, coding noise particularly affecting 
	(given the symmetry of the MURA mask) the image axes centered on the source. 
	Right: Source profiles in a decoded image along the 2 axis (black lines) 
	and the Gaussian model, approximation of the SPSF, that best fits the excess 
    (red lines) (from \citep{Gro12}).
	}
	\label{fig:ibis-spsf}
\end{figure}

An isolated significant excess in the deconvolved sky image 
may indicate the presence of a point-like source,
which will be characterized by the System Point Spread Function (SPSF), that is the spatial 
response to an isolated bright point source of the overall imaging system, 
including the deconvolution process (Fig. \ref{fig:ibis-spsf}).
The SPSF includes a possibly shift-invariant, main peak, proportional to the source intensity, 
and usually non-shift-invariant, side-lobes of the coding noise, 
also proportional to the source intensity.
For a perfect cyclic optimum coded mask system, 
the main peak is shift invariant and the side-lobes are flat within the FCFOV 
for a source in the FCFOV, but large side-lobes appear in the PCFOV (ghosts) 
along with a diffuse moderate coding noise, 
and when the source is in the PCFOV, the width of the main peak may vary depending 
on the mask pattern, and side-lobes, including the main ghosts, appear all over the field. 
In random masks, side-lobes are distributed all over the image including in the FCFOV, 
even for sources in the FCFOV,
but, generally, with low amplitude and without the strong ghosts typical of cyclic systems.

For a pixelated detector and a sky reconstruction based on the weighted
cross-correlation as described in \S~\ref{subsec:Exde}-\ref{subsec:Binn}, 
the SPSF can be described by a peak function 
correlated with a set of positive and negative delta functions of different amplitudes
(what we will call here the {\it correlation function})
that take into account the mask pattern and the decoding operation
based on correlation (see, e.g., \citep{Fen80}).
A positive $\delta$-function of maximum amplitude
of this set is of course positioned at the source location and 
will provide the main peak of the SPSF at the source position. 
The other positive and negative deltas, convolved with the peak function, 
describe the coding noise spread over the image (including ghosts).
Assuming from hereon a square geometry with square mask elements of linear dimension $m$ 
and square detector pixels of linear dimension $d$ 
(extension to rectangular geometry is trivial and analog, less trivial, 
relations can be given for the hexagonal one)
the peak function \textbf{Q} is given by the normalized correlation of four 2-d box 
functions\footnote{A 1-d box function is given by: $\prod_p(z) = \cases{ 1 \rm{~for~} |z| \le p/2 \cr
0 ~~ \rm{elsewhere} \cr} $}, 
two of mask element width $\prod_m(x,y)= \prod_m(x)\cdot \prod_m(y)$ 
and two of pixel width $\prod_d(x,y)=\prod_d(x)\cdot \prod_d(y)$ 
$$
\textbf{Q}(x,y) = Q(x) \cdot Q(y) 
~~~~~~{where}~~~ 
 Q(x) = {\prod_m(x) \star \prod_d(x) \star \prod_m(x) \star \prod_d(x) \over {d^2 m}}
$$

This function, a blurred square pyramidal function for square geometry,  
can be expressed analytically. 
The 1-d analytical function $Q$ that composes it has a peak value (at zero lag)
given by the simple equation
\begin{equation}
    Q(0) = 1 - {1 \over{3 r}}
    \label{equ:anal-Q0}
\end{equation}
where, as usual, $r$ is the ratio $r=m/d$. 
This quantity, which corresponds to the term {\it coding power} in \citep{Ski95}, is important 
because it appears in the expression of the error estimate for the source flux and location.

The 2-d function \textbf{Q}, 
can be also conveniently approximated by a 2-d Gaussian function with FWHM width 
of $\sqrt{r^2+1}$ along the two axes (Fig. \ref{fig:ibis-spsf} right).
For a continuous detector the SPSF (where now pixel is the data discrete sampling step) 
is the function above but further convolved with the detector PSF.
The explicit formulae of the SPSF for such a system, where the detector PSF is 
approximated by a Gaussian, 
were given in the description of SIGMA/GRANAT data analysis \citep{Gol95}\citep{Bou01}.

The use of the SPSF in the analysis of CMI data is important because in general 
both the detector resolution and the sampling in discrete pixels are finite. 
Then the discrete images produced by the correlations, with the same steps of the data sampling, 
do not provide the full information, unless the resolution is exactly given by the sampling, 
pixels are in integer number per mask element and the source is exactly located at the center of
a sky-projected pixel in order to project a shadow exactly sampled by the detector pixels.
Of course an artificial finer sampling can be introduced in the correlation analysis, 
but this implies rebinning of data with alteration of their statistical properties and
increase in computing time for deconvolution (dominant part of the overall processing),
and finally the precision may not be adequate to the different level of SNR of the sources, 
where the brightest ones may be located with higher accuracy than the artificial oversampling used.

Therefore, in order to evaluate source parameters, and in particular the position 
of the source, in a finer way than provided by the sky images with the sampling equivalent 
to the detector pixels, a practical method is to perform a chi-square fit 
of the detected excess in the deconvolved sky image with the continuous SPSF peak 
analytical formula or its Gaussian approximation (see Fig. \ref{fig:ibis-spsf} right). 
The procedure can also be used to disentangle partially overlapping sources \citep{Bel06}.
Once the fine position of the source is determined, a model of the projected image 
on the detector can be computed and used to evaluate the source flux, 
subtract the source contribution or its coding noise, and perform simultaneous 
fit with other sources and background models to extract spectra and light curves.
%

Even though the fit, in the deconvolved image, of the source peak with the model of the SPSF 
peak will provide a reasonable estimate of the source parameters,
the error calculation cannot be performed in the standard way directly 
using the chi-square value of the best fit and its variation around the minimum, 
because pixels are too much correlated.
Nevertheless formulae for the expected error in source flux and source localization 
can be derived from the formalism of chi-square estimation in the detector space 
and can be used to provide uncertainties, after some {\it calibration} 
on real data that will account for residual systematic biases.

\subsubsection{Flux and Location Errors} \label{subsubsec:Erro}

One can show that the correlation reconstruction for a single point-like source 
in condition of dominant Gaussian spatially flat background noise is equivalent to the minimum 
chi-square determination of source flux and position in the detector image space,
where one can determine the errors, using the minimum chi-square paradigm.

Using the notations used above for the SPSF and introducing
the terms $t$ for integration time, $A$ for detector geometrical area,
$b$ (in $cts/s/cm^2$) for a background count rate, 
and $s_0$ (in $cts/s/cm^2$) for the considered source count rate, 
both integrated within an energy band, we define 
$$ 
\sigma_{CR} = \sqrt{b \over{A \cdot t} } 
~~~~~~~ SNR_{CR} = {s_0 \over \sigma_{CR} } 
~~~~~~~ f_I(x,y) = {SPSF(x,y) \over N \cdot a} - a
$$
where $f_I$ is called the {\it image function} and is linked, 
as shown, to the shape of the SPSF, and $N$ is the average number of mask elements 
in the detector area $A$ ($N=A/m^2$) (and in the basic pattern for an optimum system).
$\sigma_{CR}$ and $SNR_{CR}$ are, respectively, the minimum error and maximal signal to noise
from purely statistical noise given by the measured count rates 
in case perfect reconstruction can be achieved 
and for a mask aperture of 1 (i.e., no mask, where all the area $A$ is used for the measure) 
with the idealistic assumption that a 
measurement of the background is available (in the same observation time t).  

Using the minimum chi-square method
applied to the detector image compared to a source shadow-gram model,
one obtains, from the inversion of the Hessian matrix of the chi-square function, 
the expression for source flux and position errors,
expressed as 1~$\sigma$ at 68.3\% confidence level in one parameter, 
which are related to the image function and to its second partial derivatives 
\citep{Coo84}\citep{Fin87}\citep{Int94}. 

The flux error is given by
\begin{equation}
\sigma_S = \sigma_{CR} \sqrt{1 \over{a \cdot f_I(0,0)}} = \sigma_{CR} {1 \over Q(0)} \sqrt{1 \over{a (1-a\cdot f_M)}} 
    \label{equ:anal-flerr}
\end{equation}
where the {\it mask function} $f_M$ is 1 for optimal masks and, in average, for random masks 
and is given by a more complex relation for the general case, which involves
the cross-correlation of the mask pattern. The SNR is then
\begin{equation}
SNR = SNR_{CR} \cdot Q(0) \cdot \sqrt{a (1-a\cdot f_M)}
    \label{equ:anal-snr}
\end{equation}

The location error along one direction also can be expressed 
by an analytical expression and involves the second derivative of the image function:
\begin{equation}
\sigma_X = {1 \over SNR_{CR}} \cdot \sqrt{1 \over{a \cdot \left| {\partial^2 f_I(0,0) \over \partial x^2 } \right|}} = K_{X} \cdot {d \over SNR} 
    \label{equ:anal-loerr}
\end{equation}

For optimal masks (URA, MURA, etc.) with $a=0.5$ as well as for random masks in average 
the following formula for the constant $K_{x}$ holds approximately:
\begin{equation}
 K_{X} = {\sqrt{r \cdot Q(0) \over 2}}
    \label{equ:anal-Kx}
\end{equation}

The error here, as for the flux is given at 1~$\sigma$ along one axis direction.
The fact that Eqs.~\ref{equ:anal-loerr}-\ref{equ:anal-Kx} hold for both URAs and random masks
does not mean that these mask types always have the same localization capability 
as their signal to noise is not the same if the aperture is different.
These expressions are equivalent to those reported in \citep{Coo84} and \citep{Fin87} 
and can be extended to the case of continuous detectors by replacing 
the pixel linear dimension $d$ with the value $2\sqrt{3}\sigma_D \approx 1.5 w_D$ 
where $\sigma_D$ is the detector spatial resolution in $\sigma$ and $w_D$ 
in FWHM\footnote{Following \citep{Ski95} the numerical factor comes from the fact that 
$1 \over{2 \sqrt{3} }$ is the rms uncertainty in a variable which is known 
to plus or minus half a pixel.}.

A more complicated  expression, involving properly computed $f_M$ 
and its second partial derivatives, can be obtained for general (or not-so-random) masks.
We do not show it explicitly here because it is too cumbersome, but it has been used to identify
which quasi-random masks that cannot be purely random in order to make them self-supporting
have optimal sensitivity-location accuracy pairs.

\subsubsection{Non-uniform Background and Detector Response} \label{subsubsec:Back}

In gamma-ray astronomy the background is generally dominant over the source contribution.
Its statistical noise, spatial structure and time variability are therefore important 
problems for any kind of instrument working in this energy range.
CMI, unlike non-imaging instruments, 
allow measurement of the background simultaneously with the sources, 
limiting the problems linked to its time variability. 
However if the background is not flat over the detector plane, 
its inherent subtraction during image deconvolution does not work properly. 
In fact any spatial modulation is even magnified by the decoding procedure
\citep{Lau88}. 
Therefore the non-uniform background shall be corrected before decoding as well as 
any non-uniform spatial detector response which may affect both the background and 
the source contributions.

Using an estimation of the detector spatial efficiency $E$ for the given observation
(spatial efficiency variations due to noisy pixels, dead times or other time-varying effects) 
and of the detector non-uniformity $U$ 
(quantum efficiency spatial variation depending on energy),
along with a measure (e.g. from empty field observations already corrected for both $E$ and $U$) 
or a model of the background shape $B$,
a correction of the detector image $D$ affected by non-uniformity can be given by
$$D^C= \frac{D}{E \cdot U} - \overline{b} \cdot B $$
and use then this corrected image $D^C$ to reconstruct the sky (Eqs.~\ref{equ:anal-sky}-\ref{equ:anal-var}). 
The background normalization factor $\overline{b}$ can be computed from the ratio of the 
averages of the input detector and background images or from their relative exposure times.
If one can neglect the variance of both $B$ and $\overline{b}$ 
and assuming the Poisson distribution 
in each detector pixel, the variance of the corrected image can be approximated  by 
$$\sigma^2_{D^C} = {\frac {D} {{E^2} \cdot {U^2{}}}} $$

This implies that in the computation for the sky variance (Eq.~\ref{equ:anal-var}), 
this detector variance shall 
be used instead of simply the detector image $D^C$.

Of course the details of the procedures, 
including other different, more sophisticated correction techniques, 
to account for spatial modulations not due to the mask, 
depend on the instrument properties, observing conditions, and calibration data 
(see, e.g., \citep{Gol03}\citep{Seg10}).
In general extensive ground and in-flight calibrations, including empty-field observations, 
will be needed in order to get the best models of the background  
and of the instrument response.

One typical contribution to a non-uniform background is the CXB, dominant at low energies, 
and whose contribution on the detector plane, despite its isotropic character, 
becomes significantly non-uniform for large FOVs.
%
\begin{figure}[h]
	\centering
 \includegraphics[width=0.98 \textwidth]{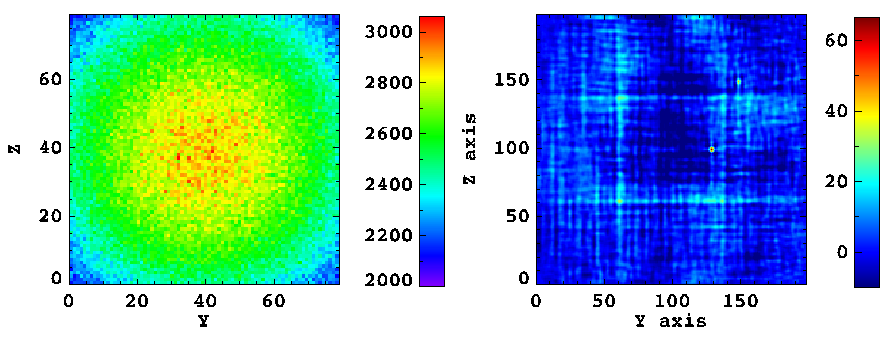}
	\caption{Non-uniform CXB in ECLAIRs. 
	Left: Simulation of the CXB intensity (counts/pix) on the ECLAIRs detector 
	during one orbit.
	Right: Decoded sky SNR map of the detector image (left), 
	when two bright sources (SNR$>$60) are also added in the simulation
	and no correction for the non-uniform background is performed.
	The sources can be barely seen in spite of their high SNR. 
    }
	\label{fig:non-unif-bkg}
\end{figure}
%
In fact the CXB is viewed by each detector pixel through all the instrument opening 
with different solid angles, dependent on the instrument geometry 
(mask holes, shield, collimator, supporting structures, etc.).
An example of such effect expected on the ECLAIRs detector plane 
is shown in Fig.~\ref{fig:non-unif-bkg}, 
which also illustrates the noise that this effect produces in the decoded sky image 
if not properly corrected before reconstruction.
In \S ~\ref{subsec:Ecla}, dedicated to ECLAIRs, we discuss the further modulation of the CXB
produced when the Earth enters the instrument FOV.

Other observational conditions can also be very important in this regard.
For satellite orbits that intersect the radiation belts or the South Atlantic Anomaly,
parts of the satellite, instrument and the detector itself
may be activated during the passage through this cloud of high-energy particles.
The non-uniform distribution of the material in or around the detector may produce
an additional non-flat time-varying background remnant that will spoil the images.
A careful study of these effects is often required in order to introduce proper
corrections in the analysis.

\subsection{Overall Analysis Procedure, Iterative Cleaning and Mosaics} \label{subsec:Iros}
Once the raw data, possibly in event list form, are calibrated and binned in detector images,
along with their weighting array,
and then background, non-uniformity and efficiency are corrected, 
the decoding can be performed by applying Eqs.~\ref{equ:anal-sky}-\ref{equ:anal-var} 
and prescriptions given in \S~\ref{subsec:Exde}-\ref{subsec:Binn}
in order to derive preliminary sky images. 
Point sources are then searched throughout them by looking for SNR significant peaks. 
The detected source is finely located by fitting the peak of the SPSF function to the detected excess. 
A localization error can be associated (Eqs.~\ref{equ:anal-loerr}-\ref{equ:anal-Kx}) 
from the source SNR which allows to select the potential candidates for the identification.
%
\begin{figure}[h]
	\centering
    \includegraphics[width=1\textwidth]{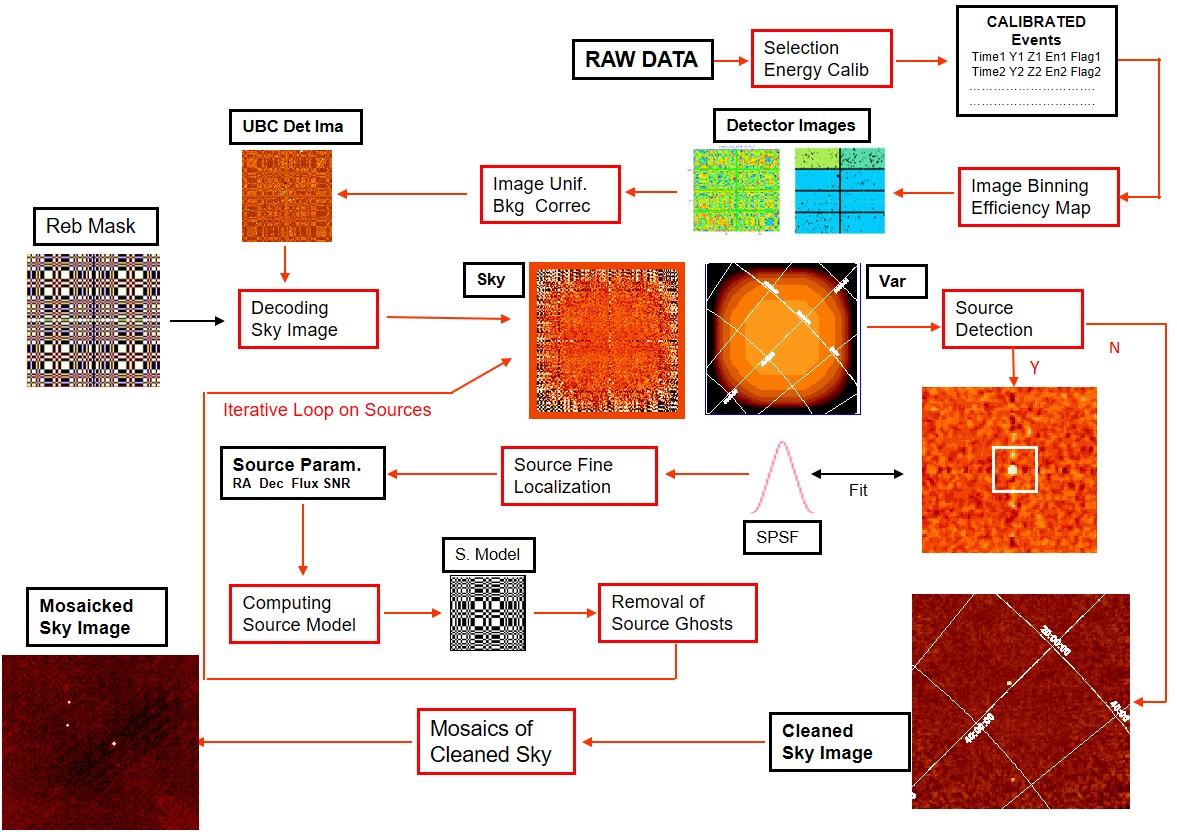}
	\caption{Simplified scheme of an overall image analysis procedure for CMI data including selection 
	and binning of corrected/calibrated events, background and non-uniformity correction, 
	decoding using the mask pattern, an IROS cleaning procedure on detected sources, and finally image mosaic, 
	illustrated using the IBIS images \citep{Gol03}.}
	\label{fig:IBIS-iros}
\end{figure}

Iterative cleaning of coding noise from detected sources is performed in order to 
search for the weaker objects.
This is done by modeling each source and subtracting its contribution, 
either in the detector image, 
which then must be decoded to look again for new sources, or directly in the deconvolved one.
Typically the procedure is iterative, starting with the most significant source in the field
and going on to the weaker sources, one by one, until no excess is found above the established 
detection threshold.
Few iterations can be implemented, by restarting the procedure with the source fluxes corrected 
by the contamination from all other sources, for a deeper search.
For close sources with overlapping main peaks a simultaneous fit of their SPSFs 
may have to be implemented.
A catalog is usually employed to identify, and even to facilitate the search, of the sources.
This iterative cleaning procedure has been sometimes called {\bf Iterative
Removal Of Sources (IROS)} \citep{Ham92}\citep{Gol03}, 
the most important element of which is the proper estimation of the source contribution 
in the recorded image which depends on a well-calibrated model of the instrument.
As for the background correction, very often the source modelling is not perfect, and the ghost
cleaning procedure leaves systematic effects which may dominate the noise in the images 
of large exposure times or on large sets of combined data.

One way to smear out background and source residual systematic noise 
is to combine reconstructed images from different pointing directions and orientations.
Overlapping cleaned sky images can be combined, 
after a normalization accounting for off-axis losses, in sky mosaics 
by a proper roto-translation to a common grid frame
and then a weighted sum using the inverse of variance as the weight. 
While this is a standard procedure in astronomy imaging, here again one has to remember that
we are treating correlation images and that the combined variance shall be computed 
including the co-variance term. 
The combination of images may take different forms depending on the scope of the mosaics 
(e.g., preserve source flux estimation versus reducing source peak smearing) \citep{Ski87b}\citep{Gol03}.

A schematic picture of the overall analysis procedure using the IBIS images as example 
is shown in Fig.~\ref{fig:IBIS-iros} (see also the procedures described by \citep{Seg10} and \citep{Kri10}).

\section{Coded Mask System Performances} \label{sec:Perf}
From the error estimations (\S~\ref{subsubsec:Erro}), one can determine the expected   
CMI performance as function of instrument parameters and design.
It is usually evaluated in terms of sensitivity, angular resolution, 
localization accuracy, field of view and shape of SPSF.
We already discussed the FOV in \S~\ref{subsec:Defi} and the SPSF in \S~\ref{subsubsec:Spsf}.
\subsection{Sensitivity and Imaging Efficiency} \label{subsec:Sens}
The sensitivity of a coded mask system is given by the minimum point-like source flux 
that can be detected above a certain significance level $n_{\sigma}$.
The lower the minimum flux, the higher the sensitivity of the instrument.

This minimum flux can be derived as function of the CMI parameters 
from the flux error estimation of Eq.~\ref{equ:anal-flerr}. 
Let $\epsilon$ be the detector efficiency (we neglect here energy redistribution) 
and $\tau_o$ and $\tau_c$, respectively, the transparencies of the open and closed mask elements,
all dependent on the energy $E$ of the incident radiation;
then for given observation conditions and detector and mask parameters, 
with symbol meaning as in previous sections, 
assuming Gaussian statistics and neglecting systematic effects 
(see \citep{Ski08} for different hypothesis),
the continuum sensitivity $F_S$, in units of ph/cm$^2$/s/keV, of a coded aperture system, 
on-axis and in the energy interval $\Delta E$ around $E$ (keV), is given by
\begin{equation}
F_S = n_{\sigma}^2  \cdot 
{ [ (1-a) \cdot \tau_o + a \cdot \tau_c ] +
 \sqrt{[ (1-a) \cdot \tau_o + a \cdot \tau_c]^2 + { 4 \cdot t \cdot b \cdot A (\tau_o-\tau_c)^2 \cdot a \cdot (1-a)  \over n_{\sigma}^2} }
\over 2 \cdot \epsilon \cdot A \cdot t  \cdot \Delta E \cdot (\tau_o-\tau_c)^2 \cdot a \cdot (1-a) }
    \label{equ:perf-sensi}
\end{equation}
In the case of dominant background the same relation holds with
the term $[(1-a) \cdot \tau_o + a \cdot \tau_c]$ at the numerator set to zero.

Equation~\ref{equ:perf-sensi} can be solved toward $n_{\sigma}$ for a given source flux $F_S$ 
providing the upper signal to noise (SNR) limit attainable on-axis for that exposure or
toward time $t$ to have the observation exposure needed to reach the desired detection 
significance $n_{\sigma}$ for a given source flux $F_S$.

This formula, and the analogous ones for SNR and exposure,
usually found in the literature (e.g., \citep{Car82}\citep{Ski08}), 
neglects both the mask pattern and the finite spatial resolution of the detector. 
Therefore it approaches the case for optimum or pure random mask systems with infinite
resolution or with integer resolution parameter $r$ and 
the source exactly located in the center of a sky pixel, that is, when  
detector pixels are all either fully illuminated or fully obscured for that source.
This is in fact the most favorable configuration and gives the highest sensitivity, 
but in the general case one must take into account the effect of the finite 
detector spatial resolution which is dependent on source position in the FOV.
As discussed in \S~\ref{subsubsec:Erro}, this gives an additional 
loss in the SNR due to {\it imaging}, 
which, averaged over source location within a pixel, is given by the term $Q(0)$ 
of the SPSF, which depends on the resolution parameter $r$ through Eq.~\ref{equ:anal-Q0}. 

We therefore define the {\bf imaging efficiency} as $\epsilon_I = Q(0) = 1 - {1 \over 3r}$.
The formula of Eq.~\ref{equ:perf-sensi} for the sensitivity can be used as it is
also when including an average imaging loss over a pixel size, 
if one replaces everywhere the value $n_{\sigma}$ with ${n_\sigma}_I 
= {n_\sigma \over \epsilon_I}$. In the same way, 
the SNR derived from Eq.~\ref{equ:perf-sensi} will be reduced to an 
{\it imaging} SNR by a factor given by the imaging efficiency, i.e.,
$SNR_I = SNR \cdot \epsilon_I$.

This formula, modified with the imaging efficiency corresponds to Eq.~\ref{equ:anal-snr}
for the SNR discussed in \S~\ref{subsubsec:Erro}, 
and approximates well the sensitivity within the FCFOV when the source position
is known and the flux evaluation is performed by fitting the SPSF at the source position, 
or, which is the same, by correlating with a rebinned mask shifted at the exact source position.
If one wants to include in the calculation the fact that the source position is not known
(e.g., to establish the detection capability of unknown sources in the images) 
then an additional loss shall be included 
which takes into account that the deconvolution is performed in integer steps (sky pixels) 
usually not matching the source position (the {\it phasing error} of \citep{FW81}). 
If the source location is not in the center of a pixel, its peak will be spread over the 
surrounding pixels in the reconstructed image and the SNR for the source will appear lower.
In this case the prescriptions given above hold, but
the expression for the imaging efficiency shall be replaced by
the integral over 1 pixel of the SPSF peak, which can be approximated by
$\epsilon_I \approx 1 - {1 \over 2.1 r}$
for pixelated detector and square geometry \citep{Gol01}. 
For example for the IBIS/ISGRI system ($r$~=~2.4), the average (over a pixel) 
imaging efficiency is $\epsilon_I$~=~0.86 for known source location (fit of detected peak) 
and 0.80 for unknown location (peak in the image).

The sensitivity formula, and its extensions for different hypothesis, 
can also be used to determine the optimum open fraction $a$ of the 
mask \citep{Fen78}\citep{Ski08}.
One can easily see that for dominant background 
($b \gg s$) the SNR is optimized for $a$~=~0.5. 
However if $b$ is not dominant or if the sky component of the background is relevant, 
or in other applications like nuclear medicine \citep{Acc01},
open fractions lower than 0.5 are optimal. 
As discussed by \citep{Int94} and \citep{Ski08} however, the optimum value 
varies slowly with parameters and remains generally close to 0.4-0.5.

Other elements of the CM imaging system have an influence on the sensitivity. 
They are the used deconvolution procedure, the background shape and its correction, 
the source position knowledge, and of course the numerous systematic effects 
that may be present and some mentioned in the subsection on "real systems".
Also a decrease of the sensitivity with the increase of the source distance 
from the optical axis is present due to reduction of mask modulation,
the vignetting effect of the mask thickness, 
and the possible variation of open and closed mask element transparency 
with the incident angle.
In this case the additional sensitivity loss dependence on source direction angle 
shall be integrated in the term of the detector efficiency $\epsilon$ 
in Eq.~\ref{equ:perf-sensi}, which then becomes dependent on energy but also 
on the source direction angle $\theta$, that is $\epsilon = \epsilon(E,\theta)$.

\subsection{Angular Resolution} \label{subsec:Anre}

The separating power of a CMI system is basically determined by the angle 
subtended at the detector of one mask element.
However the finite detector spatial resolution also affects the resolution. 
For a weighted cross-correlation sky reconstruction 
(Eqs.~\ref{equ:anal-sky} - \ref{equ:anal-var}),
the resulting width of the on-axis SPSF peak in one direction, 
which gives the angular resolution (AR) in units of sky pixels, 
is well approximated, with the usual meaning of resolution parameter $r$, 
for square geometry and a pixelated detector, by 

\begin{equation}
  AR(FWHM) = \sqrt{r^2 + 1}
  \label{equ:perf-angres}
\end{equation}

To obtain the angular resolution in angular units (radians) on-axis 
one has to take the arc-tangent of this value divided by the mask 
to detector distance $H$ (Table~\ref{tab:cm-prop}).
Of course the angle subtended by a pixel varies along the FOV because of projection 
effects, and that shall be considered for the off-axis values.
Moreover the separating power may vary along the FOV and particularly in the PCFOV 
because the coding noise may deform the shape of the SPSF 
main peak while vignetting effect of mask thickness will reduce its width.
Figure~\ref{fig:ibis-spsf-psle} left shows the fitted width (in pixel units) 
of the IBIS SPSF, along the two image axes passing by the image center.
The width is consistent with the AR value of Eq.~\ref{equ:perf-angres} and of Table~\ref{tab:cm-prop} 
within the FCFOV but changes wildly in the PCFOV \citep{Gro03}.
In any case the SPSF width of a system can be evaluated 
at any location in the image, 
and the fitting procedure applied to detected sources can either use 
the fixed computed value or let the width be a free parameter.

\subsection{Point Source Localization Accuracy} \label{subsec:Psla}
An essential characteristic of an imaging system is the quality of the localization
of detected sources.
As we have seen in the analysis section the fine localization of a detected point-like 
source within the pixels around the significant peak excess 
shall be derived by a fitting procedure.

This is usually implemented as a fit of the source peak in the decoded sky image 
with a function that describes \citep{Gol95}\citep{Bou01} or approximates 
(e.g., a bi-dimensional Gaussian function) \citep{Gol03}\citep{Gro03} the SPSF, 
but can in principle (with a more complex procedure
which for each tested location models and compares to data the shadow-gram 
of the studied source) be performed on the detector image.
For this last implementation, formal errors can be derived and can be related
to the SPSF and the source strength.
As discussed in \S~\ref{subsubsec:Erro} the uncertainty on the location is  
inversely proportional to the significance of the source. 
The typical procedure is then to express the location error 
radius for a given confidence level, as a function of the source SNR.
Once the function is defined and calibrated for a given system the error
to be associated to the fitted position of a source is derived from its SNR.

Using the relation for the location error $\sigma_x$ (1-$\sigma$ error along one direction) 
of Eqs.~\ref{equ:anal-loerr}-\ref{equ:anal-Kx}
and assuming that the joint distribution of errors in both directions is bi-variate normal 
and that they are uncorrelated, one can apply the Rayleigh distribution 
to obtain and relate to the system parameters (including $r$), 
the 90$\%$ confidence level error radius as 
$
PSLE(SNR) = \sqrt{2 \cdot ln 10} \cdot \sigma_x (SNR).
$

The error can be expressed in angular units by taking the arc-tangent of the value 
divided by the mask to detector separation $H$, 
with the usual caveat that off-axis projection effects shall be considered.
In \citep{Ski08} the location error was rather approximated
with the expression for the angular resolution (Eq.~\ref{equ:perf-angres}) 
divided by the SNR, while in \citep{Car87} with 
the angle subtended by the PSD spatial resolution divided by the SNR.
A more accurate and, for optimum or random masks, formally correct
approximation with the explicit dependence on $r$ is in fact
\begin{equation}
PSLE(SNR) \approx{ \arctan { \left({ \sqrt{ ln 10} \over SNR} \cdot {d \over H} \cdot \sqrt{ {r - {1 \over 3} } }\right) } }
\label{equ:psle-vs-snr}
\end{equation}
which gives the 90\% c.l. angular error radius of the estimation of a location of an
on-axis source with signal to noise SNR.

The SNR to use in the above expression 
is the imaging SNR$_I$ for a SPSF fit at the source location. 
If one wants to use the SNR measured in the images (in average affected by sampling)
the value that should be used is the average estimation SNR$_I$ for unknown location, 
in which case the constant of Eq.~\ref{equ:psle-vs-snr} changes.

\begin{figure}[h]
	\centering
    \includegraphics[width=0.52\textwidth]{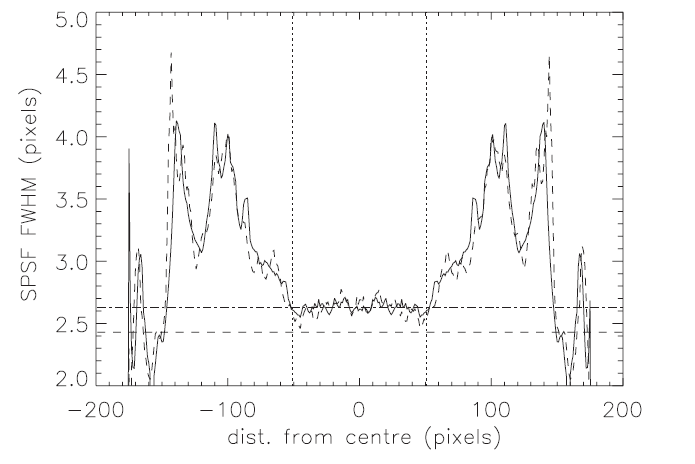}
    \includegraphics[width=0.47\textwidth]{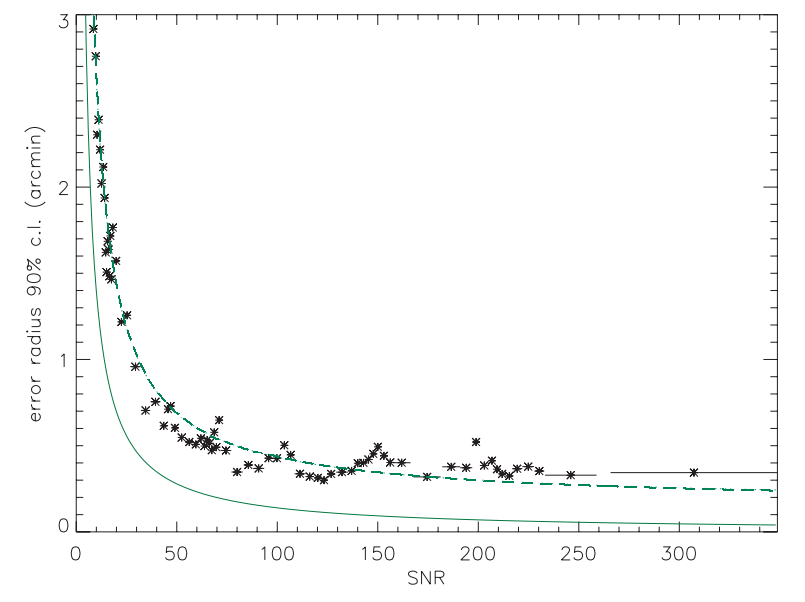}
	\caption{IBIS/ISGRI imaging performance from INTEGRAL early data
	(reproduced with permission from \citep{Gro03}, © ESO). 
	Left: FWHM of the fitted Gaussian to the SPSF 
	along the two central sky image axes: constant at $\approx$~2.6~pix 
	in the FCFOV, it changes wildly in the PCFOV.
	The upper horizontal line gives the average value in the FCFOV (delimited by 2 vertical lines), 
	the lower one the size of a mask element (2.43~pix).
    Right: PSLE radius at 90$\%$ c.l. from measured offsets of 
    known sources at different signal to noises compared to prediction (solid line). 
    Data are well fitted by a $1/SNR$ function plus a constant (dashed line).
    }
	\label{fig:ibis-spsf-psle}
\end{figure}

In any case the PSLE expression above is valid for ideal conditions
and shall be considered only as a lower limit obtainable for a given system geometry.
In real systems, the non-perfect geometry, systematic effects, and the way to measure 
the SNR induce generally larger errors in the location than predicted by 
Eq.~\ref{equ:psle-vs-snr} and can even change the expected $1/SNR$ trend. 
In fact the PSLE will generally tend to a constant value greater than 0 for high SNR,
which at the minimum includes the finite attitude accuracy.
The PSLE curve as function of SNR is therefore always calibrated with simulations
or directly on the data using known sources (Fig.~\ref{fig:ibis-spsf-psle}).
Reducing systematic effects and improving analysis techniques 
shall lead the calibrated curve to approach the theoretical one.
Figure~\ref{fig:ibis-spsf-psle} shows the measured offsets of known sources
with IBIS compared to the predicted error \citep{Gro03}. 
The SNR used to plot the data was the SNR measured in the images
and the theoretical curve is then plotted with a specific constant.
Even though the data roughly follow the 1/$SNR$ trend, 
systematic effects prevent the system to reach the "ideal" performance 
even at large SNR.

\subsection{Sensitivity versus Localization Accuracy} \label{subsec:SvsL}
\begin{figure}[h]
	\centering
    \includegraphics[width=0.51\textwidth]{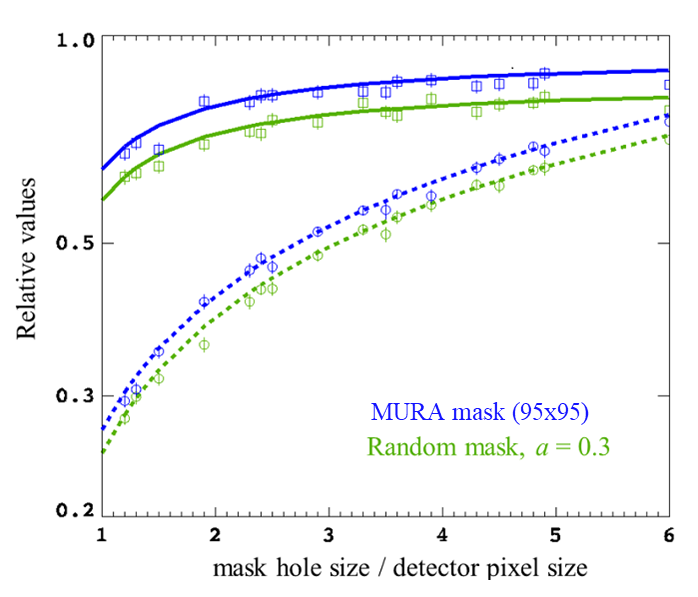}
    \includegraphics[width=0.48\textwidth]{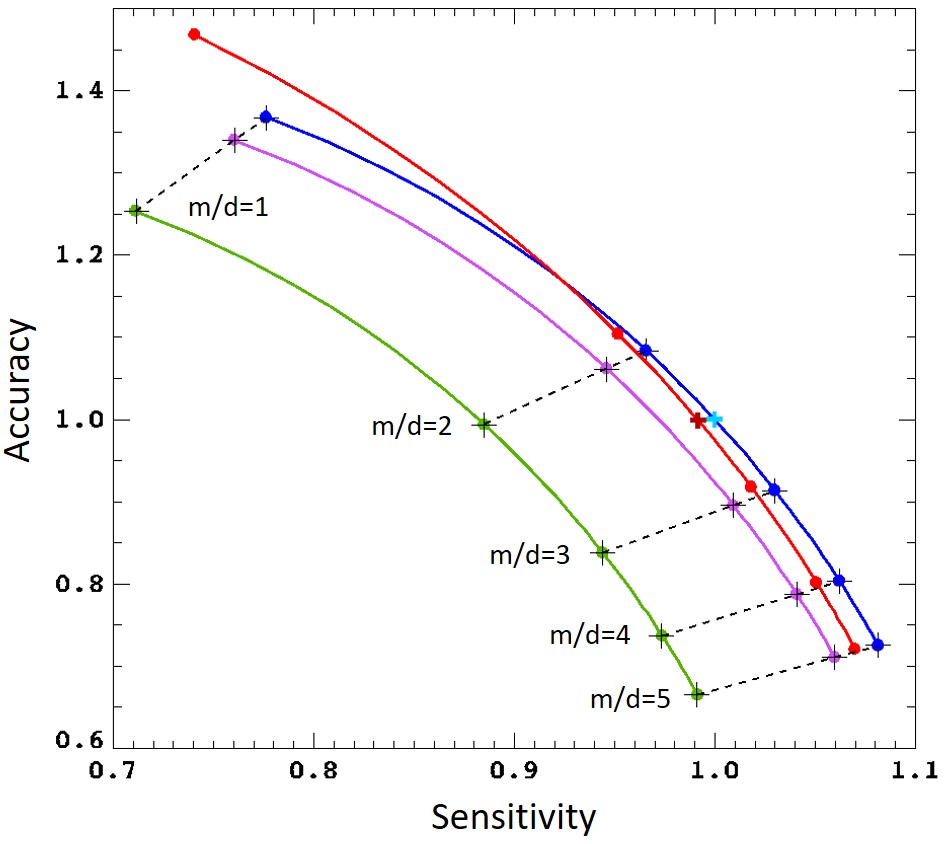}
	\caption{Theoretical, computed with full complex formulae,
	and simulated CMI performance as function of the resolution parameter $r=m/d$ assuming dominant background.
	Left: Variation of sensitivity (solid line) and location accuracy (dashed) 
	with $r$ for two types of masks, an optimum system with a replicated 
	95$\times$95 MURA of 53$\times$53 basic pattern (blue) 
	and a random mask of same dimensions and 30\% aperture (green). 
	Variation of $r$ is obtained maintaining fixed $m$ (and the mask pattern) 
	and decreasing $d$ (with $d<m$). Computed curves are compared to simulations, 
	shown by data points and their error bars with the same color code.
	Right: Normalized accuracy versus normalized sensitivity curves for different $r$, 
	where $r$ is varied by fixing detector resolution $d$ and varying $m$
	for the same masks of left panel (green, blue) plus a random mask with the same dimensions 
	and $a=0.4$ (violet). 
	Curves for random masks are obtained by computing,
	and averaging, error values of a large sample of patterns. 
	Dots give the specific values for integer $r$, 
	while gray crosses those from the approximate formulae that neglect mask pattern 
    (Eqs.~\ref{equ:anal-snr}-\ref{equ:anal-Kx}).
	The cyan cross indicates the reference value for IBIS/ISGRI ($r=2.4$) 
	in the MURA curve (blue).
	The red curve is for an optimized quasi-random (auto-sustained) mask of $a=0.4$ 
	and dimensions 46$\times$46, which is 
	the pattern chosen for ECLAIRs whose performance is positioned in this plot
	by the brown cross at $r=2.5$.
	Both sensitivity and accuracy depend on the configuration of the system 
	and therefore values for different systems are not directly comparable.
	For example, localization accuracy of IBIS/ISGRI is much higher than for ECLAIRs, 
	because sky pixels are 5$'$ wide while for ECLAIRs are 30$'$ wide, 
	even if their accuracy values appear identical in this plot.
	}
	\label{fig:perfs-cmi}
\end{figure}

In the design of CMI it is often important to make a trade-off between sensitivity 
and localization power.
Figure~\ref{fig:perfs-cmi} shows the variation of the sensitivity 
(in terms of the inverse of the flux error $\sigma_S$ of Eq.~\ref{equ:anal-flerr}) 
and of the location accuracy 
(given here by the inverse of the location error $\sigma_X$ of Eq.~\ref{equ:anal-loerr})
with the resolution parameter $r$. 
In the left panel, different $r$ values are obtained maintaining 
mask element (and mask pattern) fixed and varying the detector pixel size 
for two types of masks. 
The formulae correctly predict the performance evaluated through simulations, 
also shown in the plot: for a fixed mask element size (and then angular resolution) 
increasing detector resolution improves both sensitivity and accuracy.
The cases considered are in condition of dominant background;
thus the performance parameters with a 30\% aperture mask are slightly worse than those 
for a 50\% aperture.

In the right panel the location accuracy is plotted versus sensitivity 
for different values of $r$ where its variation
is obtained by fixing the value of $d$ and varying $m$. 
The apparent incoherence with the 
left panel plot (where accuracy increases with $r$) is due to the different way $r$ is varied. 
Clearly by increasing $m$, which determine the angular resolution, 
the location accuracy decreases from its maximum at $r=1$, in contrast to the sensitivity
which increases with $r$. 
As discussed in \citep{Ski08}, often the trade-off values of $r$ are set in the range 1.5-3
in order to have, for given detector resolution $d$, better sensitivity at 
moderate expenses of positional precision. 

This opposite trend (for a fixed and finite detector spatial resolution) 
comes from the fact that localization is determined from a measure on the detector
of the position of the boundary between transparent and opaque mask elements;
therefore the larger the total perimeter of mask holes 
(which is maximized for given $a$ when holes 
are small and are isolated), 
the better the measure of the source position.
On the other hand, signal to noise is optimum when total mask hole perimeter is minimum 
(i.e., when open elements are large and agglomerated) 
which reduces the blurring that occurs at the boundary between
open and closed mask elements. 

The above considerations explain not only the dependence on $r$ but also the one 
on the element distribution (mask pattern).
In Fig.~\ref{fig:perfs-cmi} right, the curve for the specific quasi-random pattern 
of ECLAIRs (see \S~\ref{subsec:Ecla}) is also plotted.
Considering this kind of prediction, the $r$ value for ECLAIRs was finally fixed to 2.5, 
to reach the desired localization accuracy with the highest possible sensitivity.
The formulae of Eqs.~\ref{equ:anal-flerr}-\ref{equ:anal-Kx}, 
as in general those published before,
do not include the terms related to the mask pattern 
and do not predict the performances precisely other than for optimum systems or, 
in average, for fully random masks.
But these terms can be computed using the mask auto-correlation 
in particular to select patterns which have best sensitivity/accuracy pair 
for given science objectives, as was done for ECLAIRs.
Indeed, by comparing the values at integer $r$, 
its pattern appears better in both sensitivity 
and localization accuracy than the comparable 40\% aperture random mask. 
%

\section{Coded Mask Instruments for High-Energy Astronomy}  \label{sec:Inst}
The development of coded mask imaging systems has been, 
from the beginning, linked to the prospect of employing these devices in high-energy astronomy.
We review here the implementation of CMI to this field from the first rocket experiments 
to the missions presently in operation or expected in the close future.
Even if not exhaustive,
this summary provides a chronological panorama of CMI in astronomy which 
illustrates the topics discussed above and recalls
the main achievements obtained in imaging the gamma-ray sky with these devices 
(for a complete list of hard X-ray ($>$ 10~keV) experiments including CMI see \citep{CF17}).

Specific subsections are dedicated to three major experiments successfully flown, or to be launched soon, 
on space missions, a representative set of CMI, with different and complementary characteristics.
SIGMA, the first gamma-ray CMI on a satellite, featured an Anger-type gamma-camera with a continuous spatial resolution depending on energy. Thanks to its imaging capability in the hard X-ray range it became the black hole hunter of the 90s and provided the first 20-arcmin resolution images of the Galactic bulge at energies above 30~keV, and its success opened the way to the INTEGRAL mission. IBIS, presently operating on INTEGRAL, is the most performing gamma-ray imager ever flown, reaching, for the brightest sources, better than 20$''$ location accuracy at 100~keV over a large FOV. It still provides, along with the BAT/Swift experiment, some of the most crucial results in the gamma-ray domain. ECLAIRs/SVOM is the future CMI to be mounted on an autonomously re-pointing platform dedicated to time domain astronomy. The quasi-random mask, optimized to push the threshold at low energies, shall open to the community the efficient detection of cosmological gamma-ray bursts (GRB).

\subsection{First Experiments on Rockets and Balloons}  \label{subsec:Rock}
The CMI concept was first applied to high-energy astronomy with
instruments mounted on sounding rockets or on stratospheric balloons. 
Following the first ideas on coded aperture imaging, several such projects 
were initiated mainly by American, English and Italian groups.

The first experiment that actually probed the CM concept in astronomy
was SL1501 \citep{Pro78}, built by a UK laboratory
and launched on a British Skylark sounding rocket in 1976. 
Composed of a position sensitive proportional counter (PSPC) 
and a rectangular 93$\times$11 Hadamard mask,
both of the same dimensions (box-type), delimited by the diameter of the rocket,
it provided in the few minute flight the first X-ray (2-10 keV) images of 
the Galactic Center (GC) with an angular resolution of 2.5$'\times$21$'$ 
(the higher resolution side purposely oriented along the Galactic plane) 
in a square 3.8° FOV \citep{Pro79}.
SL1501 data were combined with those of Ariel~V space mission in order 
to establish the activity of the X-ray sources of the region and 
together they even permitted to detect and localize some GC X-ray bursts.

A balloon-borne CMI which was highly successful was the US
Gamma-Ray Imaging Payload (GRIP) experiment \citep{Alt85}, 
that flew several times between 1988 and 1995 from Australia. 
Composed of an NaI(Tl) Anger camera working between 30~keV and 10~MeV 
coupled to a rotating mask \citep{Coo84} of about 2000 elements 
disposed on multiple repetition of a 127 HURA basic pattern 
(Fig.~\ref{fig:Mask-pat}), 
this telescope imaged a FOV of 14° with 1.1° angular resolution and provided 
some of the first high-quality images of the Galactic Center
at energies higher than 30~keV, in particular confirming the results obtained 
by SIGMA in the same period \citep{Coo91}. GRIP also detected and located 
gamma-ray emission from SN1997a, confirming the discovery from the Kvant Roentgen observatory.

Another American balloon experiment, 
EXITE2 \citep{Lum94}, based on a phoswich (NaI/CsI) detector 
but coupled to a fixed rectangular URA, 
with a collimator that limited the FOV to the 4.5° FCFOV,
flew several times between 1993 and 2001  
(after a preliminary 1988-1989 flight) giving some results on hard point sources.
This project was a technological preparation for a more ambitious CMI space mission, 
EXIST, not yet included in the US space program.

The American Directional Gamma-ray 
(DGT) experiment 
deserves to be mentioned because it aimed to push the CM technique to high energies \citep{Dum89}.
Using a set of 35 BGO scintillation crystals working in the range
from 150 keV to 10 MeV coupled to a 2-cm-thick lead mask with 13$\times$9 elements
disposed as a replicated 7$\times$5 URA basic pattern, 
it covered a FCFOV of 23°$\times$15° with a 3.8° angular resolution.
The instrument suffered by a large non-uniform background which
limited its performances. Nevertheless DGT could probe the CM concept
at high energies with the detection of Crab nebula and the black holes (BH)
Cyg X-1 and Cyg X-3 above 300 keV 
during a 30-hr balloon flight from Palestine (Texas) in 1984 \citep{McC89}.

These experiments and several others, only conceived, failed
or operated but for short periods,
probed the coded aperture imaging concept and paved the way to the 
implementation of CMI in space missions.

\begin{table}
\caption{Coded Mask Instruments on Satellites}
\begin{center}
\begin{tabular}{llllllcccc}
\hline
CMI & Satellite & Detector & Mask & Pattern & Energy & Ang.Res. & FOV     & Operat. & Ref \\
    &           & type     & type & or order & (keV) & (FWHM)  & (at ZR) & (years) &  \\
\hline
XRT & Challenger & PSPC & Had & 129$\times$127* & 2.5–25 & 3$'$-12$'$ & 6.8° & 1985 & \citep{Will92}\\ 
TTM & MIR Kvant & PSPC & Had & 255$\times$257 & 2–32 & 1.8$'$ & 15.8° & 87-99 & \citep{Bri85}\\  
SIGMA & GRANAT & Anger & URA & 31$\times$29 & 30–1300 & 15$'$ & 20° & 89-97 & \citep{Pau91} \\ 
ART-P & GRANAT & PSPC & URA & 43$\times$41 & 4–30 & 6$'$ & 1.8° & 89-93 & \citep{Sun90}\\ 
WFC & SAX & PSPC & Tria & 256$\times$256 & 2–30 & 5$'$ & 40° & 96-02 & \citep{Jag97}\\ 
IBIS & INTEGRAL & CdTe CsI & MUR & 53$\times$53 & 15–10000 & 12$'$ & 30° & 02-[25] & \citep{Ube03}\\ 
SPI & INTEGRAL & HPGe & HUR & 127 & 20–15000 & 2.5° & 45° & 02-[25] & \citep{Ved03}\\ 
JEM-X & INTEGRAL & MGC & HUR & 22501 & 3–35 & 3.3$'$ & 13.2° & 02-[25] & \citep{Lun03}\\ 
BAT & Swift & CdZnTe & Rand & 54000 & 15–150 & 17$'$ & 120°$\times$85° & 04-[25] & \citep{Bar05}\\ 
CZTI & ASTROSAT & CdZnTe & Had & 17$\times$15 & 20–200 & 17$'$ & 11.8° & 15-[..] & \citep{Bha17}\\ 
ECLAIRs & SVOM & CdTe & Rand & 46$\times$46 & 4–150 & 90$'$ & 90° & [24-29] & \citep{God14}\\ 
\hline
\end{tabular}
\label{tableCMI}
\end{center}
Notes: [~] foreseen at the time of writing (Jul 2022); * second module.
\end{table}

\subsection{Coded Mask Instruments on Satellites}  \label{subsec:Sate}
Table~\ref{tableCMI} reports the list of the fully 2-d imaging coded mask instruments 
successfully launched on, or securely planned for, an astronomy satellite mission.

Other CMI with 1-d only design
(or two 1-d systems disposed orthogonal to each other) 
were launched on space missions and some provided relevant results 
mainly as (all-sky) monitors of point-like sources. 
These were Gamma-1 on Gamma (URSS-Fr), XRT on Tenma (Japan), ASM on Rossi XTE (US), 
WXM on HETE2 (US) and SuperAgile on AGILE (Italy); none of them is presently in operation.
A 1-d coded mask all-sky monitor system presently in 
operation is the SSM on the Indian ASTROSAT mission \citep{Sin14}.
We will not describe them, as they are not, fully, imaging systems, even if 
the 1-d CMI concept, particularly when coupling orthogonal systems
that give locations along the two axes, 
is an interesting one and has certain advantages for which it is still
considered for some future missions.

\begin{figure}[h]
	\centering
 \includegraphics[width=0.52\textwidth]{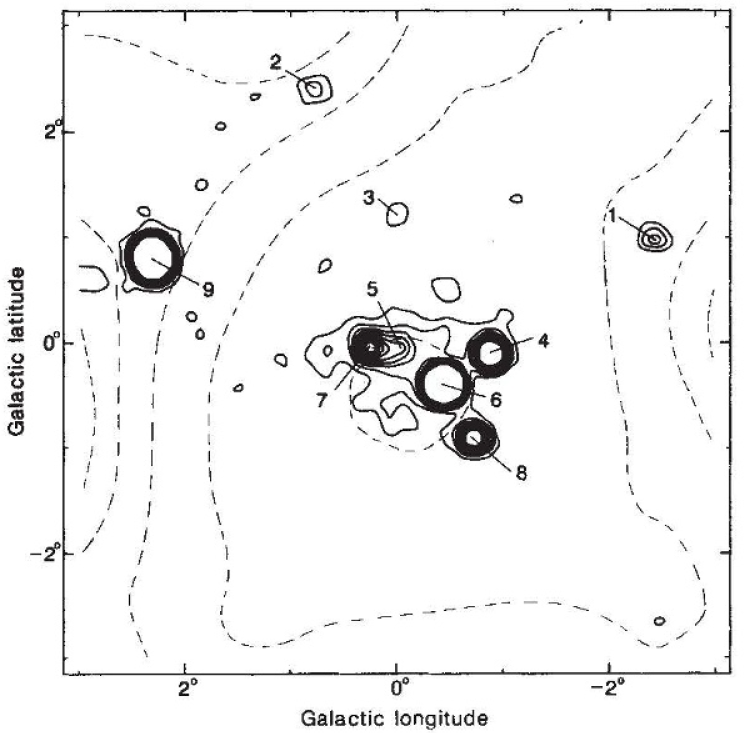}
 \includegraphics[width=0.47\textwidth]{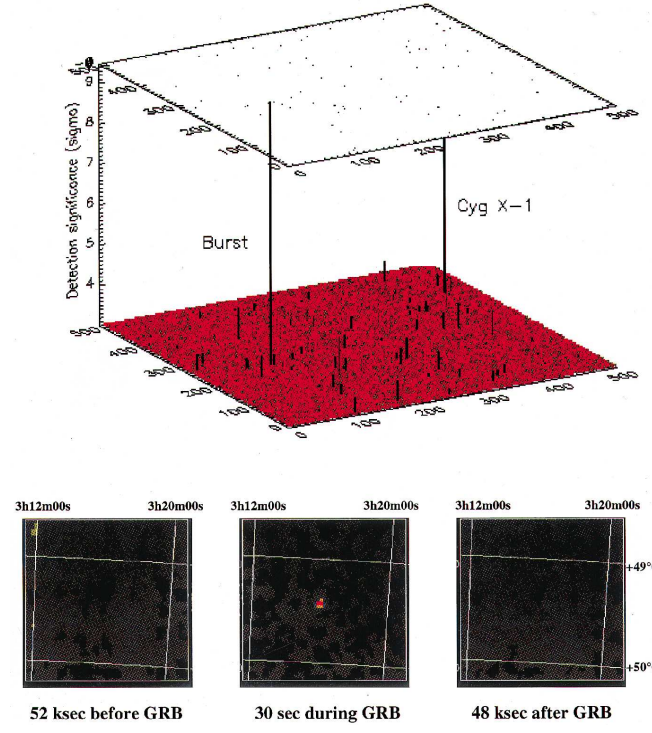}
	\caption{Left: Image of the Galactic Center obtained by the XRT/SL2 instrument in the 3-30 keV band
	(reproduced by permission from \citep{Ski87}, © Springer Nature 1987).
	Right: Detection of GRB~960720 by WFC/SAX 
	(reproduced with permission from \citep{Pir98}, © ESO).
	Top: 3D shadow picture of the WFC 40°$\times$40° FOV 
	of the observation showing the GRB peak along with the one from Cyg~X-1. 
    Bottom: Maps around the GRB before during and after the $\approx$30~s of the burst.
	}
	\label{fig:sl2-sax}
\end{figure}

The first successful CMI flown on a space mission was 
the UK XRT experiment, launched as part of Spacelab 2 (SL2) on board the NASA Space Shuttle 
Challenger for an 8-day flight in August 1985 \citep{Ski88}\citep{Will92}.
Two modules were included, equipped with the same multi-wire proportional counter
working in the 2.5-25 keV range but with two Hadamard masks of different basic pattern, 
31$\times$29 for the coarse one and 129$\times$127 for the fine one, 
and different mask element size which allowed for, respectively, coarse and high 
resolutions over the same 6.8°-wide FOV.
Remarkable results were obtained from XRT/SL2, which provided in particular the 
first GC images with few arcmin resolution at energies $>10$ keV 
(Fig.~\ref{fig:sl2-sax} left) \citep{Ski87}. 
Other XRT results concerned galaxy clusters, X-ray binaries (XRB) and the Vela supernova remnant.
%

The following CMI in space was the Coded Mask Imaging Spectrometer telescope
(COMIS/TTM) \citep{Bri85} 
flown on the Kvant module of the soviet MIR station as part of the Roentgen Observatory 
which included three other non-imaging experiments. 
The instrument, built by Dutch and UK laboratories, 
used again a Hadamard mask (Fig.~\ref{fig:Mask-pat}) coupled to a PSPC in a simple system \citep{Int92}.
It operated at different times between 1987 and 1999 and provided interesting 
hard X-ray images of the Galactic Center and upper limits of the famous SN1987a in the LMC \citep{Sun87}.

In spite of the progress obtained in the UK and US, it was finally France that built 
the first soft gamma-ray CMI to fly on a satellite, SIGMA. 
It was launched in 1989 on the Soviet satellite GRANAT
along with few other experiments: the Russian ART-S and CMI ART-P, 
the Danish rotating collimator monitor Watch and the French Phebus burst detector.
SIGMA spectacular results firmly established the superiority of CM imaging 
over collimation, on/off chopping or Earth occultation techniques for gamma-ray astronomy.
SIGMA is described in \S~\ref{subsec:Sigm}.

Soon after SIGMA, the Dutch Wide Field Camera (WFC) \citep{Jag97} working in X-rays up to 30~keV 
was launched in 1996 on the Italian space mission Beppo-SAX \citep{Boe97}. 
This instrument was based on a pseudo-random mask, with a pattern called "triadic residues",
with low open fraction (33\%), more adapted to the X-ray domain than URAs,
arranged in a simple system configuration which provided a 40°$\times$40° PCFOV \citep{Int94}.
Two such cameras were disposed orthogonal to each other and to the optical axis of the other
SAX instruments and particularly of the X-ray mirror telescope.
The WFC allowed the discovery of the first X-ray afterglow of a
GRB \citep{Cos97} when the satellite could 
re-point its X-ray mirrors on the transient source positioned with arcmin 
precision (Fig.~\ref{fig:sl2-sax} right). 
This was a crucial astrophysical discovery which allowed the community, 
with follow-up optical observations, to establish that GRBs are extra-galactic events,
which now we know are connected to explosion or coalescence of stars in external galaxies.

The heritage of SIGMA allowed Europe to maintain and consolidate the advantage in CMI.
In fact France, Germany and Italy took the lead of the development of the two main 
instruments of the INTEGRAL Mission, both based on coded aperture techniques.
The International Gamma-Ray Astrophysical Laboratory \citep{Win03} 
of the European Space Agency (ESA) 
with participation of Russia and the US was launched on the 17th October 2002 
from Baikonour by a Proton rocket on a very eccentric orbit, which allows 
long uninterrupted observations of the sky, about 3 days before entry in the 
radiation belts. The platform carries four co-axial instruments, 
the two main gamma-ray CMI, the imager IBIS, and the high-resolution spectrometer SPI,
plus the coded mask X-ray monitor JEM-X and the optical telescope OMC.
INTEGRAL performs observations in dithering mode where a set of pointing 
of $\approx$~30~min are carried out along a grid of directions about 2° apart 
around the target source. 
Data are sent to ground in real time which allows fast analysis and reaction in the case 
of detection of transient events.

The Spectrometer on INTEGRAL \citep{Ved03} working in the range 20~keV-8~MeV
is composed of 19 individual cooled Germanium detectors of hexagonal shape with 3.2~cm side 
disposed in an hexagonal array 1.7~m below a thick tungsten non-replicated and non-rotating
HURA mask of order 127, with hexagonal elements of the size of the Ge crystals. 
The very high spectral resolution (2.5~keV at 1.33~MeV) of the 6-cm-thick Ge detectors 
allows study of gamma-ray lines with moderate imaging capabilities (2.5° resolution over a 16° FCFOV 
and a 30° half coded EXFOV) thanks to the CM system 
and the dithering mode which permits a better correction of the background.
The SPI CsI anti-coincidence system is by itself a large area detector that is presently 
used also for the search of GRB events outside the FOV of the CMI.
The Danish JEM-X monitor \citep{Lun03} working in the range 3-30 keV is also a CMI. 
Composed of two identical modules with the same non-cyclic fixed HURA of more than 20,000 elements 
with 25\% aperture but rotated by 180° in order to have different ghost distribution 
and high-pressure Microstrip Gas Chamber (MGC) detectors, it provides 3$'$ angular resolution images over a 7° FWHM FOV.
IBIS is certainly the core of the CM imaging capabilities of INTEGRAL and is described in \S~\ref{subsec:Ibis}.

INTEGRAL provides the community with a large amount of excellent astrophysical results
and crucial discoveries,
in particular with the mapping of the 511 keV line of the Galaxy; the measurements of 
gamma-ray lines from SNR and close supernovae (SN); the detection and study of GRBs 
and BH sources in binary systems and in active galactic nuclei (AGN), 
of all variety of neutron star (NS) systems; and the detailed imaging of the GC region.
The mission is still operating today \citep{Kuu21} with many important results obtained each year, 
the most recent ones in the domain of time domain astronomy 
(see, e.g., \citep{Fer21} and references therein).

Meanwhile, on the side of the high-energy time domain astronomy, Beppo\footnote{
Beppo-SAX, the Italian Satellite for X-ray Astronomy was named in honor of 
Giuseppe Occhialini, the eclectic and visionary Italian physicist, 
to whom we owe, in addition to a huge scientific heritage, 
the strong involvement of Europe in astrophysical space programs.}
had, once more, showed the way.
The future of this domain would reside on agile spacecrafts 
with capability of fast (therefore autonomous) re-pointing and a set
of multi-wavelength instrumentation, including a large field imaging instrument at high energies,
based on coded mask technique, and high-resolution mirror-based telescopes at low energies.
Too late to implement these features in INTEGRAL, 
the US did not miss the opportunity by developing in collaboration with UK and Italy 
the Neil Gehrels Swift mission \citep{Geh04}, dedicated to GRBs and the transient sky.
Using a platform conceived for the military "star-wars" program, 
with unprecedented, and still unequalled, capability of fast (tens of seconds) 
autonomous re-pointing,
this mission has provided since its launch in 2004 exceptional results in the domain
of GRB science \citep{GR13} but also of the variable and transient high-energy sky \citep{GC15}.

\begin{figure}[h]
	\centering
 \includegraphics[width=0.49\textwidth]{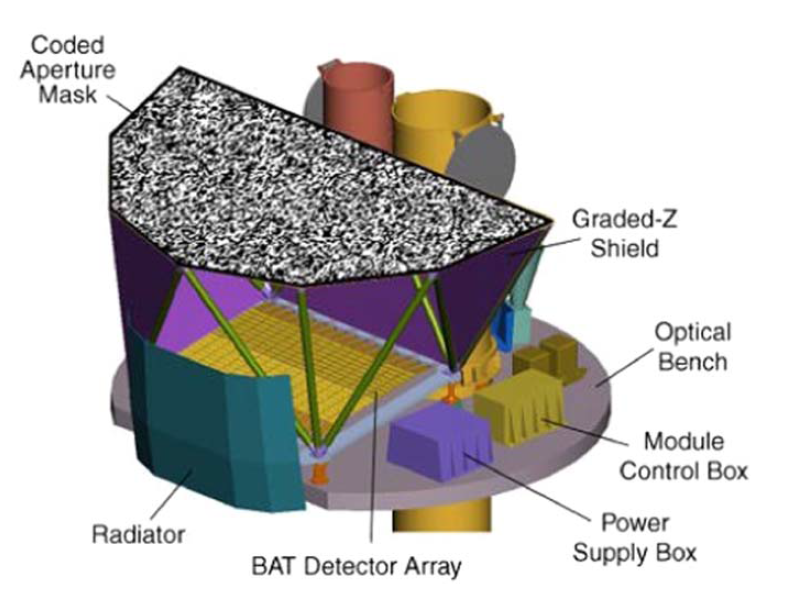}
 \includegraphics[width=0.48\textwidth]{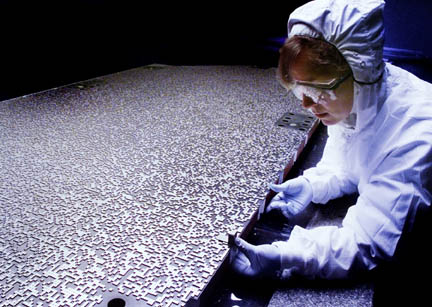} 
	\caption{Left: Scheme of the BAT CM instrument on board the Swift satellite
	(Credit NASA,
    \url{https://swift.gsfc.nasa.gov/about_swift/bat_desc.html}).
	Right: Assembly of the BAT coded mask, characterized by a "D" shape and 
	a random distribution of elements 
    (reproduced by permission from \citep{Bar05}, © Springer Nature 2005).
	}
	\label{fig:swift}
\end{figure}

Many of these are based on the Burst Alert Telescope (BAT) \citep{Bar05} 
a large coded mask instrument which is still in operation along with the two narrow-field 
telescopes of the mission, one for X-rays (XRT) and one for ultraviolet-optical frequencies (UVOT).
BAT (Fig.~\ref{fig:swift}) is the instrument that detects and localizes the GRB 
and triggers the platform re-pointing. 
It is composed of an array of 32,768 individual square CdZnTe semiconductor detectors,
for a total area of 5240~cm$^2$,
coupled to a large random mask with a "D" shape and a 50\% aperture, made of about 52,000 elements,
each of 1~mm thickness and dimensions 5$\times$5~mm$^2$ with a resolution ratio $r=1.2$
with respect to the detector pixels. 
The mask is set at 1~m from the detection plane and it is connected to the detector 
by a graded-Z shield that reduces the background.
BAT provides a resolution of about 20$'$ over a huge FOV of 1.4~sr 
(half coded), a location accuracy of 1$'$-4$'$, 
and good sensitivity in the range 15-150 keV. 
Ground BAT data analysis is described in \citep{Tue10}\citep{Seg10}\citep{Bau13}\citep{OhK18}
and it is quite similar to the standard one described above.

\begin{figure}[h]
	\centering
 \includegraphics[width=0.3\textwidth]{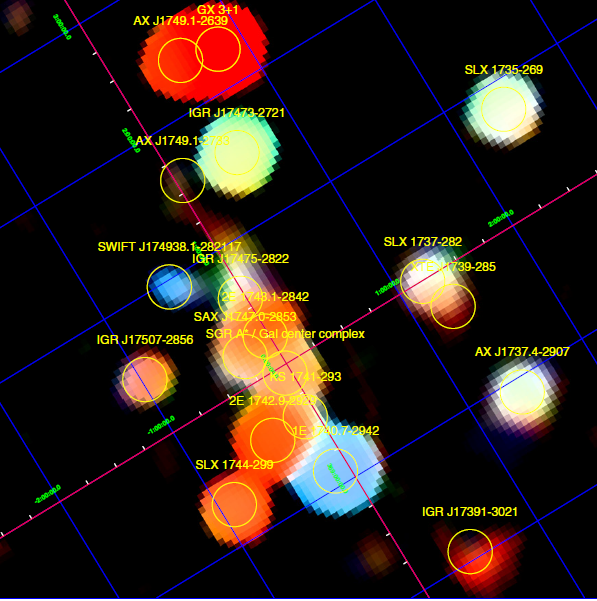}
 \includegraphics[width=0.69\textwidth]{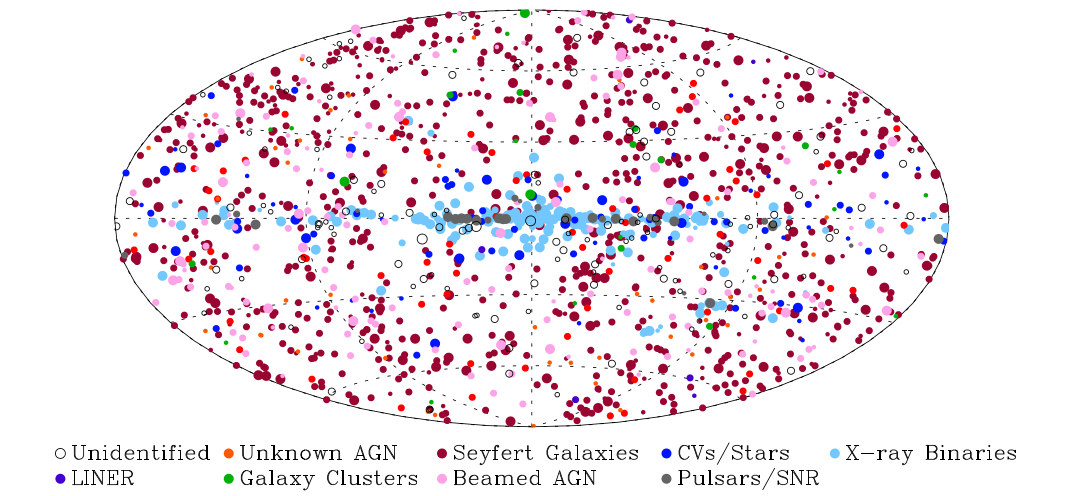}
	\caption{
	Left: Reconstructed image of the Galactic Center from BAT data 
	(reproduced with permission from \citep{Bau13}, © AAS).
	Right: The BAT/Swift catalogue of sources ($>$~15~keV) detected in the first 105 months of operations 
	(reproduced with permission from \citep{OhK18}, © AAS).
	}
	\label{fig:swift-cat}
\end{figure}

Figure~\ref{fig:swift-cat} left shows the BAT reconstructed image of the Galactic Center 
from which one can appreciate the imaging capability of this CMI
and can compare it to the IBIS one (\S~\ref{subsec:Ibis}).
A specific feature of the instrument is that a BAT data analysis is continuously 
performed on board in near real time thanks to an image processor 
which allows the detection and position of GRBs within 12~s from their start
and the rapid triggering of the platform re-pointing to the computed location. 
BAT performances are the key of the large success of the mission. 
The instrument detects and positions about 100 GRBs per year 
allowing the following red-shift determination for about 1/3 of them. 
It provides excellent results on many different variable hard X-ray sources, 
both Galactic and extra-galactic, like AGNs, magnetars, different types of binaries, and others \citep{GC15}, 
for example it allowed the discovery in 2011 of the first tidal disruption event with a relativistic jet \citep{Bur11}. 
More than 1600 non-GRB hard X-ray sources have been detected by BAT/Swift in the first 105 months 
of operations \citep{OhK18} (Fig.~\ref{fig:swift-cat} right).
%

The most recent launch of a coded aperture instrument on a space mission is  
the Cadmium Zinc Telluride Imager (CZTI) \citep{Bha17} of the Indian ASTROSAT \citep{Sin14} 
mission in operation since 2015.
The CZTI is composed of four identical and co-axial modules disposed 2$\times$2 
on the platform. The modules are based on a mask composed of 4$\times$4 arrays, 
each one following a Hadamard pattern built (in different way) from the same 255 PN sequence,
and then coupled to a pixelated CdZnTe detector in a "simple system" configuration.
Its overall parameters are given in Table~\ref{tableCMI} but for more complete and recent reports 
about the in-flight calibrations and performance of the instrument, see \citep{Vib21}.
Important results have been obtained on transient sources, pulsars and polarization measurements.

\subsection{SIGMA on GRANAT: the First Gamma-Ray Coded Mask Instrument on a Satellite}  \label{subsec:Sigm}
SIGMA ({\it Syst\`eme d'Imagerie Gamma \`a Masque Al\'eatoire}) 
is the first gamma-ray coded mask telescope flown on a satellite \citep{Pau91} 
and provided extraordinary discoveries and results in the domain of black hole astrophysics.
Launched on the 1st December 1989 from Baikonour (URSS) by a Proton rocket
on the three-axis stabilized Soviet GRANAT
satellite, it operated in pure pointing mode till 1995 and then mainly in scanning mode
for a couple of years more.
\begin{figure}[h]
	\centering
 \includegraphics[width=0.40\textwidth]{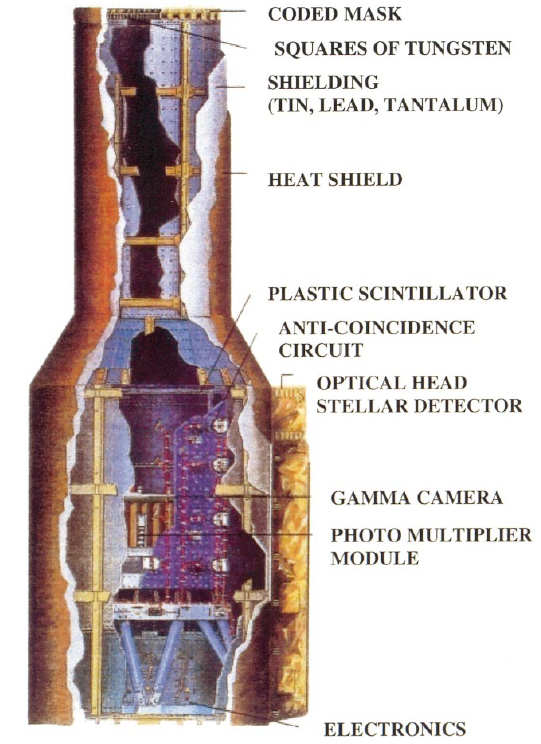}
 ~~~~
 \includegraphics[width=0.56\textwidth]{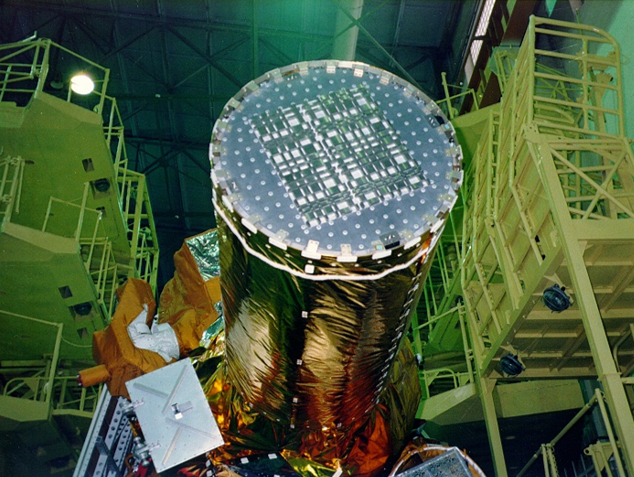}
	\caption{The SIGMA/GRANAT coded mask instrument.
	Left: Scheme of the SIGMA telescope (reproduced with permission from \citep{Bou01}, © AAS).
	Right: SIGMA, with its URA tungsten coded mask, mounted on the GRANAT spacecraft (credit CEA/IRFU).}
	\label{fig:sigma}
\end{figure}
A schematic view of SIGMA is given in Fig.~\ref{fig:sigma} where
its coded mask is also shown, with its characteristic URA pattern, 
after the instrument was mounted on the platform.
Made of a NaI(Tl) Anger camera, composed of 1.25-cm-thick circular scintillating crystal 
viewed by 61 photo-multipliers, surrounded by a CsI(Tl) anti-coincidence system, 
and set at 2.5~m from a 1.5-cm-thick tungsten coded mask, SIGMA could provide images 
in the 35~keV-1.3~MeV range with angular resolution of 20$'$-13$'$ in a FCFOV of 4.7°$\times$4.3° 
and a half-coded EXFOV of 11.5°$\times$10.9° with an on-axis 40-120~keV 3$\sigma$ sensitivity 
of the order of 100 mCrabs in 1-day observation.
The rectangular mask of 53$\times$49 elements was a replicated 31$\times$29 URA
(and not a random mask as implied by the instrument acronym)
and the events, detected in the central rectangular 725~cm$^2$ area of the NaI crystal 
corresponding to the URA basic pattern, 
were coded in 124$\times$116 pixel detector images and in 95 energy channels.

\begin{figure}[h]
	\centering
 \includegraphics[width=0.55\textwidth]{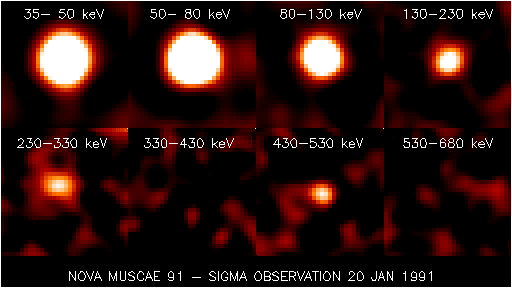}
 \includegraphics[width=0.44\textwidth]{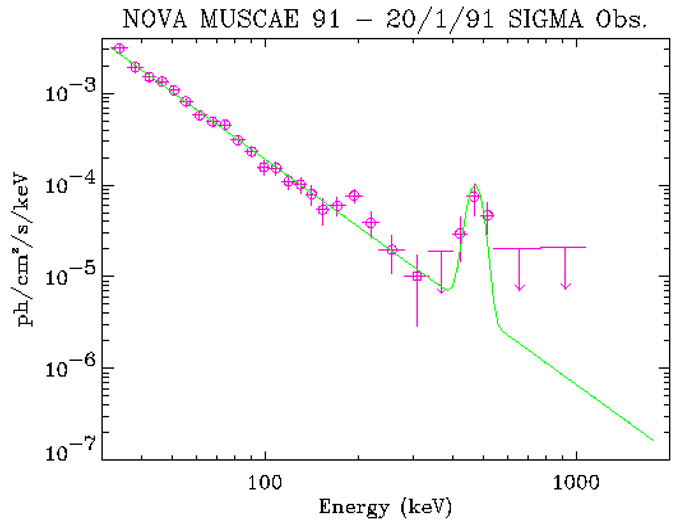}
	\caption{SIGMA Observation of Nova Muscae 91 in January 1991 \cite{Gol92}.
	Left: Sky image sectors around the source in different energy bands 
	which show the large variation of the width of the SPSF
	due to the changes in the gamma-camera spatial resolution with energy, along with 
	the reappearance of a significant excess in the band around 500~keV.
	Right: Source spectrum, derived from the deconvolved images, which shows 
	the presence of a high-energy feature.
	}
	\label{fig:Sigma-NM}
\end{figure}

Data analysis and calibration of SIGMA are reported in \citep{Gol95} 
and \citep{Bou01} and references therein.
A specific feature of the decoding process was the use of the efficiency array to take into 
account the drifts of the platform \citep{Gol95} for the extension of the 
sky image reconstruction to the PCFOV of the instrument.
Another important element was that the continuous spatial resolution of the gamma camera 
varied with energy (see Fig.~\ref{fig:Sigma-NM}) 
and time, and therefore it had to be monitored and modelled along the mission \citep{Bou01}
in order to optimize the analysis of the data.
\begin{figure}[h]
	\centering
 \includegraphics[width=0.52\textwidth]{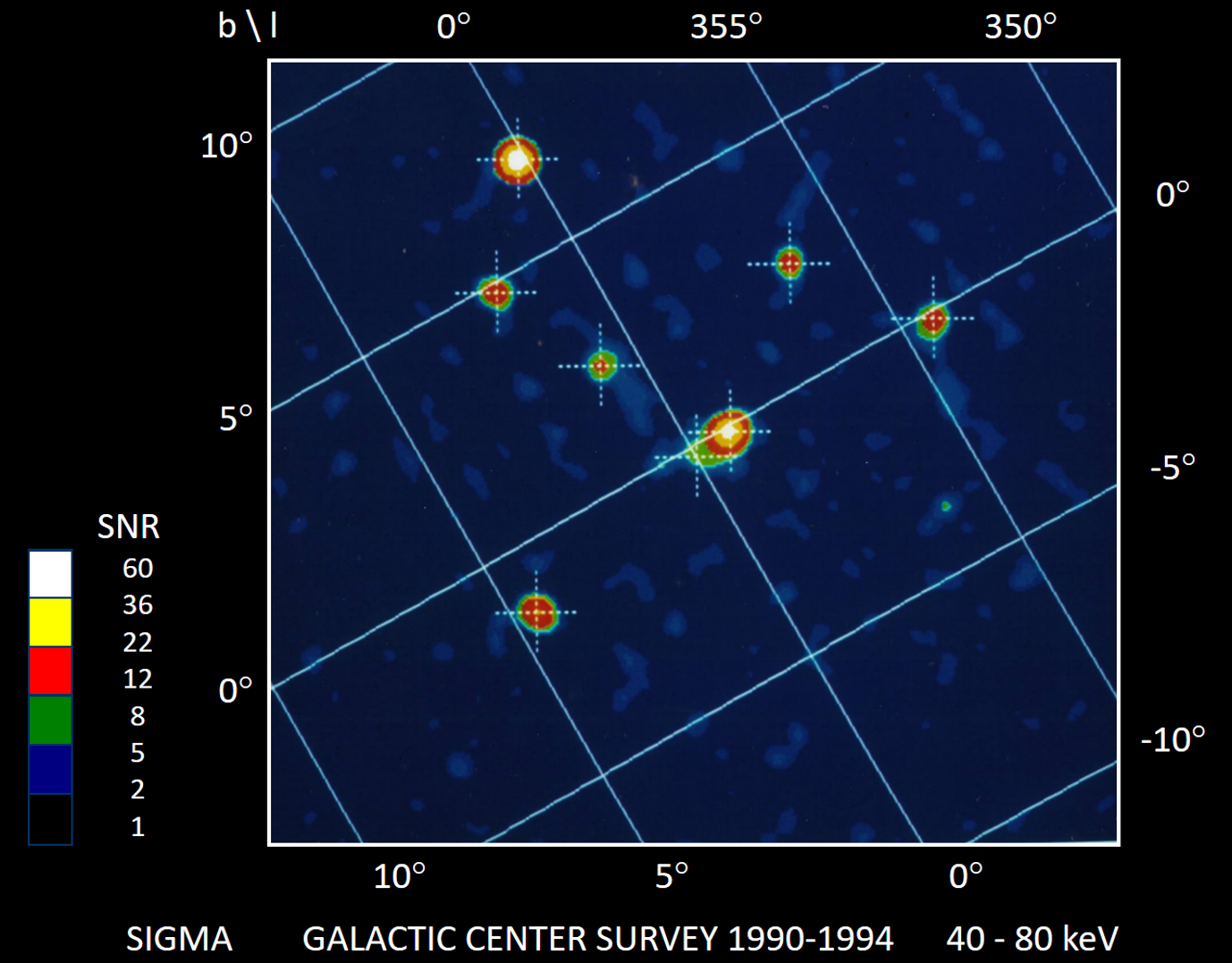}
 \includegraphics[width=0.465\textwidth]{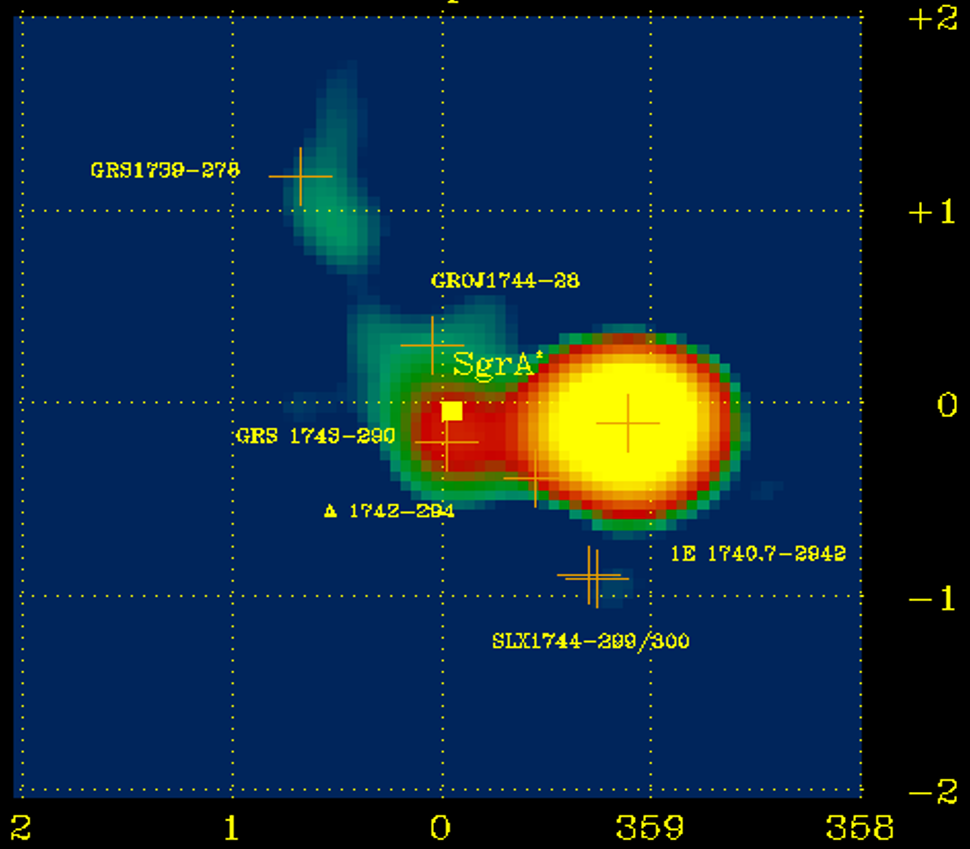}
	\caption{The Galactic Bulge and Galactic Center observed with SIGMA/GRANAT \cite{Gol94}. Left: The mosaic of the 40–80~keV sky images reconstructed from SIGMA data from observations of the GC in 1990–1994. Right: Zoom in the central GC region, dominated by the bright micro-quasar 1E~1740.7–2942. Very weak emission is present at the position of the SMBH Sgr~A*.
	}
	\label{fig:Sigma-GC}
\end{figure}

The SIGMA long and repeated observations of the Galactic Bulge (Fig.~\ref{fig:Sigma-GC}), 
allowed by the fact that its imaging capabilities could be fully exploited over its huge PCFOV, 
clarified the situation of the high-energy emission from this very active and variable region by
showing in particular that the central degrees at energies $>20$~keV are fully dominated by the
source 1E~1740.7-2942, not particularly bright at low energies,  that was soon after identified as 
the first persistent Galactic BH micro-quasar displaying extended radio jets.
They also led to the discovery of the other persistent X-ray BH binary GRS~1758-258 \citep{Man91}
of the bulge and second identified micro-quasar of the Galaxy
(too close to the NS XRB GX5-1 to be studied by previous non-imaging instruments), 
to a measure of the weakness of the central massive black hole Sgr~A$^*$ at high energies,
and to the detection of several other BH and NS X-ray persistent 
and transient bulge sources \citep{Gol94}.
Again thanks to its huge FOV and imaging capabilities SIGMA was very efficient in discovering
Galactic BH X-ray transients (or X-ray novae) which are particularly hard sources. 
It detected and positioned seven of them in its 6 years of nominal operations, 
and between them is the other famous BH micro-quasar GRS~1915+105 that was the first specimen to 
reveal super-luminal radio jets.
An important result was the detection, in the BH X-ray Nova Muscae 91, 
of a weak and transient feature around 511~keV 
(Fig.~\ref{fig:Sigma-NM}), the energy of the electron-positron annihilation line \citep{Gol92}.
These results showed how well-designed CMI can provide 
also high-quality spectra and light curves of gamma-ray sources.
SIGMA detected, in its 12~Ms 1990-1997 survey, a total of about 35 sources including 
14 BH candidates, 10 XRB, 5 AGN, 2 pulsars and 9 new sources
\citep{Rev04b}.

SIGMA data were complemented at low energies by those of the ART-P coded mask hard X-ray telescope
\citep{Sun90} which had four identical modules made of a PSPC coupled to a replicated URA mask 
and providing 6$'$ angular resolution over less than 2° FOV. 
ART-P's most relevant results were the hard X-ray images of the GC that 
complemented nicely those of SIGMA \citep{Pav94} 
and revealed a diffuse emission consistent with the molecular clouds of the region
and interpreted as scattering by the clouds of high-energy emission emitted elsewhere. 
Initially used to put limits on the activity of Sgr~A$^*$,
the detected emission was later recognized as a signal of the Galactic SMBH past activity
and was coupled to the measurements of the molecular cloud Sgr~B2 with IBIS/INTEGRAL 
to constrain the Sgr~A$^*$ ancient outbursts \citep{Rev04}.

\subsection{IBIS on INTEGRAL: the Most Performant Gamma-Ray Coded Mask Instrument}  \label{subsec:Ibis}

\begin{figure}[h]
	\centering
 \includegraphics[width=0.50\textwidth]{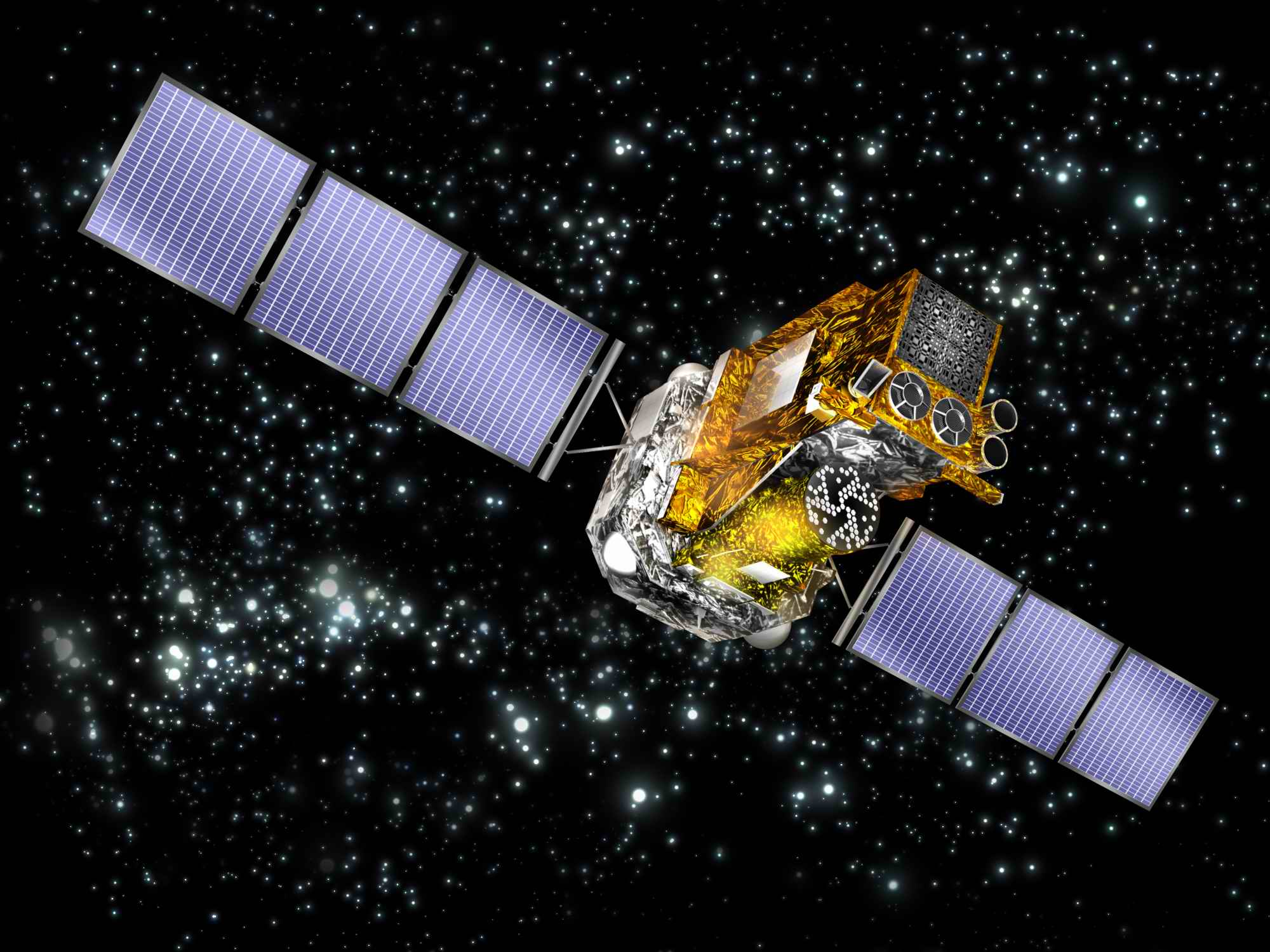}
 \includegraphics[width=0.47\textwidth]{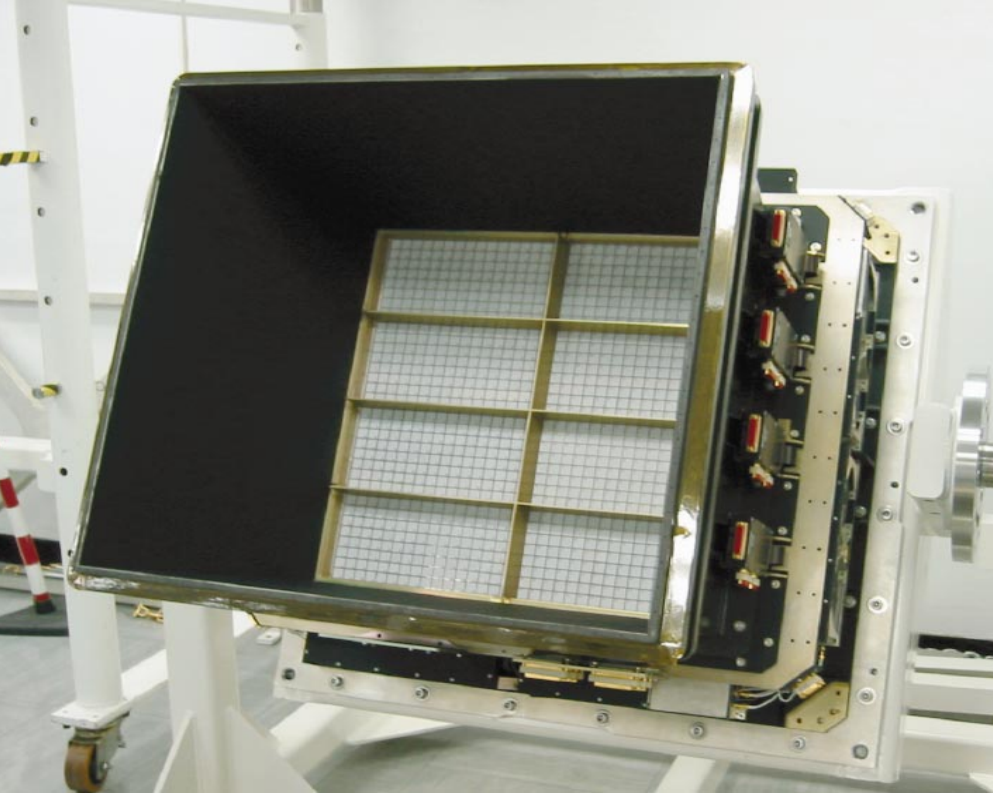}
	\caption{IBIS/INTEGRAL coded mask instrument.
	Left: Artistic view of the INTEGRAL satellite, where the 
	characteristic MURA and HURA patterns of the IBIS and SPI masks are visible (credit ESA).
	Right: The ISGRI detection plane composed of 8 modules of 2048 CdTe detectors 
	integrated in the IBIS instrument 
	(reproduced with permission from \citep{Leb03}, © ESO).
	}
	\label{fig:IBIS}
\end{figure}

The main gamma-ray imaging device on INTEGRAL (Fig.~\ref{fig:IBIS} left) is 
the Imager on board the INTEGRAL satellite, 
a hard X-ray/soft gamma-ray coded mask telescope \citep{Ube03} developed mainly 
by Italy and France.
IBIS is composed of a replicated MURA mask of 95$\times$95, 1.6-cm-thick tungsten 
square elements (see the pattern in Fig.~\ref{fig:Mask-pat}) with 50$\%$ open fraction 
coupled to two position sensitive detectors, the Integral Soft Gamma-Ray Imager (ISGRI) and 
the Pixellated Imaging Caesium Iodide Telescope (PICsIT), 
both of the same dimension of the central MURA basic pattern of 53$\times$53 elements. 
ISGRI \citep{Leb03} is made of 128$\times$128 individual 2-mm-thick cadmium telluride (CdTe) 
semiconductor square detectors each of dimensions 4$\times$4~mm$^2$ 
(for a total area of 2600~cm$^2$) (Fig.~\ref{fig:IBIS} right),
works in the range 15~keV-1000~MeV and is placed 3.2~m below the mask.
The overall IBIS/ISGRI sensitivity is of the order of a mCrab 
for 1~Ms exposure at 80 keV with typical spectral resolution of 7$\%$ (FWHM).

PICsIT \citep{DiC03} is placed 10~cm below ISGRI and is composed of 64$\times$64 CsI bars, 
each exposing 
a collecting area 4 times of an ISGRI pixel and working in the range 175~keV-10~MeV.
The detector planes are surrounded by an active anti-coincidence system of BGO blocks, 
and an absorbing tube connects the unit with the mask allowing for reduction of
un-modulated sky radiation. 
Data of both instruments are recorded, transmitted, and analyzed independently,
but coincident events from the two detector layers are combined to provide the so-called 
Compton mode data which are particularly useful to study polarimetry properties of the
incident radiation.
In the following we will refer to the IBIS/ISGRI system only, given that it provides the best 
imaging performances of the telescope and we will neglect the Compton mode.

\begin{figure}[h]
	\centering
 \includegraphics[width=0.52\textwidth]{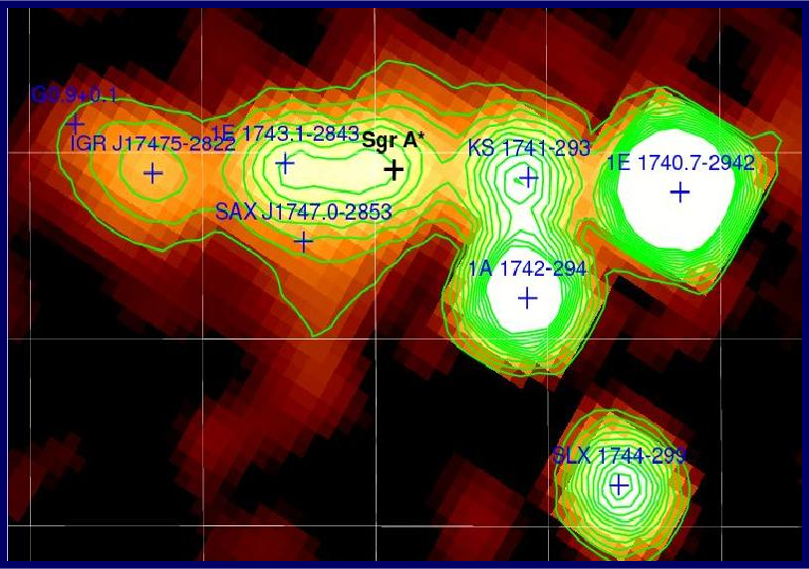}
 \includegraphics[width=0.465\textwidth]{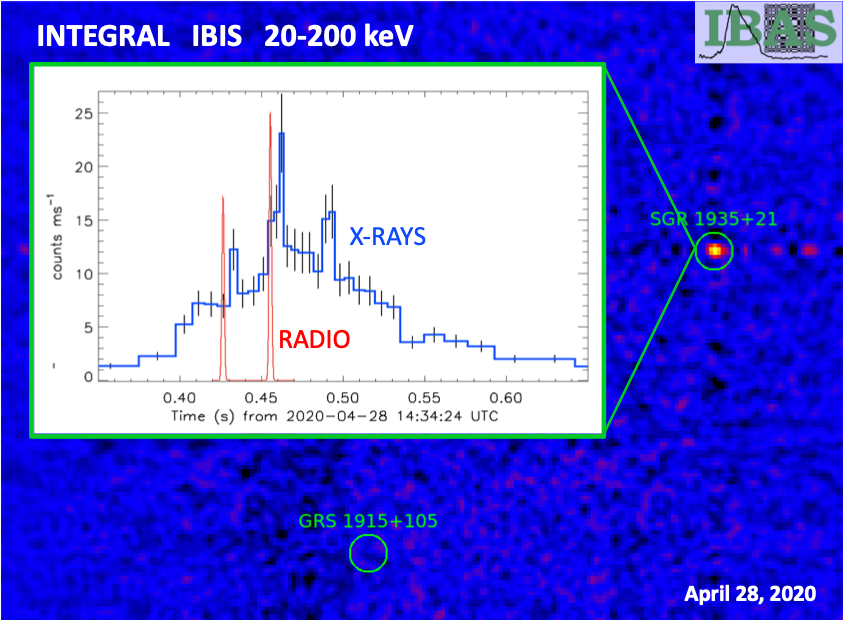}
	\caption{IBIS/INTEGRAL results.
	Left: Image of the Galactic Center (2.5°$\times$1.5°) in the 20-40 keV from 
	the mosaic of 5~Ms of observations of the field 
	(reproduced with permission from \citep{Bel06}, © AAS). 
	Right: IBIS detection of the magnetar SGR~1935+2154 flares coincident 
	with a FRB \citep{Mer20}: sky image, source location and light curve (blue) 
	compared to the radio bursts (red) 
	(INTEGRAL POM 07/2020, credits: S. Mereghetti and ESA).
	}
	\label{fig:IBIS-results}
\end{figure}

IBIS has provided the most precise images of the GC (Fig.~\ref{fig:IBIS-results} left) 
at $>20$~keV
before the recent extension of the grazing incidence technique to 70-80~keV with NuSTAR, 
and it is still the best imager that can cover such large FOV ($>2$°) at high energies.
The most recent and remarkable discoveries of this telescope have been the detection of
emission of a close supernova (SN2014j) in the Na lines 
and the identification of a magnetar flare (Fig.~\ref{fig:IBIS-results} right) 
with a fast radio burst (FRB) \citep{Mer20}.


\subsubsection{IBIS Data Analysis and Imaging Performance}  \label{subsubsec:IbisAn}
The IBIS coded mask system and the standard analysis procedures of the data are described in 
\citep{Gol03} and \citep{Gro03}, but see also \citep{Kri10} for sky surveys at high energies
and \citep{Ren06} for analysis of extended sources.
The instrument analysis software is integrated in the 
Integral Science Data Center (ISDC) \citep{Cou03} 
through which it is distributed to users as Off-line Scientific Analysis (OSA) packages. 
After 20 years of operations the instrument is still providing excellent data
and several new features have been integrated in the analysis procedures \citep{Kuu21}.
We have already largely used characteristics and data from this system
in order to illustrate CMI design, analysis, and performance concepts: 
in Table~\ref{tab:cm-prop} for imaging design/performance, 
in Fig.~\ref{fig:Mask-pat} for the mask pattern,
in Fig.~\ref{fig:decon} for decoding process, in Fig.~\ref{fig:cmi-SNR} for 
distribution of peaks in reconstructed image, in
Figs.~\ref{fig:ibis-spsf} and \ref{fig:ibis-spsf-psle} for the resulting SPSF and PSLE,
and in Fig.~\ref{fig:IBIS-iros} for the overall analysis process.    

Indeed IBIS represents a typical CMI with a cyclic optimum (MURA) mask
coupled to a pixelated detector.
The detector spatial resolution is just given by the geometrical 
dimension of the square pixels, independent from energy. 
CdTe square pixels have size of 4~mm, but the pitch between them 
is 4.6~mm with 0.6~mm of dead area.
The mask elements are not integer number of pixel pitch also in order to avoid 
ambiguity in source position due to the dead zones.
This does introduce a non-perfect coding even for sources in the FCFOV; however other
factors break the perfect coding, and noise is anyway introduced.
%
\begin{figure}[h]
	\centering
 \includegraphics[width=0.48\textwidth]{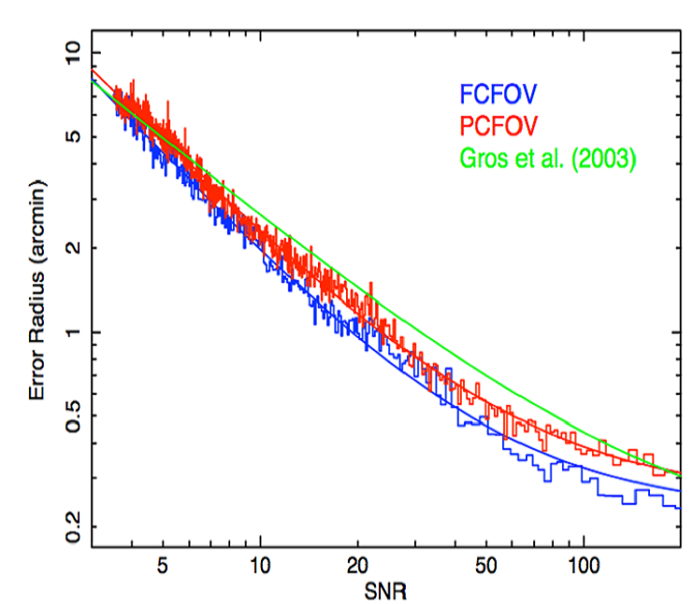}
 ~~
 \includegraphics[width=0.49\textwidth]{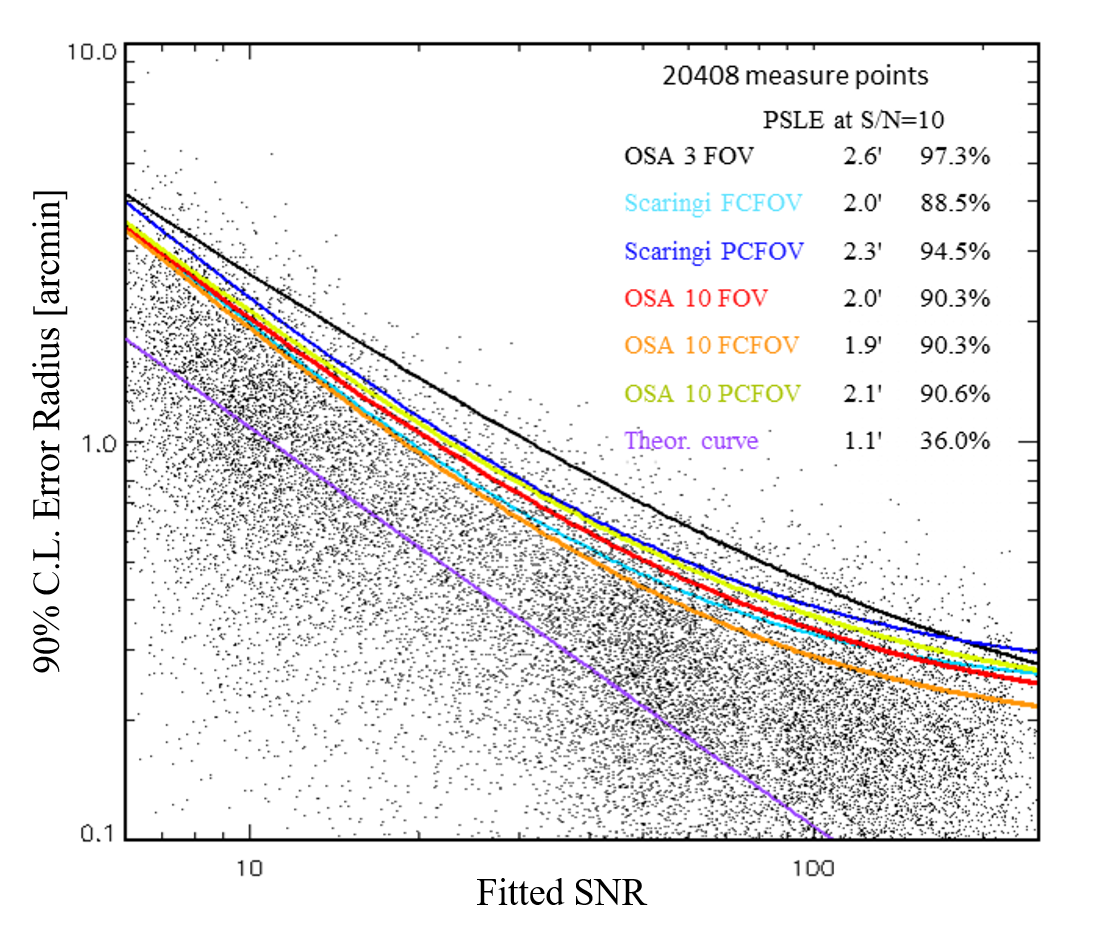}
	\caption{IBIS/ISGRI imaging performance: recent determination of the PSLE.
	Left: PSLE vs SNR in FCFOV and PCFOV compared to early measurements \citep{Sca10}
	(credit: IBIS Observer’s Manual, 2017, ESA SOC).
	Right: Recent PSLE measurements (dots) and derived curves (red, orange, yellow) using refined analysis, 
	compared to previous results and theoretical trend (violet) \citep{Gro12}.
	}
	\label{fig:IBIS-perf}
\end{figure}
\begin{figure}[h]
	\centering
 \includegraphics[width=0.93\textwidth]{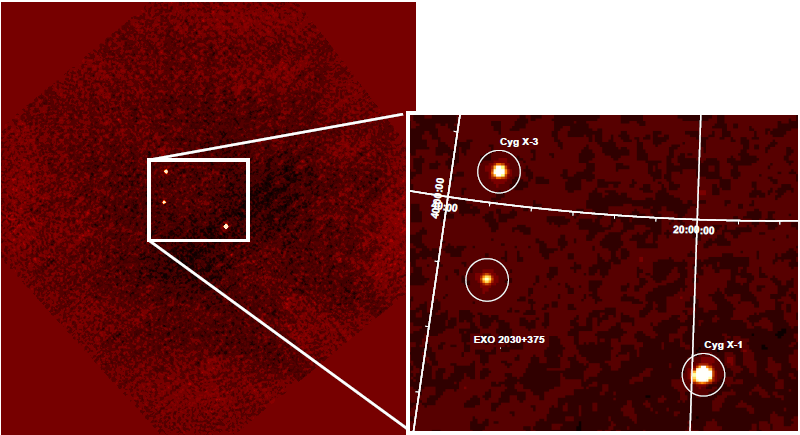}
	\caption{IBIS/ISGRI imaging performance: 
	mosaic in the Cygnus Galactic region 
	after the decoding, analysis, cleaning, roto-translation and sum
	of several individual sky images \citep{Gol03}.
	}
	\label{fig:IBIS-mosa}
\end{figure}
%
Imaging performances were studied on different data sets of bright and weak known point-like 
sources along the years \citep{Gro03}\citep{Sca10}\citep{Gro12}.
The FCFOV is 8°$\times$8°, the half-coded EXFOV 19°$\times$19°, and the zero-response one 29°$\times$29°. 
The detector pixel pitch 
(and therefore the reconstructed sky pixel for the decoding process described above) 
subtends an angle of 5$'$ on the sky, while the mask element an angle of 12$'$.
With a ratio mask element to pixel pitch of 2.43, IBIS is expected to have an average 
image efficiency (at source location) of 86$\%$; an angular resolution, for weighted reconstruction, 
of 13$'$~FWHM in the FCFOV; and a localization better than 0.5$'$ at SNR~$>30$. 
The width of the SPFS however varies wildly along the PCFOV due to the secondary lobes 
as shown in Fig.~\ref{fig:ibis-spsf-psle} left \citep{Gro03}.
The localization error as measured at the beginning of the mission 
(Fig.~\ref{fig:ibis-spsf-psle} right) \citep{Gro03},  
even if it followed well the expected 1/SNR trend and 
reached values of less than 1~arcmin at SNR~$>30$, 
was not as good as the theoretical curve and stalled at a constant level of 20$''$
even for very high SNR. 
With the improvement of the analysis software and the reduction of systematic effects, 
the PSLE was significantly reduced \citep{Sca10}\citep{Gro12} and reaches now about 40$''$ at SNR 30.
Figure~\ref{fig:IBIS-perf} reports the PSLE as a function of the source SNR obtained 
at later stages of the mission.

Systematic effects like pixel on/off, absorption by different detector structures, 
mask vignetting, and absorption by mask elements including screws and glue and in the  
mask support honeycomb structure have been studied along the years. 
All shall be accurately accounted for in source modelling. In fact while the 
MURA optimum system provides clean and narrow SPSF in the FCFOV, 
it also creates strong ghosts and coding noise in particular along the image axis 
passing through the source position, 
which must be removed in order to search for weaker excesses.
An iterative algorithm of search, modelling, and removal of sources is implemented 
in OSA (Fig.~\ref{fig:IBIS-iros}) in order to clean the images before summing
them in sky mosaics (Fig.~\ref{fig:IBIS-mosa}).

\begin{figure}[h]
	\centering
 \includegraphics[width=0.40\textwidth]{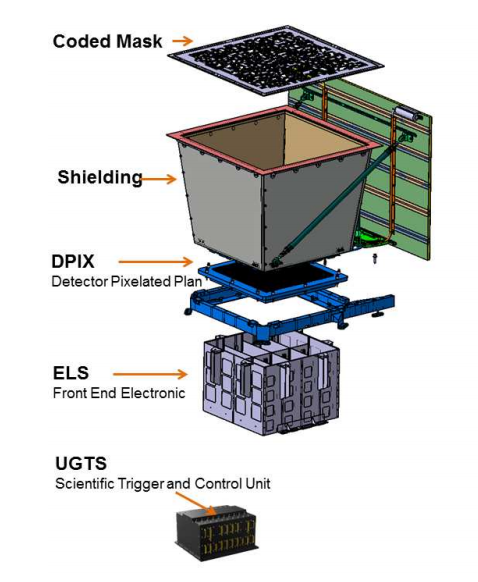}
 \includegraphics[width=0.59\textwidth]{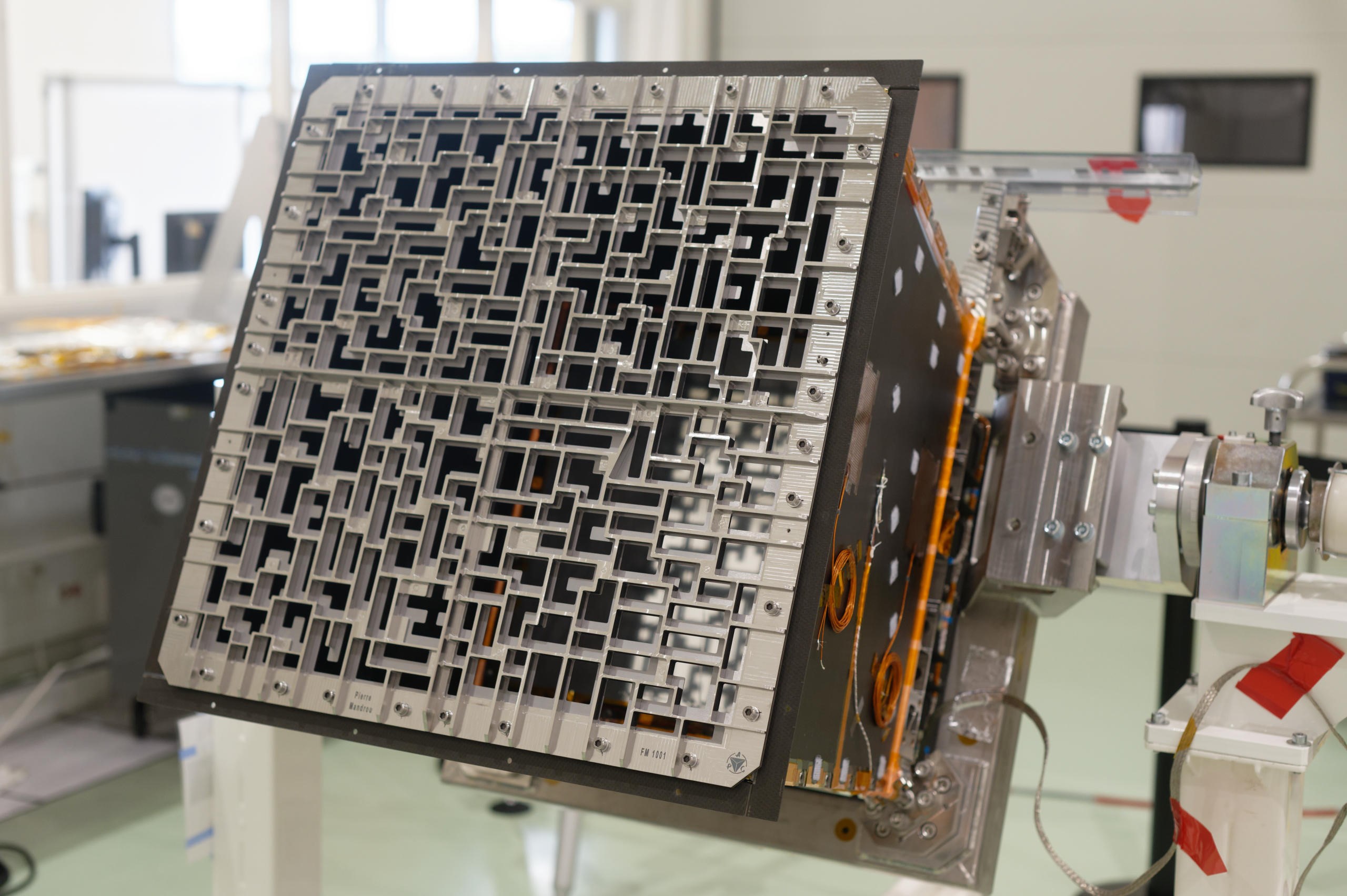}
	\caption{The ECLAIRs/SVOM coded mask instrument.
	Left: A scheme of the instrument showing the different elements (from \citep{God14}).
	Right: The ECLAIRs Mask mounted on the instrument at the CNES premises (credit CST/CNES).}
	\label{fig:ECLAIRs}
\end{figure}

\subsection{ECLAIRs on SVOM: the Next Coded Mask Instrument in Space}  \label{subsec:Ecla}
The Chinese-French SVOM (Space-based multi-band astronomical Variable Object
Monitor) space mission \citep{Wei16}\citep{Cor15}, planned, today, for a launch in 2023-2024, 
is a multi-wavelength observatory dedicated to the astrophysics of GRBs
and of the high-energy variable sky.
Between the four instruments of the payload, 
the hard X-ray coded mask imager ECLAIRs \citep{God14} (Fig.~\ref{fig:ECLAIRs}),
operating in the 4-150~keV energy range, 
will autonomously detect onboard GRBs and other high-energy 
transients providing their localization to the ground (through a fast VHF system)
and triggering on board,
under certain criteria, the slew of the platform in order to point in few minutes the 
SVOM narrow-field telescopes working in X-rays 
(Micro X-ray channel plate Telescope, MXT) and in the optical (VT) 
toward the event.

The ECLAIRs detection plane is made of 6400 pixels of Schottky type CdTe 
(4$\times$4~mm$^2$, 1~mm thick) for a total geometrical area (including dead zones)
of $\approx$ 1300~cm$^2$.
A 54$\times$54~cm$^2$ coded mask with 40$\%$ open fraction 
is located 46~cm above the detection plane
to observe a FOV of 2~sr (zero coded) with an angular resolution of 90~arcmin (FWHM). 
A passive lateral Pb/Al/Cu-layer shield blocks radiation and particles
coming from outside the aperture.
Sky images will be reconstructed in maps of 199$\times$199 square pixels 
with angular size ranging from 34$'$ on-axis down to 20$'$ at the edges of the FOV. 
ECLAIRs provides a sensitive area of $\approx$~400~cm$^2$, 
a point source localization error better than 12$'$ 
for 90$\%$ of the sources at the detection limit and is expected to detect each year about 70~GRBs, 
several non-GRB extra-galactic transients, 
dozens of AGNs and hundreds of Galactic X-ray transients and persistent sources. 
Its low energy threshold of 4~keV will open to SVOM the realm of extra-galactic soft X-ray transients, 
such as X-Ray flashes or SN shock breakouts, which are still poorly explored,
and in particular will allow the detection and study of cosmological GRBs 
whose emission peak is red-shifted in the X-ray band.
%
\begin{figure}[h]
	\centering
 \includegraphics[width=0.46\textwidth]{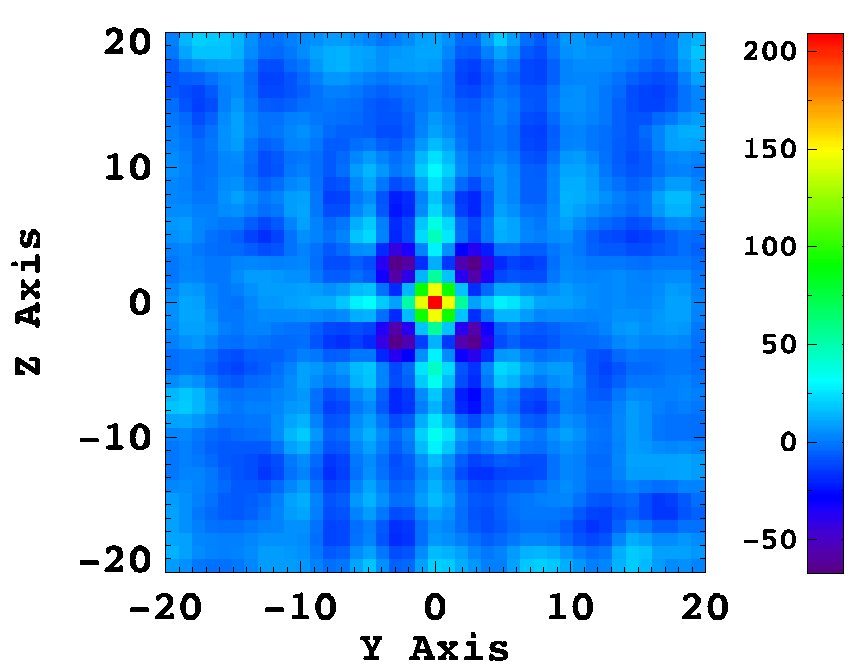}
 \includegraphics[width=0.53\textwidth]{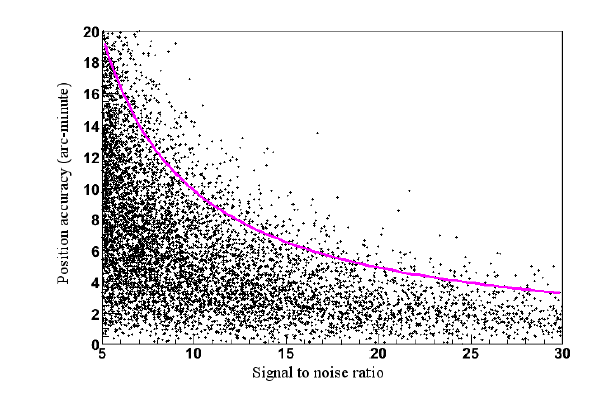}
	\caption{ECLAIRs/SVOM imaging performance.
	Left: Central part of the SPSF for on-axis position. 
	Lobes close to the central source peak are present.
	Right: Expected PSLE radius at 90\% c.l. versus the source signal to noise. 
	Source offsets from  true positions obtained from simulations are indicated by black dots and 
	the fitted PSLE, with 1/SNR trend, by the solid line (from \citep{God14}).
	}
	\label{fig:ECL-perf}
\end{figure}

The 46$\times$46 square mask elements have linear size 2.53 times the detector pixel pitch,
and their distribution follows an optimized quasi-random 
pattern chosen by requiring connection between elements in order to allow
the mask to be auto-sustained.
Thousands of quasi-random patterns of this kind with a 40\% aperture, 
which optimizes performance at these energies, were generated and studied,
using the formulae of error estimation for general masks (mentioned 
but not explicitly given in \S~\ref{subsubsec:Erro}),
in order to select the one presenting the best 
compromise between sensitivity and source localization for the GRB science,
compatible with the mechanical criteria.
The performance as function of the resolution parameter $r$
for the specific chosen mask pattern is shown in Fig.~\ref{fig:perfs-cmi} right 
along with the relative values,
at the selected resolution factor which was chosen in order to optimize 
the system for the scientific objectives of the mission.

For the chosen design, the predicted imaging performances are shown 
in Fig.~\ref{fig:ECL-perf}. Left panel shows the peak of the SPSF that, 
given the non-optimum system based on a quasi-random mask,
does present relevant side-lobes even in the center of the FCFOV.
Once the source is detected and positioned, the lobes must be cleaned 
by an IROS procedure in order to search for weaker sources.
Right panel shows the localization error curve from simulations of sources at different SNR.
The accuracy is expected to be within half the size of the VT FOV 
($\approx$~26$'$) in order to always have the event within both the optical and the X-ray telescope FOVs
after the slew of the platform to the ECLAIRs measured GRB position.

\begin{figure}[h]
	\centering
 \includegraphics[width=1.0\textwidth]{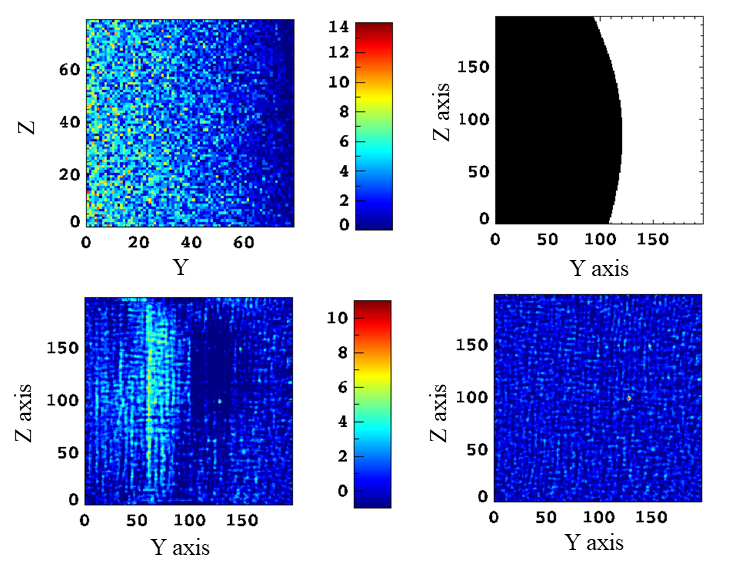}
	\caption{Effect of CXB modulation by partial Earth occultation of the ECLAIRs FOV.
	From left to right and top to bottom: 
	Simulated detector image of the Earth-modulated CXB in 20~s exposure
	(during which the Earth can be considered stable in the FOV). 
	Configuration of Earth occultation of ECLAIRs FOV considered in the simulation.
	Decoded SNR sky image of the simulated detector image and including 
	two not-obscured sources 
	when background is not corrected: large modulation is present, 
	and the sources are not easily detected.
	Decoded SNR sky image when proper model of CXB modulated by the Earth is used for
	the background correction: the reconstructed image is flat, 
	and the two sources are detected as the highest peaks.
	}
	\label{fig:ECLAIRs-Earth}
\end{figure}

The actual mask, integrated in the ECLAIRs instrument, is shown in 
Fig.~\ref{fig:ECLAIRs} right.
It is composed by a Ti-Ta-Ti sandwich with the tantalum providing
the main absorbing power and the titanium the mechanical strength. 
A central opaque cross, of width 1.4 times the mask element size, is added,
along with fine titanium supporting structures running along the mask elements 
on the side which avoids vignetting of off-axis sources,
to make the overall structure resistant to the expected vibration amplitudes of the launch. 
The design of the ECLAIRs mask has been optimized in this way 
in order to allow the instrument to be sensitive at energies as low as 4~keV.
That requires to have a solid self-sustained mask without a support structure 
that would absorb the radiation passing through the mask open elements. 
The multi-layer thermal coating insulation that envelops the telescope in order 
to protect the camera from light and micro-meteoroids will stop however
X-rays below 3-4~keV and determines the low-energy threshold of the instrument.

One particular feature of the SVOM mission is that the general program observations,
during which data on other sources are collected while waiting to detect GRB events, 
will be scheduled giving priority to an attitude law that 
optimizes the search of GRB. SVOM will generally point opposite to the sun, 
toward the Earth night, so that detected GRB can be rapidly observed with ground-based 
observatories, and also, to reduce noise, will avoid the bright Galactic plane 
and rather observe the sky Galactic poles.
These constraints and the low Earth orbit of the satellite ($\approx$~650~km) 
will lead to a frequent and variable occultation of the instrument FOVs by the Earth.
ECLAIRs will then often experience partial Earth occultation of its large FOV
during which the CXB will be modulated in a variable way during the $\approx$~90~min orbit. 

Figure~\ref{fig:ECLAIRs-Earth} shows a simulation of the expected spatial modulation 
on the detector by this effect, the impact on the imaging performance,
and the expected correction results.
Given the uncertainties of the CXB model and the additional components of Earth albedo 
and reflection, the background correction of ECLAIRs images affected by the Earth 
in its FOV in real conditions will certainly be challenging.

However CMI have been proven to be robust and effective imaging systems, 
and new exciting results are expected from this novel coded mask-based high-energy mission, 
dedicated to the transient sky, that will be launched soon.

\section{Summary and Conclusions}

In this review we have described the concept of coded mask instrument for gamma-ray astronomy, 
discussed the mask patterns, and introduced definitions and terminology useful
to understand the large literature on the subject.
We have illustrated the correlation analysis procedure to apply to the data of 
standard CM systems, along with some practical recipes for the analysis. 
We have provided the formulae to evaluate errors and performance 
of the systems from their design parameters and illustrated them 
with data, simulations, or calculations for some of the CMI presently in operation 
or in preparation.
Finally we have described the historical development of the field and
the main CMI implementations in space missions and recalled some of the important 
results obtained by these devices in imaging the soft gamma-ray sky.

Even if focusing techniques, with their power to reduce background and to reach
arcsec scale angular resolutions, are more and more extended to high energies 
and are definitely performing well in exploring the most dense sky regions 
like the Galactic center, they are limited to narrow fields,
and CMI remain the best options for the simultaneous monitoring of large sky regions.

Since the X-/gamma-ray sky is dominated by compact objects which are, most of the time, 
very variable and even transient, these surveys are crucial to explore this realm, 
especially in the new era of time-domain and multi-messenger astronomy.
Indeed rapid localization at moderate resolutions of high-energy electromagnetic counterparts
of gravitational wave or neutrino burst events, which have large positional uncertainties,
can trigger the set of high-resolution observations with narrow-field instruments
which finally lead to identification of the events.
This is what happened for the very first identified GW source (GW170817),
and this is the strategy envisaged for the next multi-messenger campaigns of observations.
The reaction time for the follow-up of fast transients is obviously 
very important; therefore the way to go is to couple imaging wide-field monitors
with a multi-wavelength set of space or/and ground-based narrow-field telescopes
with fast autonomous capability to point the sky positions provided by the monitors. 
This strategy first implemented by Swift and ready to be used by SVOM is still 
based on CMI.

Another technique that is emerging to design large field of view X-ray telescopes 
is the so-called \textbf{lobster-eye} or \textbf{micro-pore} optics (MPO). 
The concept, taken from the optical system of the eyes of crustaceans,
is to use grazing reflection by the walls of many, very small
channels to concentrate X-rays toward the focal plane PSD. 
By disposing a wide micro-channel plate with a large number 
of micro-holes with very polished and flat reflecting walls 
over a properly curved surface, a focusing system with a large FOV can be obtained.
MPO is used for the MXT of SVOM \citep{Goe14} 
that will obtain X-ray images with arcmin angular resolution over 1° FOV, 
but larger systems are now being developed, e.g. for the Einstein Probe mission \citep{Yua18}.

However MPO technique is for now limited to low X-ray energies,
and projects for future high-energy missions dedicated to variable sky 
still plan to implement coded mask wide-field cameras, 
as, for example, the set of orthogonal 1-d cameras in the Chinese-European eXTP 
(enhanced X-ray Timing and Polarimetry) mission \citep{Zha19},
or the two full 2-d cameras of the XGIS (X-Gamma ray Imaging Spectrometer) 
instrument \citep{Lab20} of the Theseus 
(Transient High-Energy Sky and Early Universe Surveyor) \citep{Ama21} project, 
proposed recently to ESA for a medium size mission (M7) of the Cosmic Vision program. 
This last instrument, based on a random mask coupled to Silicon Drift Detectors,
features a combined ZR-EXFOV of 117°$\times$77°, 
an AR of 120$'$ and a localization accuracy of 15$'$ at SNR~=~7 in the 2-150 keV range.
A rather complex CMI devoted to soft gamma-rays is also
proposed for a future NASA explorer mission, GECCO, an observatory
working in the 50~KeV-10~MeV range that combines coded mask imaging at 
low energies and a Compton mode system for the high energies \citep{Orl21}.
A specific feature of this CMI is its capability of deploying after launch 
a mast, which can extend the mask to any distance from the detector up to 20~m, 
in order to reach the desired imaging performances by tuning this 
system parameter ($H$ in Table~\ref{tab:cm-prop}).

In conclusion, 
coded mask instruments are very efficient devices to carry out imaging surveys of the 
hard X-ray/soft gamma-ray sky over large fields of view
with moderate angular resolution and localization power and are still
considered for future missions dedicated to the time-domain astronomy, 
for which lighter and more agile system are now designed.
While several such instruments are still presently in operation 
on INTEGRAL, Swift and ASTROSAT,
the next CMI to fly soon is ECLAIRs, on the SVOM multi-wavelength mission, 
which pushes the technique to cover a large energy band 
from X to hard X-rays and is expected to provide exceptional
results on GRB and the transient sky science.

\section*{Acknowledgments}

We thank all the colleagues who have shared with us (and some still do) the exciting
adventure of exploiting coded aperture imaging in high-energy astronomy.


\end{document}